\documentclass[preprintnumbers,nofootinbib,
superscriptaddress,amsmath,floatfix,prd]{revtex4}

\usepackage{amssymb}  
\usepackage{graphicx,subfigure}
\usepackage{exscale}
\usepackage{textcomp}
\usepackage{enumerate}

\usepackage{color}

\usepackage[normalem]{ulem}  

\newcommand{\non}{\nonumber\\}

\newcommand{\be}{\begin{equation}}
\newcommand{\ee}{\end{equation}}
\newcommand{\bea}{\begin{eqnarray}}
\newcommand{\eea}{\end{eqnarray}}
\newcommand{\ba}[1]{\begin{array}{#1}}
\newcommand{\ea}{\end{array}}

\newcommand{\Tr}{{\rm Tr}}

\newcommand{\Del}{\nabla}
\newcommand{\eqn}[1]{(\ref{#1})}

\begin{document}

\title{Role reversal in first and second sound in a relativistic superfluid}

\author{Mark G.\ Alford}
\email{alford@wuphys.wustl.edu}
\affiliation{Department of Physics, Washington University St Louis, MO, 63130, USA}

\author{S. Kumar Mallavarapu}
\email{kumar.s@go.wustl.edu}
\affiliation{Department of Physics, Washington University St Louis, MO, 63130, USA}

\author{Andreas Schmitt}
\email{aschmitt@hep.itp.tuwien.ac.at}
\affiliation{Institut f\"{u}r Theoretische Physik, Technische Universit\"{a}t Wien, 1040 Vienna, Austria}

\author{Stephan Stetina}
\email{stetina@hep.itp.tuwien.ac.at}
\affiliation{Institut f\"{u}r Theoretische Physik, Technische Universit\"{a}t Wien, 1040 Vienna, Austria}

\date{18 March 2014}

\begin{abstract}

Relativistic superfluidity at arbitrary temperature, chemical potential and (uniform) superflow is discussed within a self-consistent field-theoretical approach. 
Our starting point is a complex scalar field with a $\varphi^4$ interaction, for which we calculate the 2-particle-irreducible effective action in the Hartree approximation. With this underlying microscopic theory, we can 
obtain the two-fluid picture of a superfluid, and compute properties
such as the superfluid density and 
the entrainment coefficient for all temperatures below the critical temperature for superfluidity. 
We compute the critical velocity, taking into account the full self-consistent effect of the temperature and 
superflow on the quasiparticle dispersion. We also discuss first and second sound modes and how first (second) sound evolves from a density (temperature) wave at low temperatures to a
temperature (density) wave at high temperatures. This role reversal is investigated for ultra-relativistic and near-non-relativistic systems for 
zero and nonzero superflow. For nonzero superflow, we also observe a role reversal as a function of the direction of the sound wave. 

\end{abstract}


\maketitle


\section{Introduction}
\label{intro}

Superfluid matter is likely to exist in the interior of compact stars. Neutrons in the core and/or the inner crust of a neutron star
as well as quarks in the core of a hybrid star may become a superfluid through Cooper pairing. While Cooper-paired neutron matter spontaneously breaks the $U(1)$
symmetry associated with baryon number conservation, Cooper-paired quark matter may or may not break this symmetry, depending on the pairing pattern. 
The color-superconducting phases that do break 
this $U(1)$ and thus are expected to behave as a superfluid are the color-flavor locked (CFL) \cite{Alford:1998mk,Alford:2007xm} and color-spin locked 
(CSL) \cite{Schafer:2000tw,Schmitt:2004et} phases. (In the kaon-condensed CFL phase, another $U(1)$ associated with strangeness is broken additionally,
suggesting a two-component superfluid. This $U(1)$, however, is only an approximate symmetry because of the weak interactions.) 
There are several observable phenomena in the physics of compact star 
that are sensitive to the hydrodynamics of these superfluids, such as pulsar glitches \cite{Anderson:1975} and the $r$-mode 
instability \cite{Andersson:1997xt}. Consequently, it is important to develop the superfluid hydrodynamics of nuclear and quark matter, in a relativistic framework. 
While superfluid quark matter must be treated relativistically, relativistic corrections to nuclear matter are smaller, but, at least for large densities, 
not negligible.

For a hydrodynamic description of a superfluid one usually employs a two-fluid approach \cite{tisza38,landau41} whose microscopic input is obtained by 
computing the response of the system to, and the behavior in the presence of, a superflow. By superflow we shall always mean a relative flow between the 
superfluid and the normal 
fluid. In field-theoretical language, a superflow is generated by a nonzero spatial gradient of the phase of the condensate. 
The microscopic input to the two-fluid formalism was derived from field theory in our recent work \cite{2013PhRvD..87f5001A};
see Refs.\ \cite{Carter:1995if,Comer:2002dm,Mannarelli:2008jq,Nicolis:2011cs,Andersson:2013jga} for similar studies. Instead of starting from a fermionic theory 
that describes neutrons or quarks, 
we considered the simpler, and more general, situation of a complex scalar field with quartic interactions. This model, which we also use in this paper, can be viewed
as an effective description of a more microscopic, fermionic theory.

In Ref.\ \cite{2013PhRvD..87f5001A} we restricted ourselves to the low-temperature, weak-coupling limit of a dissipationless, homogeneous superfluid. The present paper is an extension of that work in that we now consider arbitrary temperatures up to the critical temperature, and go beyond the weak-coupling limit
by resumming certain contributions to all orders in the coupling constant.  We still neglect dissipation and keep the superflow uniform in time and space.
The extension to high temperatures requires a more elaborate treatment than the standard one-loop effective action of Ref.\ \cite{2013PhRvD..87f5001A}. The 
reason is that the condensate has to be determined self-consistently, whereas in the simplest low-temperature approximation the temperature dependence
of the condensate can be neglected. We shall use the 2-particle irreducible (2PI) formalism \cite{Luttinger:1960ua,Baym:1962sx,Cornwall:1974vz}
(also called Cornwall-Jackiw-Tomboulis (CJT) formalism or $\Phi$-derivable approximation scheme) in the Hartree approximation at two-loop level. 
This formalism is particularly well suited to systems with spontaneously broken symmetries. It has been used previously, among many other applications, to describe 
meson condensation in the CFL phase \cite{Andersen:2006ys,Alford:2007qa,Alford:2008pb,Andersen:2008tn}, but without
including the effects of a superflow. To our knowledge, this is the first time that a superflow has been implemented in this formalism. 

The extension to all temperatures below the critical temperature for superfluidity is relevant in the context of compact stars because temperatures in the star may 
well be of the order of the critical temperature or higher. Typically, a compact star is born with a temperature of order $10\,{\rm MeV}$ and cools down quickly 
to temperatures in the keV range. For neutron matter, recent observations suggest a (density-dependent) critical temperature for superfluidity of 
about $10\,{\rm keV}$ at most \cite{Page:2010aw,Shternin:2010qi}. 
For the CSL phase of quark matter, the critical temperature may be even lower \cite{Schmitt:2002sc}, so again it is important to understand it at all
temperatures up to the critical temperature.
For CFL quark matter, on the other hand,
the critical temperature 
is expected to be higher, of the order of $T_c\sim 10\, {\rm MeV}$,
so after the very early stages of the proto-neutron star
any CFL matter will be well described by a low- or even zero-temperature 
approximation.

Since our general model makes no particular reference to nuclear or quark matter, it is also interesting to consider the application of our results to 
non-relativistic systems such as liquid helium \cite{khala,pines} or ultra-cold atoms \cite{2010NJPh...12d3040H,2013arXiv1302.2871S}.  By 
varying the mass of the bosons we can continuously extrapolate between the ultra-relativistic and non-relativistic limits.
We shall elaborate 
on the speeds of first and second sound and
the nature of first and second sounds as density or entropy waves or a mixture of the two.
In superfluid $^4$He, the first (second) sound is predominantly a density (entropy) wave, for almost all temperatures. The situation is more complicated
in both ultra-cold atoms and our case of weakly-interacting bosons with four-point interaction, where by increasing the temperature
the roles of first and second sound may be almost completely reversed. We shall discuss this role reversal in detail and point out that it
may also happen as a function of the direction of the sound wave, if there is a nonzero superflow.
We can also compare our results to those obtained from holographic 
models which are being used to explore the strong-coupling limit  of nonzero-temperature superfluidity \cite{Herzog:2008he,Herzog:2009md}. For instance, we shall 
discuss the phase diagram in the plane of temperature and superfluid velocity that can be viewed as a field-theoretical analogue of the one recently discussed in an 
AdS/CFT approach \cite{Amado:2013aea}.

Our paper is organized as follows. In Sec.\ \ref{sec:setup} we explain the setup, i.e., we write down the effective action and the stationarity equations, we 
explain the renormalization (details in appendix \ref{AppA}) and a modification of the stationarity equations that ensures the validity of the Goldstone theorem. 
Sec.\ \ref{sec:basic} is independent of the field-theoretical calculation and introduces the basic properties of a superfluid. These properties are then 
computed numerically and discussed in Sec.\ \ref{sec:results}. We have divided this section into four parts: 
an explanation of the calculation in Sec.\ \ref{sec:calc}; the condensate and critical velocity in Sec.\ \ref{sec:results1}; the superfluid and normal-fluid densities 
and the entrainment coefficient in Sec.\ \ref{sec:results2}; the sound velocities in Sec.\ \ref{sec:results3}. 
We summarize our results and give an outlook in Sec.\ \ref{sec:sum}.

\section{Setup}
\label{sec:setup}

\subsection{Effective 2PI action and stationarity equations}

We consider the following Lagrangian for a complex scalar field $\varphi$,
\be
{\cal L} = \partial_\mu\varphi\partial^\mu\varphi^* - m^2|\varphi|^2 -\lambda |\varphi|^4 \, ,
\ee
with mass parameter $m>0$ and coupling constant $\lambda>0$. This Lagrangian is $U(1)$ symmetric, and we are interested in the superfluid
state, which spontaneously breaks this $U(1)$. In such a state a Bose-Einstein condensate is formed whose modulus and phase we denote by $\rho$ and $\psi$, respectively. 
We are interested in a homogeneous superfluid where $\rho$ and $\partial_\mu\psi$ are constant in space and time. The constant gradient
of the phase can be identified with the superfluid four-velocity, 
\be
v^\mu = \frac{\partial^\mu\psi}{\sigma} \, , 
\ee
where 
\be
\sigma \equiv \sqrt{\partial_\mu\psi\partial^\mu\psi} = \mu\sqrt{1-{\bf v}^2} 
\ee
plays the role of the chemical potential in the rest frame of the superfluid, and 
\be
{\bf v} = -\frac{\nabla\psi}{\mu} 
\ee
is the superfluid three-velocity with $\mu$ being the chemical potential in the frame where the fluid moves with velocity ${\bf v}$. (In our finite-temperature 
calculation, this will be the rest frame of the normal fluid.) The chemical potential and the superfluid velocity are treated as external 
parameters, i.e., we are free
to choose the value for $\partial^\mu\psi$ as a boundary condition, while the condensate $\rho$ has to be determined as a function of the 
thermodynamic parameters $\mu$, ${\bf v}$, and the temperature $T$. As explained in Ref.\ \cite{Alford:2007qa}, computing the condensate for all temperatures
up to the critical temperature for superfluidity $T_c$ requires a self-consistent scheme: for a fixed $m<\mu$ one of the quasiparticle energies acquires unphysical 
negative values if the condensate is sufficiently small. Since the condensate is expected to melt away at the critical temperature, this problem will necessarily 
occur for sufficiently large temperatures. Therefore, the quasiparticle dispersions need to be computed self-consistently, and thus, in addition to computing the 
condensate, we need a self-consistency equation for the masses that enter the dispersion relations. This equation is the Dyson-Schwinger equation that 
can be derived from the 2PI effective action, which we shall use in the following. 

To write down the 2PI effective action, we first determine the tree-level potential 
$U(\rho)$ and the inverse tree-level
propagator $S_0^{-1}$. They are obtained by replacing the field $\varphi$ by the condensate plus fluctuations,
\be
\varphi \to \frac{\rho e^{i\psi}}{\sqrt{2}} + \varphi  \, .
\ee
Then, the tree-level potential is obtained by neglecting all fluctuations, 
\be
U(\rho) = -\frac{\rho^2}{2}(\sigma^2-m^2)+\frac{\lambda}{4}\rho^4 \, , 
\ee
while the inverse tree-level propagator in momentum space is obtained from the terms quadratic in the fluctuations, 
\be \label{S0inv}
S_0^{-1}(K) = \left(\begin{array}{cc} -K^2 -\sigma^2+m^2+3\lambda\rho^2  & 2iK_\mu\partial^\mu\psi  \\[2ex]
-2iK_\mu\partial^\mu\psi & -K^2 -\sigma^2+m^2+\lambda\rho^2 \end{array}\right)  \, . 
\ee
This $2\times 2$ matrix is given in the basis of real and imaginary parts of the transformed fluctuations $\varphi'=e^{-i\psi}\varphi$. This transformation 
is useful since otherwise the propagator would be non-diagonal in momentum space due to the space-time dependent phase $\psi$.
The four-momentum is $K=(k_0,{\bf k})$, $k_0=-i\omega_n$, with the bosonic Matsubara frequencies $\omega_n=2\pi n T$, $n\in \mathbb{Z}$. 

The 2PI effective action depends on the modulus of the condensate $\rho$ and the full propagator $S$. The effective action density ($T/V$ times the effective action) is
\be \label{Psifull}
\Psi[\rho,S] = -U(\rho) - \frac{1}{2}\frac{T}{V}\sum_K\Tr \ln \frac{S^{-1}}{T^2} - \frac{1}{2}\frac{T}{V}\sum_K\Tr[S_0^{-1}(\rho)S-1]-V_2[\rho,S] \, , 
\ee
where the trace is taken over the internal $2\times 2$ space and $V$ is the three-volume. For convenience, we first discuss the unrenormalized effective action and include
a counterterm $\delta\Psi$ later, see Sec.\ \ref{sec:renorm} and appendix \ref{AppA}.  
We work with the two-loop truncation, i.e., the potential $V_2[\rho,S]$ includes all two-loop, two-particle-irreducible, diagrams. Due to the condensate, 
there is an induced cubic interaction, whose vertex is given by $\lambda\rho$. We shall work in the Hartree approximation
in which the contribution of the corresponding diagram (``sunset diagram'') to $V_2$ is neglected. We are thus left with a single two-loop diagram 
(``double bubble diagram'') which is generated by the quartic interactions and whose algebraic expression is  
\be \label{V2}
V_2[S] \simeq \frac{\lambda}{4}\left(\frac{T}{V}\right)^2\sum_{K,Q}\Big\{\Tr[S(K)]\,\Tr[S(Q)] + \Tr[S(K)S(Q)] + \Tr[S(K)S(Q)^T]\Big\} \, .
\ee
Had we included the cubic interactions, $V_2$ would also depend explicitly on $\rho$. Moreover, the self-energy would depend on momentum. Therefore, neglecting the
contribution from the cubic interaction is a tremendous simplification, even though only an explicit calculation can show whether its contribution 
is indeed small. Naively, the additional factor of the condensate at the cubic vertex suggests that for chemical potentials only slightly 
above the mass $m$ our simplification is a good approximation. However, it was that shown the contribution we neglect is important to obtain a second order phase 
transition, i.e., 
the Hartree approximation shows, unphysically, a first order phase transition \cite{Baacke:2002pi,Marko:2012wc,Marko:2013lxa}.
We shall come back to this issue when we present our results in Sec.\ \ref{sec:results1}.

In our approximation the self-energy does not depend on momentum and is given by 
\be \label{Sigma}
\Sigma \equiv 2\frac{\delta V_2}{\delta S} \simeq \lambda\frac{T}{V}\sum_K \Tr[S(K)] + \lambda\frac{T}{V}\sum_K [S(K)+S(K)^T] \, , 
\ee
where the first term is proportional to the unit matrix. One can now easily confirm the useful relation  
\be \label{V24}
V_2[S] = \frac{1}{4}\frac{T}{V}\sum_K \Tr[\Sigma\, S(K)] \, .
\ee
To determine the ground state of the system, we need to find the stationary points of the effective action. To this end, we take the (functional) derivatives 
of the effective action with respect to $\rho$ and $S$ and set these to zero,
\begin{subequations}\label{stats}
\bea
0 &=& \frac{\partial U}{\partial \rho} + \frac{1}{2}\frac{T}{V}\sum_K\Tr\left[\frac{\partial S_0^{-1}}{\partial\rho} S\right] \, , \label{stat01}\\[2ex]
S^{-1} &=& S_0^{-1} + \Sigma \, . \label{DS}
\eea
\end{subequations}
With an ansatz for the full propagator we can bring these equations into a more explicit form. Within the present approximation, the most general form of the propagator 
is \cite{Alford:2007qa}
\be \label{Sinv}
S^{-1}(K) = \left(\begin{array}{cc} -K^2 -\sigma^2+ M^2+\delta M^2  & 2iK_\mu\partial^\mu\psi  \\[2ex]
-2iK_\mu\partial^\mu\psi & -K^2 -\sigma^2+ M^2- \delta M^2 \end{array}\right) \, , 
\ee
with two mass parameters $M$, $\delta M$, that have to be determined self-consistently. 
With this ansatz, the off-diagonal components of the Dyson-Schwinger equation (\ref{DS}) are automatically fulfilled. We are left with the scalar 
equation (\ref{stat01}) and the two diagonal components of Eq.\ (\ref{DS}). Inserting the first of the diagonal components into Eq.\ (\ref{stat01}), 
and adding and subtracting the two diagonal components to/from each other, yields the following (yet unrenormalized) three equations for the three variables 
$\rho$, $M$, and $\delta M$,
\begin{subequations} \label{stat1}
\bea
M^2+\delta M^2-\sigma^2 &=& 2\lambda \rho^2 \, , \label{stat11} \\[2ex]
M^2&=& m^2+2\lambda\rho^2 +2\lambda\frac{T}{V}\sum_K[S_{11}(K)+S_{22}(K)] \, ,  \label{stat12}\\[2ex]
\delta M^2 &=& \lambda\rho^2+\lambda\frac{T}{V}\sum_K[S_{11}(K)-S_{22}(K)] \label{stat13}\, ,
\eea
\end{subequations}
where $S_{11}(K)$ and $S_{22}(K)$ are the diagonal elements of the full propagator $S$, 
and where we have already assumed that the condensate $\rho$ is nonzero (there is also the trivial solution $\rho=0$ which we briefly discuss in the 
context of renormalization, see appendix \ref{AppA}).  
With the help of Eqs.\ (\ref{V24}) and (\ref{DS}), the pressure at the stationary point can be written as 
\be\label{Psistat}
\Psi_{\rm stat} = -U - \frac{1}{2}\frac{T}{V}\sum_K\Tr \ln \frac{S^{-1}}{T^2} - \frac{1}{4}\frac{T}{V}\sum_K\Tr[S_0^{-1}S-1] \, . 
\ee

\subsection{Renormalized stationarity equations and pressure}
\label{sec:renorm}

Renormalization in the 2PI formalism 
has been discussed in numerous works in the literature, for instance in  
Refs.\ \cite{Jackiw:1974cv,Baym:1977qb,AmelinoCamelia:1997dd,Lenaghan:1999si,vanHees:2001ik,Blaizot:2003br,Blaizot:2003an,Ivanov:2005bv,Berges:2005hc,Andersen:2006ys,Fejos:2007ec,Andersen:2008tn,Seel:2011ju,Pilaftsis:2013xna,Marko:2013lxa}. For our purposes, the methods developed and used in 
Refs.\ \cite{Andersen:2006ys,Fejos:2007ec,Andersen:2008tn} are most useful. While Ref.\ \cite{Fejos:2007ec} introduces
counterterms ``directly'' in the effective action, Refs.\ \cite{Andersen:2006ys,Andersen:2008tn} use an iterative method, based on 
Refs.\ \cite{Blaizot:2003br,Blaizot:2003an}, where the counterterms are introduced order by order in the coupling. Both methods are equivalent.
We shall follow the ``direct'' approach of Ref.\ \cite{Fejos:2007ec}. 
All details of the renormalization are discussed in appendix \ref{AppA}. Here we simply summarize the main steps and give the results. 

The renormalization requires to add appropriate counterterms to the effective action (\ref{Psifull}), proportional to the (infinite) parameters $\delta m^2$, 
$\delta \lambda_1$, $\delta\lambda_2$. In the condensed phase, two different parameters $\delta \lambda_1$, $\delta\lambda_2$
for the renormalization of the coupling are necessary. Then one can show, after regularizing the ultraviolet divergent integrals in the action and the 
stationarity equations, that the parameters $\delta m^2$, $\delta \lambda_1$, $\delta\lambda_2$ can be expressed in terms of the (finite) renormalized parameters $m^2$, 
$\lambda$, an ultraviolet cutoff $\Lambda$, and a renormalization scale $\ell$. And, importantly, these parameters do not depend on the medium, i.e., 
on $\mu$, $T$, and $\nabla\psi$. The relation between the cutoff dependent quantities and the renormalized ones becomes a bit more compact if we introduce (infinite) 
bare parameters via $m_{\rm bare}^2 = m^2+\delta m^2$, $\lambda_{1/2,{\rm bare}} = \lambda+\delta \lambda_{1/2}$. Then, we can write the renormalized parameters as 
\be 
\frac{1}{\lambda} = \frac{1}{\lambda_{1,{\rm bare}}}+\frac{1}{4\pi^2}\ln\frac{\Lambda^2}{\ell^2}
= \frac{1}{\lambda_{2,{\rm bare}}}+\frac{1}{8\pi^2}\ln\frac{\Lambda^2}{\ell^2} \, , \qquad 
\frac{m^2}{\lambda} = \frac{m_{\rm bare}^2}{\lambda_{1,{\rm bare}}}+\frac{\Lambda^2}{4\pi^2} \, .
\ee
For the regularization of the divergent momentum integrals we use Schwinger's proper time regularization \cite{Schwinger:1951nm}, where the cutoff $\Lambda$ is 
introduced by setting the lower boundary of the proper time integral to $1/\Lambda^2$. More precisely, we separate a ``vacuum'' contribution 
from each of the divergent integrals such that a finite integral remains and the ``vacuum'' term can be regularized. This term is not exactly a vacuum term because 
the ultraviolet divergences depend on the self-consistent mass $M$ (and thus implicitly on $\mu$, $T$, and $\nabla\psi$), and therefore the subtraction term must 
be (implicitly) medium dependent. In the presence of a superflow, we even find that the ultraviolet divergences depend explicitly on $\nabla\psi$, see discussion 
in Sec.\ \ref{sec:condsuper}.

To write down the result of the renormalization procedure we first introduce the following abbreviations for the momentum sums,
\be
I^\pm \equiv \frac{T}{V}\sum_K[S_{11}(K)\pm S_{22}(K)] \, , \qquad J \equiv  - \frac{1}{2}\frac{T}{V}\sum_K\Tr \ln \frac{S^{-1}}{T^2}  \, . 
\ee 
Then, the renormalized stationarity equations (\ref{stat1}) are 
\begin{subequations} \label{stat1ren}
\bea
M^2+\delta M^2-\sigma^2 &=& 2\lambda \rho^2 \, , \label{stat11ren} \\[2ex]
M^2&=& m^2+2\lambda\rho^2 +2\lambda I_{\rm finite}^+ \, ,  \label{stat12ren}\\[2ex]
\delta M^2 &=& \lambda\rho^2+\lambda I_{\rm finite}^- \label{stat13ren}\, ,
\eea
\end{subequations}
where the finite parts of the momentum sums are   
\begin{subequations} \label{Ifin}
\bea
 \label{Ip}
I^+_{\rm finite} &=& \frac{M^2}{8\pi^2}(\gamma-1) + \frac{M^2+\delta M^2}{16\pi^2}\ln\frac{M^2+\delta M^2}{\ell^2}
+ \frac{M^2-\delta M^2}{16\pi^2}\ln\frac{M^2-\delta M^2}{\ell^2}\non[2ex]
&&+\sum_{e=\pm}\int\frac{d^3{\bf k}}{(2\pi)^3}\left\{\frac{2[(\epsilon_{\bf k}^e)^2-k^2-M^2+\sigma^2][1+2f(\epsilon_{\bf k}^e)]}
{(\epsilon_{\bf k}^e+\epsilon_{-{\bf k}}^e)(\epsilon_{\bf k}^e+\epsilon_{-{\bf k}}^{-e})(\epsilon_{\bf k}^e-\epsilon_{\bf k}^{-e})}-\frac{1}{2\omega_{\bf k}^e}\right\} 
\, , \\[2ex]
\label{Im}
I^-_{\rm finite} &=& \frac{\delta M^2}{8\pi^2}(\gamma-1) + \frac{M^2+\delta M^2}{16\pi^2}\ln\frac{M^2+\delta M^2}{\ell^2} 
- \frac{M^2-\delta M^2}{16\pi^2}\ln\frac{M^2-\delta M^2}{\ell^2} \non[2ex]
&&+ \sum_{e=\pm}\int\frac{d^3{\bf k}}{(2\pi)^3}\left\{\frac{2\delta M^2[1+2f(\epsilon_{\bf k}^e)]}
{(\epsilon_{\bf k}^e+\epsilon_{-{\bf k}}^e)(\epsilon_{\bf k}^e+\epsilon_{-{\bf k}}^{-e})(\epsilon_{\bf k}^e-\epsilon_{\bf k}^{-e})}
-\frac{e}{2\omega_{\bf k}^e}\right\} \, , 
\eea
\end{subequations}
with the Euler-Mascheroni constant $\gamma\simeq 0.5772$, the Bose distribution function $f(x)=1/(e^{x/T}-1)$, and the quasiparticle excitations $\epsilon_{\bf k}^e$ 
that are given by the positive solutions of ${\rm det}\,S^{-1}=0$. The energies 
\be
\omega_{\bf k}^e = \sqrt{({\bf k}+e\nabla\psi)^2+M^2+e\delta M^2} 
\ee
appear in the ``vacuum'' subtractions whose regularized versions give rise to the (medium dependent) finite  
terms in the first lines of Eqs.\ (\ref{Ip}) and (\ref{Im}) and to (medium independent) infinite terms which are absorbed in the renormalized coupling constant and the 
renormalized mass.  

The renormalized version of the pressure at the stationary point is
\be \label{Psistatren}
\Psi_{\rm stat} = \frac{\rho^2}{2}(\mu^2-m^2) - \frac{\lambda}{4}\rho^4 + J_{\rm finite} 
+ \frac{(M^2-m^2-2\lambda\rho^2)^2}{8\lambda}+ \frac{(\delta M^2-\lambda\rho^2)^2}{4\lambda} \, ,
\ee
with $M$, $\delta M$, and $\rho$ being solutions of the stationarity conditions \eqn{stat1ren}, and
the finite part of the momentum sum  
\bea \label{Jfin}
J_{\rm finite} &=& \frac{M^4+\delta M^4}{64\pi^2}(3-2\gamma)-\frac{(M^2+\delta M^2)^2}{64\pi^2}
\ln\frac{M^2+\delta M^2}{\ell^2}-\frac{(M^2-\delta M^2)^2}{64\pi^2}\ln\frac{M^2-\delta M^2}{\ell^2} \non[2ex]
&&-\frac{1}{2}\sum_{e=\pm}\int\frac{d^3{\bf k}}{(2\pi)^3}\left[\epsilon_{\bf k}^e-\omega_{\bf k}^e +2T\ln\left(1-e^{-\epsilon_{\bf k}^e/T}\right)\right] \, .
\eea

\subsection{Goldstone mode}
\label{sec:gold}

The quasiparticle dispersion relations $\epsilon_{\bf k}^e$ are determined by the zeros of ${\rm det}\,S^{-1}$. In general, the dispersions  
are very complicated expressions because ${\rm det}\,S^{-1}$ is a quartic polynomial in $k_0$ which contains a linear term in the presence of a superflow 
$\nabla\psi$. Since condensation breaks the global U(1) symmetry of the Lagrangian spontaneously, we expect one massive mode and one Goldstone mode. For the Goldstone mode
we expect $k_0=0$ at ${\bf k}=0$. Setting ${\bf k}=0$ in the inverse propagator (\ref{Sinv}), we see that $k_0=0$ is only a zero of ${\rm det}\,S^{-1}$
if $M^2-\sigma^2-\delta M^2 =0$ (or if $M^2-\sigma^2+\delta M^2 =0$). 
However, this condition for the existence of a Goldstone mode is in contradiction to Eqs.\ (\ref{stat11ren}) 
and (\ref{stat13ren}) which imply $M^2-\sigma^2-\delta M^2 =-2\lambda I^-_{\rm finite}$, where $I^-_{\rm finite}$ might be small but does not vanish. 
Consequently, the Goldstone theorem is violated in our approach 
\cite{Baym:1977qb,AmelinoCamelia:1997dd,Lenaghan:1999si,Ivanov:2005bv,Ivanov:2005yj,Berges:2005hc,Andersen:2006ys,Alford:2007qa,Fejos:2007ec,Andersen:2008tn,Seel:2011ju,Pilaftsis:2013xna}. For our discussion of 
the superfluid properties it is crucial to work with an exact Goldstone mode. Therefore, we shall ignore the contribution from the momentum sum in 
Eq.\ (\ref{stat13ren}), thereby giving up the exact self-consistency of our approach \cite{Baym:1977qb,Alford:2007qa}. 
This is an ad hoc modification of the stationarity equations, i.e., we do not consider the true minimum of the 
full self-consistency equations, but a point away from this minimum. The benefit of this modification is that the Goldstone theorem is built into our 
calculation. Of course, our choice of enforcing the Goldstone theorem is not unique, and there are infinitely many ``Goldstone points'' once the exact 
self-consistency is sacrificed. A similar, but not identical, procedure is followed in Ref.\ \cite{Pilaftsis:2013xna}, where  
the stationary point in the constrained subspace given by the condition of an exact Goldstone mode is determined. Our
modification results in a particularly simple set of equations, because the three stationarity equations now reduce to two trivial ones and only one 
that still contains a momentum integral,
\begin{subequations}
\bea \label{statapp}
\delta M^2&=&\lambda\rho^2=M^2-\sigma^2 \, , \label{statapp1}\\[2ex]
2\sigma^2-m^2 &=& M^2+2\lambda I_{\rm finite}^+ \, . \label{statapp2}
\eea
\end{subequations}
The two dispersion relations $\epsilon_{\bf k}^\pm$ are then determined from 
\be \label{detS}
0 = {\rm det}\,S^{-1} = K^2[K^2-2(M^2-\sigma^2)]-4(K_\mu\partial^\mu\psi)^2 \, ,
\ee
where Eq.\ (\ref{statapp1}) has been used to eliminate $\delta M$.
Since we are interested in arbitrary temperatures below $T_c$, we shall need the full dispersion of both modes. Their explicit form is too lengthy 
to write down, but it is instructive to write down the linear part of the Goldstone mode, which is the only relevant excitation for sufficiently low temperatures, 
\bea \label{slope}
\epsilon_{\bf k}^+ &=& \frac{\sqrt{(M^2-\sigma^2)(M^2+\sigma^2+2[(\nabla\psi)^2-(\hat{\bf k}\cdot\nabla\psi)^2])}-2\partial_0\psi\,\hat{\bf k}\cdot\nabla\psi}
{M^2+\sigma^2+2(\nabla\psi)^2}\, k +\ldots  
\eea
For vanishing superflow we have $\sigma=\mu$ and obtain 
\bea \label{goldslope}
\epsilon_{\bf k}^+(\nabla\psi=0) &=& \sqrt{\frac{M^2-\mu^2}{M^2+\mu^2}}\, k + \ldots \, . 
\eea
We shall see that for low temperatures the slope of the Goldstone dispersion is identical to the speed of first sound. This is no longer true for larger temperatures.

If we work at a point that is not exactly the stationary point, we cannot use the pressure (\ref{Psistatren}). We rather have to evaluate the 
effective action density at the ``Goldstone point''. For the renormalization it is crucial that we only modify the finite part of the stationarity equations. 
Therefore, all infinities cancel in the same way as above, see appendix \ref{AppGold} for a more detailed discussion; for the pressure at the ``Goldstone point'' 
we then find 
\be \label{PsiGold}
\Psi_{\rm Gold} = \frac{(M^2-\sigma^2)(3\sigma^2-M^2-2m^2)}{4\lambda} + J_{\rm finite} +\frac{(M^2-2\sigma^2+m^2)^2}{8\lambda}
 - \frac{\lambda}{4}(I^-_{\rm finite})^2\, . 
\ee
Here we have already eliminated $\rho$ and $\delta M$ with the help of Eq.\ (\ref{statapp1}). 
The stationarity equation (\ref{statapp2}) and the pressure (\ref{PsiGold}) are the starting point for our calculations. We shall solve the stationarity
equation numerically for the self-consistent mass $M$, which in turn gives the condensate via $\lambda\rho^2 = M^2-\sigma^2$ as well as the dispersion relations of the 
Goldstone mode and the massive mode. We need the pressure for various thermodynamic derivatives that are needed to compute for instance the sound velocities, as we 
shall explain in the next section.

\section{Basic hydrodynamic quantities of a superfluid}
\label{sec:basic}

\subsection{Two-fluid picture, entrainment, and superfluid density}

Let us briefly recapitulate the basic hydrodynamic quantities of a superfluid, in particular their definitions in terms of field theory, as worked out
in Ref.\ \cite{2013PhRvD..87f5001A} (for short summaries see Refs.\ \cite{Alford:2012mv,Alford:2013ota}).
At nonzero temperature, a superfluid can be viewed as a system of two interacting fluids 
\cite{tisza38,landau41,2013PhRvD..87f5001A,1982PhLA...91...70K,1982ZhETF..83.1601L,Nicolis:2011cs,1992PhRvD..45.4536C,carter89,Carter:1995if,Son:2000ht,Comer:2002dm,Andersson:2006nr,Andersson:2013jga}. The corresponding 
two currents are the charge current $j^\mu$ and the entropy current $s^\mu$. While the charge current is exactly conserved due to the 
exact symmetry $U(1)$, the entropy current is in general not conserved due to dissipative effects. Here we neglect dissipation, such that both  currents are conserved.
Each current has an associated conjugate momentum. In the case of the charge current, this is the gradient of the phase of the condensate $\partial^\mu\psi$ (which 
follows directly from the field-theoretical definition of the Noether current). In the case of the entropy current, this is the thermal four-vector $\Theta^\mu$ whose 
temporal component is the temperature. These four four-vectors can be combined to a generalized, covariant thermodynamic relation between the generalized pressure 
$\Psi$ and the generalized energy density $\Lambda$,
\be
\Lambda = -\Psi + j_\mu\partial^\mu\psi + s_\mu\Theta^\mu \, , 
\ee
and the stress-energy tensor can be written as
\be
T^{\mu\nu} = -g^{\mu\nu}\Psi + j^\mu \partial^\nu\psi + s^\mu\Theta^\nu \, , 
\ee
with the Minkowski metric $g^{\mu\nu} = (1,-1,-1,-1)$. Only two of these four-vectors are independent of each other, and one is free to choose any two of them 
as the basic hydrodynamic variables. For instance, if one chooses the two momenta $\partial^\mu\psi$, $\Theta^\mu$ as the basic variables, the two currents are 
obtained via
\begin{subequations} \label{js}
\bea
j^\mu &=& \overline{\cal B} \,\partial^\mu\psi +\overline{\cal A} \,\Theta^\mu \, , \\[2ex]
s^\mu &=& \overline{\cal A} \,\partial^\mu\psi + \overline{\cal C} \,\Theta^\mu \, . 
\eea
\end{subequations}
The coefficients $\overline{\cal A}$, $\overline{\cal B}$, $\overline{\cal C}$ contain information about the 
microscopic physics\footnote{The notation is chosen to be consistent with Ref.\ \cite{2013PhRvD..87f5001A}
where the coefficients of the inverse transformation are denoted by ${\cal A}$, ${\cal B}$, ${\cal C}$.}. 
In particular, $\overline{\cal A}$ is a measure for the interaction between the two fluids and is thus called entrainment coefficient. The reason is that,
if  $\overline{\cal A}$ is nonzero, each current is not four-parallel to its own conjugate momentum (which would be the case in a single-fluid system), 
but also receives an admixture from the momentum associated with the other current. 
In the given choice of basic variables, the microscopic information is encoded in the generalized pressure which is, in the two-fluid formalism,
a function of the Lorentz scalars $\sigma^2$, $\Theta^2$, and $\Theta_\mu\partial^\mu\psi$, such that 
\be \label{entrainbar}
\overline{\cal A}\equiv \frac{\partial \Psi}{\partial(\Theta_\mu\partial^\mu\psi)} \, , \qquad 
\overline{\cal B}\equiv 2\frac{\partial \Psi}{\partial \sigma^2}
\,, \qquad \overline{\cal C}\equiv 2\frac{\partial \Psi}{\partial\Theta^2} \, .
\ee
Such a function $\Psi(\sigma^2,\Theta^2,\Theta_\mu\partial^\mu\psi)$ is usually not the starting point in field theory and thus Eqs.\ (\ref{entrainbar})
are not very useful for computing $\overline{\cal A}$, $\overline{\cal B}$, $\overline{\cal C}$. Nevertheless, one can compute these 
coefficients in field-theoretical terms. This ``translation'' was worked out in detail 
in Ref.\ \cite{2013PhRvD..87f5001A}, and we quote the main results that we need in our present context. The generalized pressure is identified with the 
effective action density, hence the notation $\Psi$ in the previous section. The coefficients $\overline{\cal A}$, $\overline{\cal B}$, $\overline{\cal C}$ are best
computed in terms of the superfluid density $n_s$, the normal-fluid density $n_n$, and elementary thermodynamic equilibrium quantities,
\be
\overline{\cal A} = \frac{n_n s}{w} \, ,\qquad \overline{\cal B} = \frac{n_n^2}{w}+\frac{n_s}{\sigma} \, ,\qquad \overline{\cal C} = \frac{s^2}{w} \, ,
\ee
where $s$ is the entropy density, $w=\mu n_n + sT$ the enthalpy density of the normal fluid, with $\mu=\partial_0\psi$, $T=\Theta_0$, $s=s_0$, and $n_n$
all measured in the rest frame of the normal fluid. As discussed in Ref.\ \cite{2013PhRvD..87f5001A}, this is the frame where the field-theoretical calculation is performed. 
The superfluid density $n_s$ is measured in the 
rest frame of the superfluid  -- such that $\mu/\sigma\,n_s$ is the superfluid density measured in the rest frame of the normal fluid -- and is computed from 
\be \label{ns}
n_s = -\sigma \frac{\nabla\psi\cdot{\bf j}}{(\nabla \psi)^2} \, . 
\ee
Here, ${\bf j}$ is the spatial part of the charge current $j^\mu$ that is given by the usual field-theoretical definition,
\be
j^\mu = \frac{\partial \Psi}{\partial(\partial_\mu\psi)} \, .
\ee
The normal fluid density is then computed from $n_n = n - \mu/\sigma\,n_s$, where $n=j^0$ is the charge density, measured in the normal fluid rest frame. 
In the original, non-relativistic context, superfluid and normal fluid densities (there: mass densities $\rho_n$, $\rho_s$) play a fundamental role 
since the current is divided into a superfluid and a normal part, ${\bf j}=\rho_n {\bf v}_n + \rho_s {\bf v}_s$. 
In the relativistic version of that decomposition, we can write the four-current as
\be
j^\mu = n_n u^\mu + n_s\frac{\partial^\mu\psi}{\sigma} \, , 
\ee
where the four-velocity of the normal fluid is related to the entropy current via $u^\mu = s^\mu/s$. Using the above classification into currents and conjugate momenta,
this formalism uses one current, namely $s^\mu$, and one momentum, namely $\partial^\mu\psi$, as its basic variables. 
This is different from the formalism introduced above and originally used in the relativistic context \cite{1982PhLA...91...70K,1982ZhETF..83.1601L}, where it is
more natural to work either with the two currents or the two momenta. Both formalisms are equivalent and can be 
translated into each other \cite{Herzog:2008he,2013PhRvD..87f5001A}. 

We shall compute $n_s$, $n_n$, $\overline{\cal A}$, $\overline{\cal B}$, $\overline{\cal C}$ within the 2PI formalism. The above definitions 
show that, to this end, we need the first derivatives of the pressure $\Psi$ with respect to $\mu$, $T$ and $|\nabla\psi|$. Even though we are 
also interested in the general case of a non-vanishing superflow, let us briefly discuss how the calculation simplifies in the limit $|\nabla\psi|\to 0$. In that case,
when we compute derivatives with respect to $T$ and $\mu$ we can set $\nabla\psi= 0$ straightforwardly. But, when we compute $n_s$ and $n_n$ we have to be more
careful. These quantities describe the response of the system to a superflow, i.e., even if we are eventually interested in the case $\nabla\psi\to 0$, we need to work 
initially with a nonzero superflow. We can write the superfluid density (\ref{ns}) for $\nabla\psi\to 0$ as 
\be \label{ns1}
n_s\Big|_{\nabla\psi=0} = -\mu\left(\frac{\partial^2 \Psi}{\partial|\nabla\psi|^2}\right)_{\nabla\psi=0}  \, ,
\ee
where we have expanded $\Psi$ in a Taylor series for small $|\nabla\psi|$. In this series we have dropped the linear term because the first derivative (i.e., the
current ${\bf j}$) vanishes for $|\nabla\psi|=0$, which is obvious physically and can also be checked explicitly. It seems that the derivatives with respect to 
$|\nabla\psi|$ are very complicated to compute because they involve the derivatives of the dispersion relations $\epsilon_{\bf k}^e$ which are 
contained in the momentum integrals in $I^\pm_{\rm finite}$ and $J_{\rm finite}$, see Eqs.\ (\ref{Ifin}), (\ref{Jfin}). However, 
since we know that $\epsilon_{\bf k}^e$ are the solutions to the quartic equation (\ref{detS}), we can simplify the calculation significantly by taking the first 
and second derivatives of Eq.\ (\ref{detS}). This yields
\begin{subequations}\label{deps2}
\bea
\left.\frac{\partial\epsilon_{\bf k}^e}{\partial|\nabla\psi|}\right|_{\nabla\psi=0} &=& \frac{2\mu k_\parallel}{(\epsilon_k^e)^2-k^2-M^2-\mu^2} \,, \label{d1eps}\\[2ex]
\left.\frac{\partial^2\epsilon_{\bf k}^e}{\partial|\nabla\psi|^2}\right|_{\nabla\psi=0} &=& 
\frac{(\epsilon_k^e)^2+2k_{\parallel}^2-k^2}{\epsilon_k^e[(\epsilon_k^e)^2-k^2-M^2-\mu^2]}
+ \frac{8\mu^2k_\parallel^2}{\epsilon_k^e[(\epsilon_k^e)^2-k^2-M^2-\mu^2]^2} - \frac{4\mu^2k_\parallel^2[3(\epsilon_k^e)^2-k^2-M^2-\mu^2]}
{\epsilon_k^e[(\epsilon_k^e)^2-k^2-M^2-\mu^2]^3} \, , \hspace{0.5cm}
\eea
\end{subequations}
where $k_{\parallel} = k\cos\theta$ with $\theta$ being the angle between $\nabla\psi$ and ${\bf k}$, and 
\be
\epsilon_k^e\equiv
\epsilon_{\bf k}^e(\nabla\psi=0) = \sqrt{k^2+M^2+\mu^2-e\sqrt{4k^2\mu^2+(M^2+\mu^2)^2}}   
\ee
are the excitation energies at vanishing superflow. Eqs.\ (\ref{deps2}) are very useful for the explicit calculation which is explained in Sec.\ \ref{sec:calc},
see also the tree-level calculation of the sound velocities in appendix \ref{AppB}.

\subsection{Sound velocities}    
\label{sec:sound}

The sound velocities are computed from the basic hydrodynamic equations. They can either be written as conservation equations for the 
energy-momentum tensor and the charge current, or alternatively as conservation equations for the two currents and the vorticity equation,  
\be \label{hydroeqs}
\partial_\mu j^\mu = 0 \, , \qquad \partial_\mu s^\mu = 0 \, , \qquad s_\mu (\partial^\mu\Theta^\nu - \partial^\nu\Theta^\mu) = 0  \, .
\ee
Starting from these equations, one considers small harmonic deviations from equilibrium and linearizes the equations in the amplitudes of these deviations. 
For example, one can choose to work with oscillations in chemical potential, temperature, and normal-fluid velocity, 
$\delta\mu = \delta\mu_0 e^{i(\omega t-{\bf k}\cdot{\bf x})}$, $\delta T = \delta T_0 e^{i(\omega t-{\bf k}\cdot{\bf x})}$, $\delta {\bf v}_n = 
\delta {\bf v}_{n,0} e^{i(\omega t-{\bf k}\cdot{\bf x})}$ with (complex) amplitudes $\delta\mu_0$, $\delta T_0$, and $\delta {\bf v}_{n,0}$. 
After eliminating $\delta {\bf v}_n$ algebraically, the equations can be reduced to two equations for $\delta\mu_0$, $\delta T_0$, which we can 
write compactly as \cite{2013PhRvD..87f5001A}  
\begin{subequations} \label{muTeqs}
\bea
0 &=& \Big[a_1 u^2+(a_2+a_4|\nabla\psi|^2\cos^2\theta)+a_3|\nabla\psi| u \cos\theta  \Big]\delta\mu_0 
+\Big(b_1 u^2 +b_2+b_3|\nabla\psi| u \cos\theta \Big)\,\delta T_0 \, , \\[2ex]
0 &=& \Big[A_1 u^2+(A_2+A_4|\nabla\psi|^2\cos^2\theta)+A_3|\nabla\psi| u \cos\theta\Big]\delta\mu_0 
+\Big(B_1u^2+B_2+B_3|\nabla\psi|u\cos\theta  \Big)\,\delta T_0 \, , 
\eea
\end{subequations}
where $u = \omega/k$ is the speed of sound, and $\theta$ is the angle between the direction of the sound wave and the superflow. In general, the coefficients of this 
system of equations are complicated combinations of first and second derivatives of the pressure. Their explicit expressions for the general case of a non-vanishing 
superflow are given in Eqs.\ (D16) of Ref.\ \cite{2013PhRvD..87f5001A}. Requiring the equations (\ref{muTeqs}) to have nontrivial solutions for $\delta\mu_0$, 
$\delta T_0$ yields a quartic equation for $u$ with two physical solutions\footnote{The wave equations, as derived here from Eqs.\ (\ref{hydroeqs}),
allow for exactly two physical solutions. Nevertheless, there are more possible sound waves in a superfluid. They can be found by starting from a certain 
subset of the conservation equations. For instance the so-called fourth sound \cite{1959PhRv..113..962A} can be excited by fixing the normal fluid by an external 
force. It is thus calculated after dropping momentum conservation \cite{khala,Yarom:2009uq,Herzog:2009md}.}, the velocities of first and second sound $u_1$ and $u_2$. 
The ratio of the amplitudes themselves are then computed from 
\be
\frac{\delta T_0}{\delta\mu_0} = -\frac{a_1 u^2+(a_2+a_4|\nabla\psi|^2\cos^2\theta)+a_3|\nabla\psi| u \cos\theta}{b_1 u^2 +b_2+b_3|\nabla\psi| u \cos\theta} 
\, , 
\ee
For each sound mode, the one-dimensional space of solutions of Eqs.\ (\ref{muTeqs}) is a straight line through the origin in the $\delta\mu_0$-$\delta T_0$ plane. 
It is convenient to define the angle of that line with the $\delta\mu_0$ axis, 
\be \label{alpha}
\alpha \equiv \arctan\frac{\delta T_0}{\delta\mu_0}  \, .
\ee
The sign of this angle tells us whether chemical potential and temperature oscillate in phase ($\alpha>0$) or out of phase ($\alpha<0$). 
The magnitude of $\alpha$ characterizes the mixture of oscillations in 
temperature and chemical potential with $\alpha=0$ corresponding to a pure oscillation in chemical potential and $|\alpha|=\pi/2$ to a pure oscillation in 
temperature. We can also translate this into the amplitudes in density and entropy. With the help of the thermodynamic relation for the pressure
$P$ 
\bea \label{diffP}
dP =  nd\mu +sdT -\frac{n_s}{\sigma}\nabla\psi\cdot d\nabla\psi  
\eea
we can derive (see also appendix D of Ref.\ \cite{2013PhRvD..87f5001A})
\begin{subequations}
\bea
\frac{\delta n_0}{\delta s_0} = \left[\frac{\partial n}{\partial \mu}+\frac{|\nabla\psi|\cos\theta}{u}\frac{\partial(n_s/\sigma)}{\partial\mu}
+\frac{\partial s}{\partial\mu}\frac{\delta T_0}{\delta \mu_0}\right]\left[\frac{\partial n}{\partial T}+\frac{|\nabla\psi|\cos\theta}{u}
\frac{\partial(n_s/\sigma)}{\partial T}+\frac{\partial s}{\partial T}\frac{\delta T_0}{\delta \mu_0}\right]^{-1} \, .   
\eea
\end{subequations}
In general, the sound modes and the corresponding amplitudes are very complicated. Let us therefore discuss the case of vanishing 
superflow, $|\nabla\psi|\to 0$. In this case, the coefficients $a_3$, $a_4$, $b_3$, $A_3$, $A_4$, $B_3$ become irrelevant, and 
\begin{subequations}\label{a1b1}
\bea
a_1 &=& \frac{w}{s}\frac{\partial n}{\partial T} \, , \qquad a_2 = -n_n \, , \qquad  
b_1 = \frac{w}{s}\frac{\partial s}{\partial T} \, , \qquad b_2=-s \, , \\[2ex]
A_1 &=& \mu\frac{\partial n}{\partial\mu}+T\frac{\partial n}{\partial T} \, , \qquad A_2 = -n \, , \qquad 
B_1 = \mu\frac{\partial s}{\partial\mu}+T\frac{\partial s}{\partial T} \, , \qquad B_2=-s \, .
\eea
\end{subequations}
We thus have the following simple quadratic equation for $u^2$,
\bea \label{au}
0 &=& (a_1 u^2+a_2)(B_1u^2+B_2)-(A_1 u^2+A_2)(b_1 u^2 +b_2) \, .
\eea
It is instructive to solve this equation in the limit where
there are no other energy scales than $\mu$ and $T$. In our context, this will be the case when we set the supercurrent and the mass parameter to zero,
$\nabla\psi=m=0$. Then, we can write the pressure as $\Psi=T^4h(T/\mu)$ with a dimensionless function $h$, and the sound velocities assume a simple
form \cite{Herzog:2008he}. The reason is that now there are simple relations between first and second derivatives of the pressure, for instance
we find $A_1=3n$, $B_1=3s$. Then, one computes the following two solutions of Eq.\ (\ref{au}) for $u^2$,
\be \label{conformal}
\mbox{scale-invariant limit:} \qquad u_1^2 = \frac{1}{3} \, ,\qquad  u_2^2 = \frac{n_ss^2}{w}\left(n\frac{\partial s}{\partial T} 
- s\frac{\partial n}{\partial T}\right)^{-1} \, .
\ee
We see that one solution is constant while the other depends on the thermodynamic details of the system. 
The ratios of the amplitudes become particularly simple in this limit. We find
\be \label{ampconf}
\left.\frac{\delta T_0}{\delta\mu_0}\right|_{u_1} = -\left.\frac{\delta n_0}{\delta s_0}\right|_{u_2} = \frac{T}{\mu} \, , \qquad 
\left.\frac{\delta n_0}{\delta s_0}\right|_{u_1} = -\left.\frac{\delta T_0}{\delta\mu_0}\right|_{u_2} =  \frac{n}{s} \, . 
\ee 
This result shows that, for a given pair of amplitudes, $\delta T_0$ and $\delta\mu_0$ or $\delta n_0$ and $\delta s_0$, first sound is always an in-phase 
oscillation while second sound is 
always an out-of-phase oscillation. Moreover, we can make an interesting observation regarding the magnitude of the amplitudes. At $T=0$, where also $s=0$, first sound 
is a pure chemical potential (and pure density) wave, while second sound is a pure temperature (and pure entropy) wave. This is no longer true for nonzero 
temperatures. If at the critical temperature $T\gg\mu$ and $s\gg n$, the roles of first and second sound completely reverse upon heating the superfluid from 
$T=0$ to $T=T_c$. We shall discuss this role reversal in more detail when we present our numerical results, see Sec.\ \ref{sec:results3}.   

Before we come to the numerical evaluation, let us compute the sound velocities in the low-temperature, weak-coupling approximation. In this case,
we can restrict ourselves to the tree-level propagator $S_0^{-1}$, and there is no need to solve any 
self-consistency equation. Moreover, only the Goldstone mode is relevant since the massive mode only becomes populated for sufficiently high temperatures. 
This is the approximation that was discussed in Ref.\ \cite{2013PhRvD..87f5001A}, where the sound velocities have been computed for non-vanishing superfluid
velocity $v$, but 
for a vanishing mass parameter, $m=0$. Here we keep $m$ since its effect as an additional energy scale will turn out to be interesting, and we can use large
values of $m$ to approach the non-relativistic limit. Since the expressions become very complicated if both $m$ and $v$ are nonzero, we present the results for $v=0$. 
We defer all details of the calculation to appendix \ref{AppB}. The final result for the two sound velocities up to quadratic corrections in the temperature is 
\begin{subequations}\label{utree}
\bea
u_1 &=& \sqrt{\frac{\mu^2-m^2}{3\mu^2-m^2}} +{\cal O}(T^4)\, , \\[2ex]
u_2 &=& \frac{1}{\sqrt{3}}\sqrt{\frac{\mu^2-m^2}{3\mu^2-m^2}} +\left(\frac{\pi T}{\mu}\right)^2
\frac{20\sqrt{3}\mu^6}{7(3\mu^2-m^2)^{3/2}(\mu^2-m^2)^{3/2}} +{\cal O}(T^4)\, .
\eea
\end{subequations}
The speed of first sound is identical to the slope of the low-energy dispersion of the Goldstone mode (\ref{goldslope}), if we replace $M$ by its tree-level
result $M^2 = 2\mu^2-m^2$ in that expression. 
The speed of second sound at zero temperature is simply $1/\sqrt{3}$ times the speed of first sound, even in the presence of a mass
$m$. The temperature corrections are positive, even though we expect the speed of second sound to decrease eventually and vanish at the critical temperature. This 
is indeed the case in the full calculation, see next section.

\section{Superfluid properties in the 2PI formalism}  
\label{sec:results}

\subsection{Explaining the calculation}
\label{sec:calc}

As we have seen in the previous section, besides solving the stationarity equation we need to compute the first and second 
derivatives of the pressure with respect to $T$, $\mu$, and $|\nabla\psi|$. The most direct way to do so is via brute force numerical evaluation, 
for instance with the method of 
finite differences. In order to obtain results less prone to numerical uncertainties, we compute the derivatives in the following semi-analytical way.  
First we note that the pressure 
depends explicitly as well as implicitly via $M$ on the relevant variables,
\be
\Psi = \Psi[M(T, \mu, |\nabla\psi|),T, \mu, |\nabla\psi|] \, ,
\ee
and each thermodynamic derivative we are interested in also sees the implicit dependence in $M$. (Had we only been interested in first derivatives {\it and} had we 
considered the exact solution of the stationarity equations -- sacrificing the Goldstone theorem -- we could have restricted ourselves to the explicit dependence, since 
then the derivative of the pressure with respect to the self-consistently determined mass would have vanished by construction at the stationary point.)
Denoting the variables by $x,y \in  \{T, \mu, |\nabla\psi|\}$, we can thus write 
\begin{subequations} \label{dxdy}
\bea
\frac{d\Psi}{dx} &=& \frac{\partial M}{\partial x}\frac{\partial\Psi}{\partial M} + \frac{\partial\Psi}{\partial x} \, , \\[2ex]
\frac{d^2\Psi}{dxdy} &=& 
\frac{\partial^2 M}{\partial x\partial y}\frac{\partial\Psi}{\partial M}+\frac{\partial M}{\partial x}\frac{\partial M}{\partial y}\frac{\partial^2\Psi}{\partial M^2} 
+\frac{\partial M}{\partial x}\frac{\partial^2\Psi}{\partial y\partial M}+\frac{\partial M}{\partial y}\frac{\partial^2}{\partial x\partial M}
+\frac{\partial^2\Psi}{\partial x\partial y} \, . 
\eea
\end{subequations}
The derivatives of $M$ can be obtained from taking the first and second derivatives of the stationarity equation (\ref{statapp2}). Writing this 
equation as $0=g(M,T,\mu,|\nabla\psi|)$, we find
\begin{subequations} \label{dMdx}
\bea
\frac{\partial M}{\partial x} &=& -\frac{\partial g}{\partial x}\left(\frac{\partial g}{\partial M}\right)^{-1} \, , \allowdisplaybreaks \\[2ex]
\frac{\partial^2 M}{\partial x\partial y} &=& -\left(\frac{\partial g}{\partial M}\right)^{-1} \left[\frac{\partial^2g}{\partial M\partial y}\frac{\partial M}{\partial x}
+\frac{\partial^2g}{\partial M\partial x}\frac{\partial M}{\partial y} +\frac{\partial^2g}{\partial M^2}\frac{\partial M}{\partial x}\frac{\partial M}{\partial y}
+\frac{\partial^2g}{\partial x\partial y}\right] \, .
\eea
\end{subequations}
We can thus use the following algorithm to compute the properties of the superfluid:
\begin{enumerate}

\item Choose values for the thermodynamic parameters $\mu$ and $|\nabla\psi|$ as well as the parameters $\lambda$, $m$, 
and the renormalization scale $\ell$.

\item Determine the critical temperature $T_c$ by solving the stationarity equation (\ref{statapp2}) for $T$ at the point 
$M^2=\sigma^2-2(\nabla\psi)^2$  [we shall explain below that this is indeed the critical value of $M$, see discussion around Eq.\ (\ref{Mcrit})]. 

\item Find the solution for $M$ of the stationarity equation (\ref{statapp2}) for various values of the temperature $0<T<T_c$ (for the results of Sec.\ \ref{sec:results1}
stop here; for the results of Secs.\ \ref{sec:results2} and \ref{sec:results3} continue with all remaining steps).

\item Compute the first and second derivatives of the integrands of $I^\pm_{\rm finite}$ and $J_{\rm finite}$ with respect to $M$, $T$, $\mu$, and $|\nabla\psi|$;
for the case without superflow use the simplification explained in Sec.\ \ref{sec:sound}. This is done algebraically, i.e, before choosing numerical values.
Nevertheless, it is useful to do all this with a computer because the results get very complicated. 

\item Perform the three-momentum integrals numerically over all expressions obtained in the previous step. Since there are three integrands ($I^+_{\rm finite}$, 
$I^-_{\rm finite}$, $J_{\rm finite}$) and four variables ($M$, $T$, $\mu$, $|\nabla\psi|$), 
we have to perform $3\times 4=12$ integrals for the first derivatives and $3\times 10=30$ integrals for the second derivatives at each temperature. In the presence
of a superflow, each of the integrals contains a non-trivial integration over the polar angle; without superflow, only the integrals needed for the superfluid density 
contain such an angular integral.

\item Use Eqs.\ (\ref{dxdy}) and (\ref{dMdx}), the results of the previous step, and some trivial derivatives of terms outside the 
momentum integrals to put together the first and second derivatives of $\Psi$ with respect to $T$, $\mu$, and $|\nabla\psi|$. There are many terms to handle
but this is a trivial task for a computer since the non-trivial numerical calculation has already been done in the step before. We have checked that the derivatives 
thus obtained are much cleaner in terms of numerical errors compared to a brute force numerical calculation using finite differences.  

\item Insert the obtained derivatives into the definitions of the physical quantities, here $n_s$, $n_n$, $\overline{\cal A}$, $\overline{\cal B}$, $\overline{\cal C}$,
$u_1$, $u_2$.

\end{enumerate}

\subsection{Results I: condensate and critical velocity}
\label{sec:results1}

\begin{figure}[t] 
\begin{center}
\hbox{\includegraphics[width=0.5\textwidth]{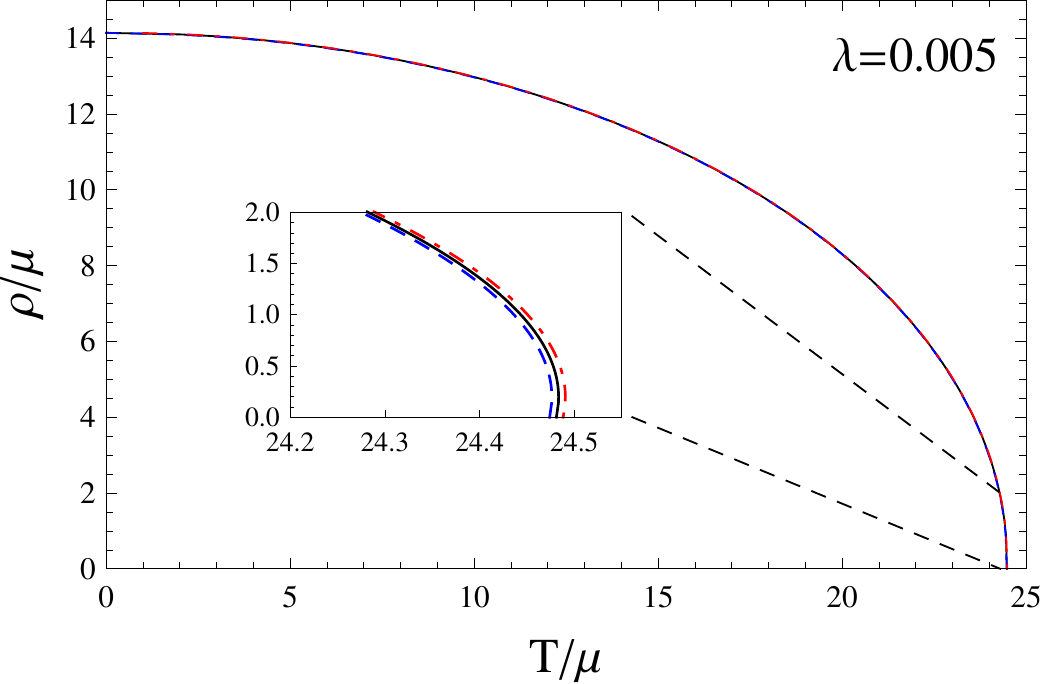}\includegraphics[width=0.5\textwidth]{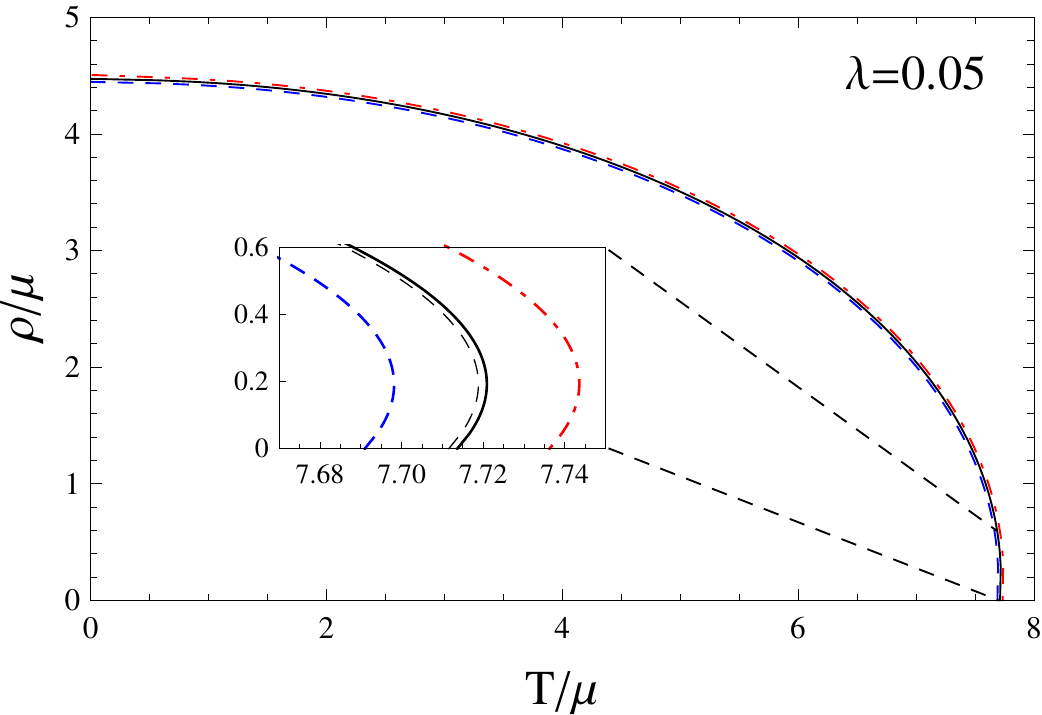}}
\caption{(Color online) Condensate $\rho$ as a function of temperature for $m=\nabla\psi=0$ and coupling constants $\lambda=0.005$ (left panel) and 
$\lambda=0.05$ (right panel). Even though barely visible on the large scale, the phase transition is first order, and the results depend on the renormalization scale 
$\ell$. For the small value of the coupling, even large variations of the renormalization scale are barely visible, while for the larger coupling, the result is 
more, but still only mildly,  sensitive to variations in $\ell$. In both panels, $\ell=0.1\mu$, $\mu$, $10\mu$ for the dashed (blue), solid (black), and dashed-dotted 
(red) lines, respectively. The thin (black) dashed line in the inset of the right panel is obtained with the approximation (\ref{Ip0Im0}), where the dependence of the 
renormalization scale drops out.} 
\label{figorder}
\end{center}
\end{figure}

Once we have determined $M$ from the stationarity equation we obtain the condensate $\rho$ through $\lambda\rho^2=M^2-\sigma^2$. In Fig.\ \ref{figorder} 
we show the condensate as a function of temperature for the simple case without superflow and for two different coupling strengths. We have set $m=0$, but the 
conclusions we draw from this figure 
are valid for all values of $m$. As mentioned below Eq.\ (\ref{V2}), the phase
transition to the non-superfluid phase turns out to be of first order, although this is barely visible if we plot the condensate for all temperatures. 
Moreover, there is a dependence on the renormalization scale $\ell$ through the 
logarithmic terms discussed in Sec.\ \ref{sec:renorm}. 
This dependence and the first order transition are very weak because of the smallness of the coupling constants chosen here.
As the figure shows, the stronger the coupling, the stronger the dependence on the renormalization scale and the stronger the first order transition.

Since we know that the first-order nature of the phase transition is an artifact of the Hartree approximation, we shall restrict ourselves to 
sufficiently weak coupling constants. We shall work with the two couplings chosen in Fig.\ \ref{figorder}. 
In this case we find that we can, to a very good approximation, work with a simplified stationarity equation and a simplified pressure, 
using\footnote{In the notation of appendix \ref{AppA}, this means that we approximate $I^\pm_{\rm finite}(T,\mu,\ell) = I_{\rm vac,finite}^\pm(\ell{}) 
+ I^\pm_\mu(0{}) + I^\pm_T(\mu{})\simeq I^\pm_T(\mu{})$ and  $J_{\rm finite}(T,\mu,\ell) \simeq J_T(\mu{})$.}  
\begin{subequations} \label{Ip0Im0}
\bea \label{Ip0}
I^+_{\rm finite} &\simeq& 4\sum_{e=\pm}\int\frac{d^3{\bf k}}{(2\pi)^3}\frac{(\epsilon_{\bf k}^e)^2-k^2-M^2+\sigma^2}
{(\epsilon_{\bf k}^e+\epsilon_{-{\bf k}}^e)(\epsilon_{\bf k}^e+\epsilon_{-{\bf k}}^{-e})(\epsilon_{\bf k}^e-\epsilon_{\bf k}^{-e})}\, f(\epsilon_{\bf k}^e) \, , 
\allowdisplaybreaks \\[2ex] 
\label{Im0}
I^-_{\rm finite}&\simeq& 4 \sum_{e=\pm}\int\frac{d^3{\bf k}}{(2\pi)^3}\frac{\delta M^2}
{(\epsilon_{\bf k}^e+\epsilon_{-{\bf k}}^e)(\epsilon_{\bf k}^e+\epsilon_{-{\bf k}}^{-e})(\epsilon_{\bf k}^e-\epsilon_{\bf k}^{-e})}\,f(\epsilon_{\bf k}^e) \, , 
\eea
\end{subequations}
and 
\be \label{J0}
J_{\rm finite} \simeq -T\sum_{e=\pm}\int\frac{d^3{\bf k}}{(2\pi)^3}\ln\left(1-e^{-\epsilon_{\bf k}^e/T}\right) \, .
\ee
In this approximation, the contributions from loop diagrams that
do not depend on temperature explicitly are neglected. As a consequence,
the zero-temperature results are identical to the tree-level results.
All dependence on the renormalization scale $\ell$ is gone and thus we do not have to specify $\ell$.
We have checked numerically that the subleading terms that we have dropped do not visibly change any of the curves we show, see also right panel of Fig.\ \ref{figorder},
where the approximation is compared to the full result.

\begin{figure}[t] 
\begin{center}
\hbox{\includegraphics[width=0.49\textwidth]{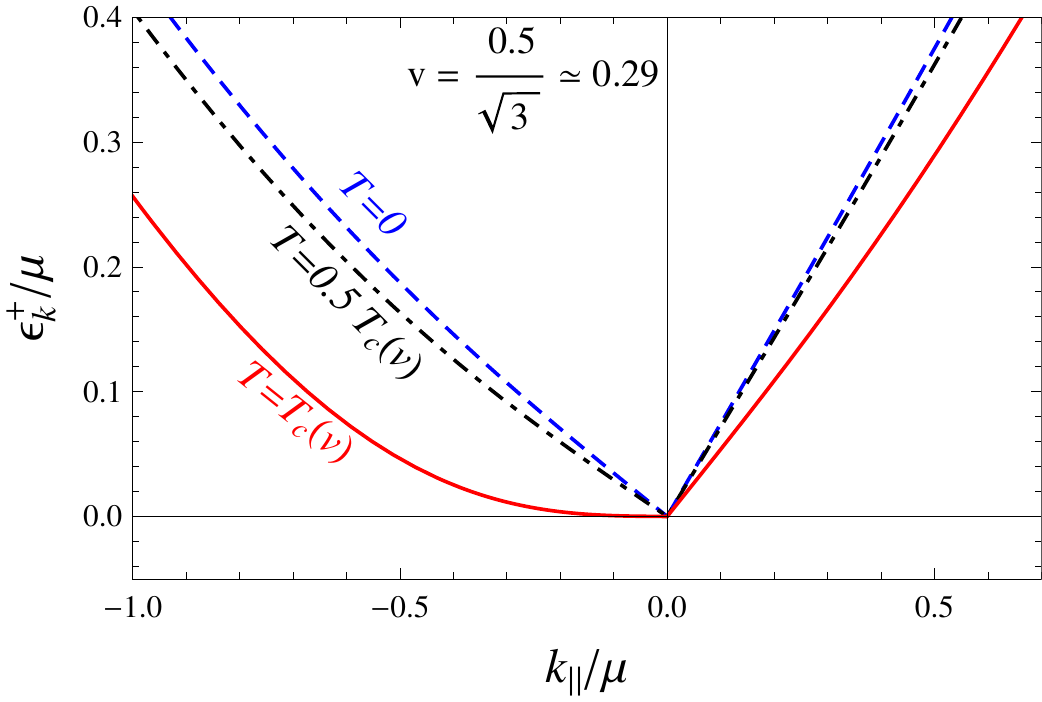}\hspace{0.3cm}\includegraphics[width=0.48\textwidth]{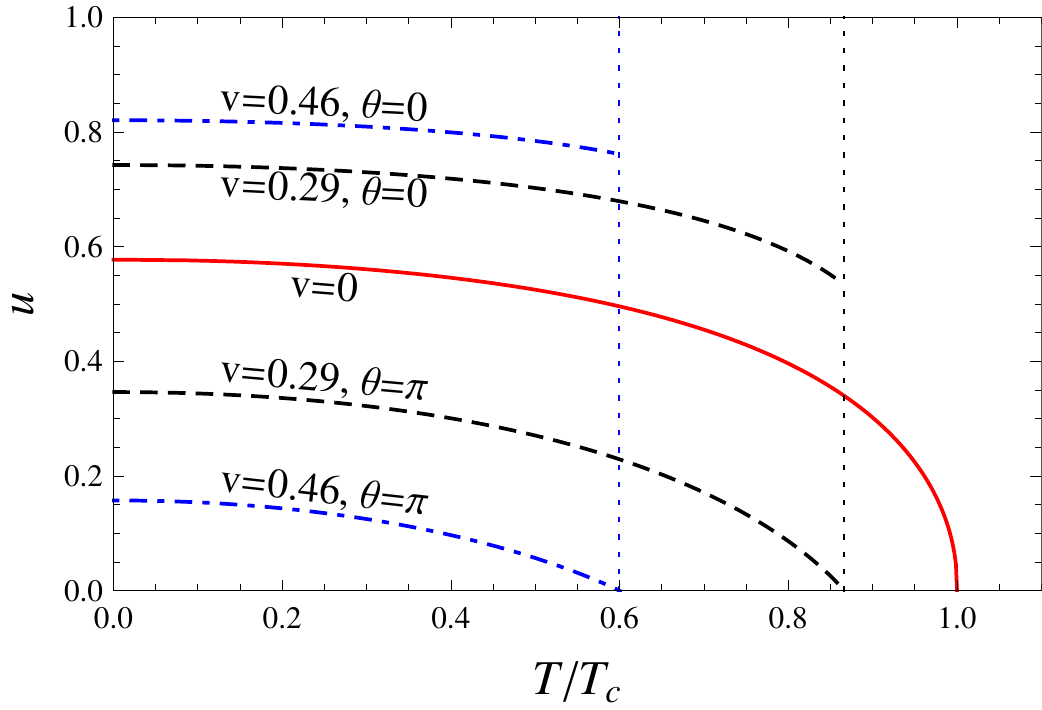}}
\caption{(Color online) 
Left panel: dispersion of the Goldstone mode parallel ($k_\parallel>0$) and anti-parallel ($k_\parallel<0$) to the superflow for a superfluid velocity 
$v=0.5/\sqrt{3}$ and three
different temperatures with $T_c(v)$ being the temperature beyond which $\epsilon_{\bf k}^+$ would become negative for small momenta. Right panel: slope $u(\theta)$ of the 
low-energy dispersion of the Goldstone mode, $\epsilon_{\bf k}^+\simeq u(\theta)k$, in the directions parallel, $\theta=0$, and antiparallel, $\theta=\pi$, to the superflow as a function of 
temperature for three different values of the superfluid velocity. 
The vertical dotted lines indicate the critical temperatures beyond which there is a negative 
excitation energy. Only for the case of vanishing superflow is the critical temperature the point where the condensate has completely melted away. This temperature is denoted by $T_{c}\equiv T_c(v=0)$.  
The case of the intermediate superfluid velocity $v\simeq 0.29$ corresponds to the left panel, i.e., $T\simeq 0.86 \, T_{c}$ in the right panel is identical to $T_c(v)$ in the left panel. 
We have set $\lambda=0.005$ and $m=0$.}
\label{figslopes}
\end{center}
\end{figure}

\begin{figure}[t] 
\begin{center}
\includegraphics[width=0.5\textwidth]{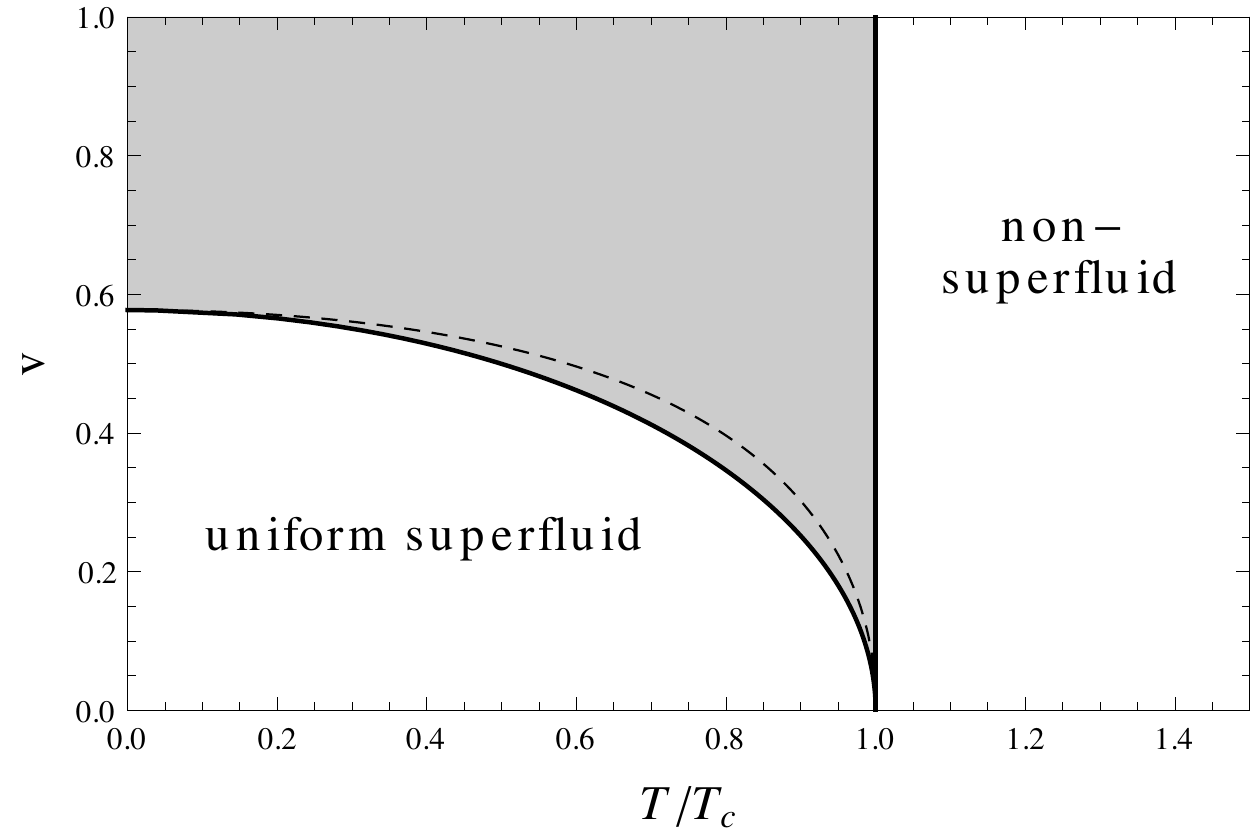}
\caption{ 
Phase diagram resulting from the instability shown in Fig.\ \ref{figslopes}, in the plane of superfluid velocity $v$ and temperature. Within our ansatz, that only allows for spatially homogeneous 
condensates, there is no stable phase in the shaded region. The dashed line is the slope of the Goldstone dispersion at $v=0$ and thus shows -- for comparison -- the would-be
critical velocity if the superflow only acted as a Lorentz transformation on the dispersion, and not also on the self-consistently determined condensate.}
\label{figphase}
\end{center}
\end{figure}

The effects of a nonzero superflow on the solution are shown in Fig.\ \ref{figslopes}, for the parameters $\lambda=0.005$ and $m=0$.
In the left panel of this figure we show the Goldstone mode dispersion 
relation $\epsilon_{\bf k}^+$, evaluated 
using the value of $M$ that solves
the stationarity equation. We see that for a given 
superfluid velocity, there is a temperature at which the dispersion becomes flat in the direction opposite to the superflow. For higher temperatures, there would be negative 
energies, indicating an instability of the system. Therefore, this particular temperature is a critical temperature, even though the condensate has not yet melted completely.
Only at vanishing superflow is the critical temperature the same as the point where the condensate has become zero (if our approach gave
an exact second-order phase transition).
The plot shows that the low-energy part of the dispersion, where $\epsilon_{\bf k}^+$ is linear in $k$, is sufficient to locate the instability. Therefore, in the right panel of the figure, we 
show the slope of the linear part, see Eq.\ (\ref{slope}), as a function of temperature for three different values of the superflow. The superfluid state breaks down when the slope in the anti-parallel direction vanishes. This defines a critical temperature for any given velocity, or a critical velocity for any given temperature. 
With the help of the low-energy dispersion (\ref{slope}) we can derive a semi-analytical result for the critical velocity. We find that the 
linear part of the dispersion is positive for all angles between the momentum ${\bf k}$ and the superflow $\nabla\psi$ only if $M^2-\sigma^2 > 2(\nabla\psi)^2$. 
This shows explicitly that the condensate $\rho^2=(M^2-\sigma^2)/\lambda$ cannot become arbitrarily small for nonzero superflow. Using  
$\sigma^2=\mu^2-(\nabla\psi)^2$ and ${\bf v}=-\nabla\psi/\mu$, we can rewrite the condition for the positivity of the excitation energy in the 
equivalent, but more instructive, form 
\be \label{Mcrit}
v< \sqrt{\frac{M^2-\sigma^2}{M^2+\sigma^2}}  \, .
\ee
This is an implicit condition for allowed values of the superfluid velocity $v$. (Remember that $M$ is a complicated function of this velocity.)
We plot the critical line, given implicitly by Eq.\ (\ref{Mcrit}), in the plane of superfluid velocity and temperature in Fig.\ \ref{figphase}.

The right-hand side of the inequality (\ref{Mcrit}), evaluated at $v=0$, is simply the slope of the Goldstone mode at $v=0$, see Eq.\ (\ref{goldslope}). 
This is related to Landau's
original argument for the critical velocity \cite{khala}: according to this argument, based on a Lorentz transformation of the excitation energy
(in the original non-relativistic context a Galilei transformation), the critical velocity is determined by the slope of the Goldstone mode at $v=0$ (unless there are non-trivial 
features such as rotons in superfluid helium, which is not the case here). 
One might thus think that we would just have to do the $v=0$ calculation to determine the critical line in the phase diagram. However, switching on a superflow is not 
equivalent to a Lorentz transformation of the excitation energy; it is a Lorentz transformation {\it plus} a change in the self-consistently determined condensate, 
which in turn back-reacts on the excitation energies. This additional effect is contained in the $v$ dependence of $M$ in Eq.\ (\ref{Mcrit}). 
For comparison, we have plotted the (incorrect) critical curve obtained from the $v=0$ dispersion in the phase diagram as a dashed 
line. We see that the full result is smaller for nonzero temperatures.  In the weak-coupling case considered here, the effect of the superflow 
(in addition to being a Lorentz boost) on the Goldstone dispersion becomes negligibly small for low temperatures. 
Therefore, at $T=0$ the critical velocity is $\frac{1}{\sqrt{3}}$, 
in exact agreement with the slope of the low-energy dispersion at $v=0$. For a similar recent discussion in a holographic approach see Ref.\ \cite{Amado:2013aea} 
and in particular the phase diagram in Fig.\ 6 of this reference.

The critical line seems to suggest a (strong) first-order phase transition to the non-superfluid phase at the critical velocity. However, at temperatures below $T_{c}$ ($T_{c}$ being the 
critical temperature in the absence of a superflow), the system still ``wants'' to condense, even for velocities beyond the critical value. In other words, 
the uncondensed phase also turns out to be unstable. In our calculation, this is seen as follows. First we note that the stationarity equation for $M$ in the 
case $\rho=0$, see Eq.\ (\ref{statnormalR}), does not depend on $\nabla\psi$. This is clear since $\psi$ is the phase of the condensate, so the uncondensed phase must be independent
of $\nabla\psi$. For supercritical temperatures the solution
to the $\rho=0$ stationarity equation gives a value of
$M$ that is greater than $\mu$, but at subcritical temperatures $M$ is less than $\mu$. 
The excitation energies are simply given by $\epsilon_{k}^e = \sqrt{k^2+M^2}-e\mu$, so $\epsilon_k^+$ becomes negative for certain momenta if $M<\mu$. 
Therefore, the non-superfluid phase is unstable below $T_{c}$. 
As a consequence, within our non-dissipative, uniform ansatz, we cannot construct a stable phase for sufficiently large superfluid velocities and low temperatures 
(shaded area in Fig.\ \ref{figphase}).  
Beyond the critical velocity there may be no stable phase, if
dissipative effects such as vortex creation arise in that regime.
There is some evidence for this in liquid helium \cite{1949Phy....15..285G,1977RSPTA.284..179A} and ultra-cold bosonic \cite{1999PhRvL..83.2502R} and fermionic
\cite{2007PhRvL..99g0402M} gases. Possibly, a more complicated, dissipative and/or inhomogeneous phase already replaces the homogeneous superfluid for superfluid 
velocities below our critical line. In this sense we have only determined an upper limit for the critical velocity as a function of temperature below which the 
homogeneous superfluid is stable. This limit can for instance be reduced by the onset of unstable sound modes due to the two-stream instability \cite{Schmitt:2013nva}.


\subsection{Results II: superfluid density and entrainment}
\label{sec:results2}

\begin{figure}[t] 
\begin{center}
\hbox{\includegraphics[width=0.5\textwidth]{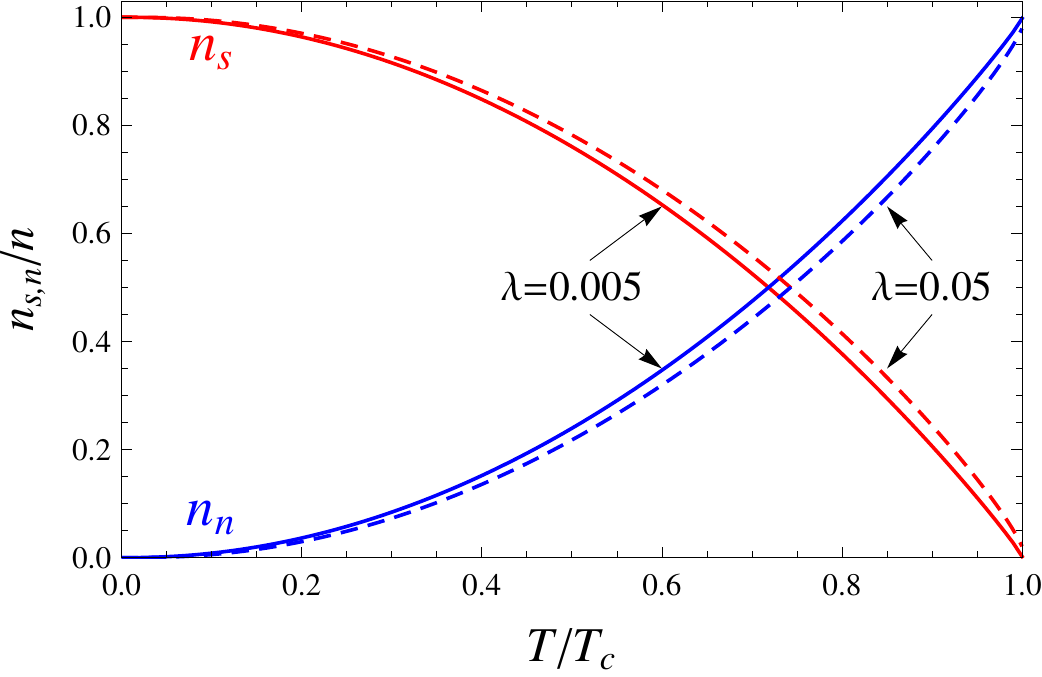}\includegraphics[width=0.48\textwidth]{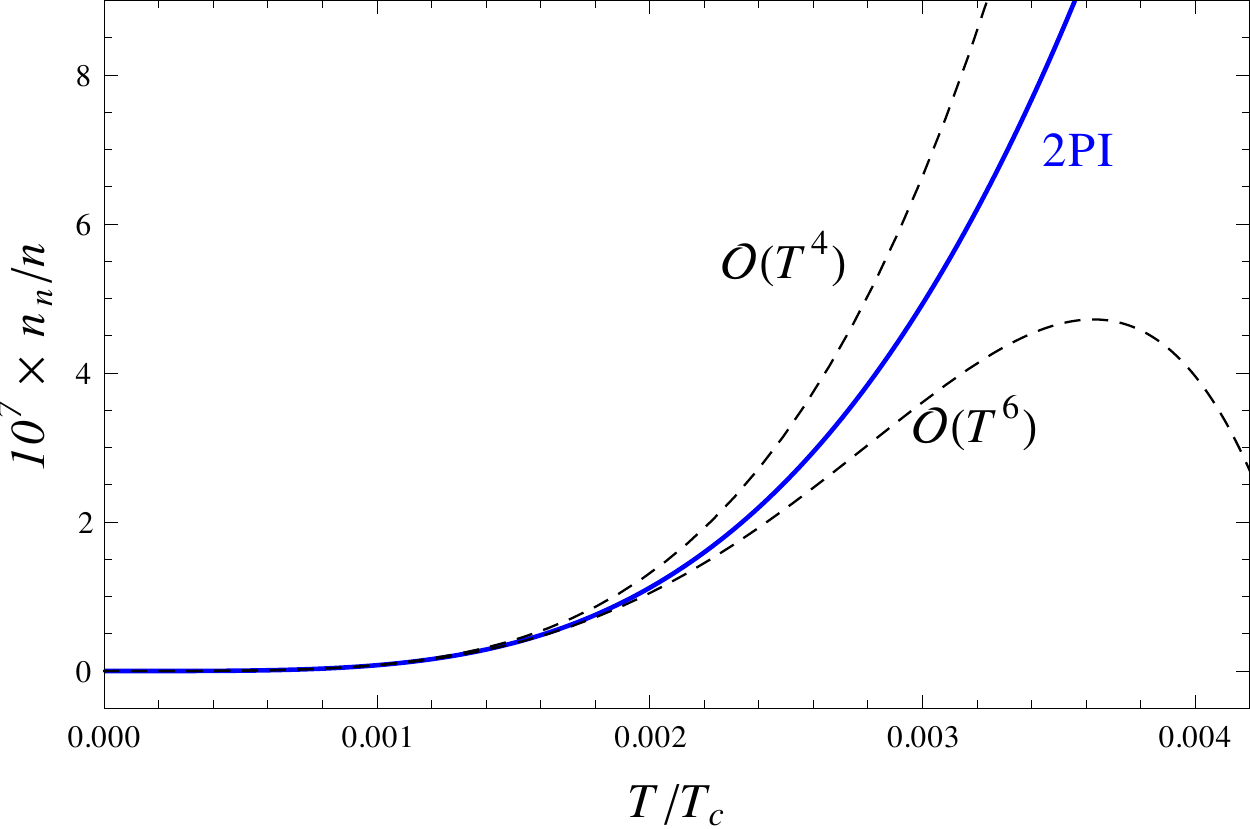}}
\caption{(Color online) Left panel: superfluid and normal fluid charge densities, normalized by the total charge density $n$, 
as a function of temperature for all temperatures up to the critical temperature 
and two different couplings $\lambda=0.005$ (solid lines) and $\lambda=0.05$ (dashed lines). Since different couplings lead to different critical temperatures, 
a given point on the horizontal axis $T/T_c$ corresponds to different {\it absolute} temperatures $T$ for solid and dashed lines. 
Right panel: comparison of the full 2PI calculation 
with the analytical low-temperature approximations from Eq.\ (\ref{nsnn}) for $\lambda=0.005$. We have set the superflow and the mass parameter to zero, 
$\nabla\psi=m=0$. }
\label{fignsnn}
\end{center}
\end{figure}

The superfluid and normal-fluid densities for all temperatures up to the critical temperature are shown in Fig.\ \ref{fignsnn}. Here we consider the case 
without superflow. As expected, the superfluid density is identical to the total density at $T=0$ and decreases monotonically with the temperature
until it goes to zero continuously at the critical temperature. 
The plot shows the densities for two different values of the coupling constant. Different coupling strengths lead to different 
critical temperatures. In the given plot, $T_c\simeq 24.5\,\mu$ for the weaker of the two chosen couplings, $\lambda=0.005$, while 
$T_c\simeq 7.71\, \mu$ for the stronger coupling, $\lambda=0.05$ (the stronger the coupling, the stronger the repulsive force between the bosons and hence the 
lower the critical temperature). 
Therefore, the {\it absolute} value of the temperature is different for the two curves 
at a given point on the horizontal axis. This has to be kept in mind for all following plots. We see that the stronger coupling tends to favor the superfluid component, i.e., 
for a given {\it relative} temperature with respect to $T_c$ an increase of the coupling leads to a (small) increase of the superfluid density fraction. 

In the right panel of the figure we compare the full 2PI result with the tree-level approximation for low temperatures.
This approximation was discussed in Ref.\ \cite{2013PhRvD..87f5001A}, and reads for $m=v=0$
\bea\label{nsnn}
n_s &\simeq&  \frac{\mu^3}{\lambda}-\frac{4\pi^2T^4}{5\sqrt{3}\,\mu}+\frac{152\pi^4T^6}{21\sqrt{3}\,\mu^3} \, .  \qquad n_n \simeq 
\frac{4\pi^2T^4}{5\sqrt{3}\,\mu}-\frac{48\pi^4T^6}{7\sqrt{3}\,\mu^3} \, . 
\eea
(Generalizations to nonzero $m$ and nonzero $v$ are read off from the results in appendix \ref{AppB} and Ref.\ \cite{2013PhRvD..87f5001A}, respectively.)
In the right panel of Fig.\ \ref{fignsnn} we have plotted the curves where the expansion is truncated at order $T^4$ and where it is truncated at order $T^6$. 
It is already clear from the comparison of these two truncations that the series in $T$ converges very slowly. Both truncations are only good approximations to the 
full result for very low temperatures compared to the critical temperature, in this case for $T\lesssim 0.002 \, T_c$. 

\begin{figure}[t] 
\begin{center}
\hbox{\includegraphics[width=0.5\textwidth]{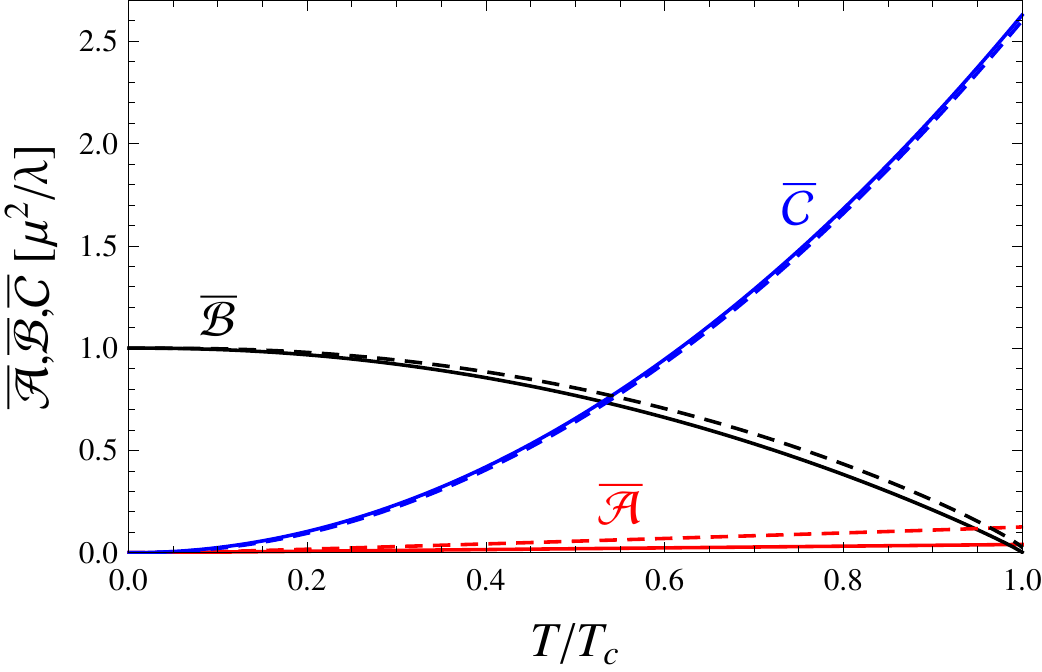}\includegraphics[width=0.48\textwidth]{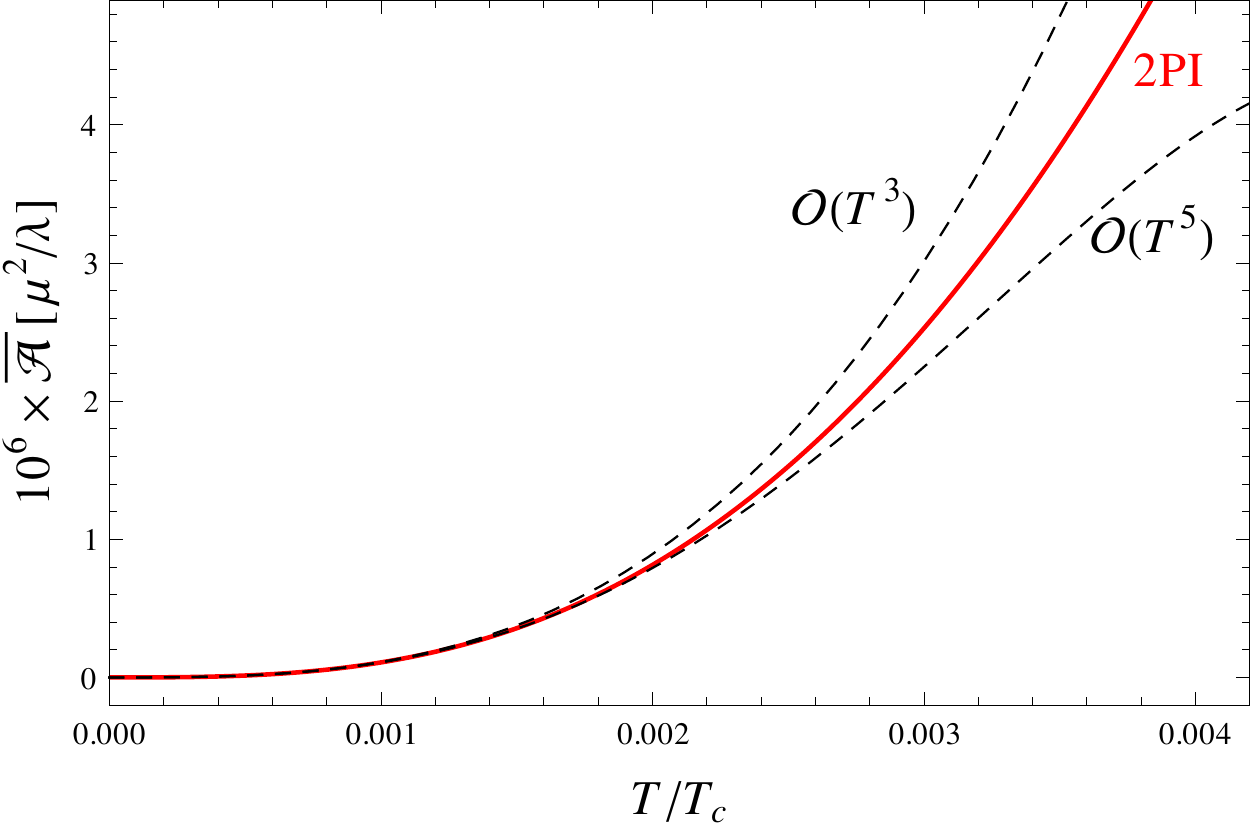}}
\caption{(Color online) Left panel: coefficients $\overline{\cal A}$, $\overline{\cal B}$, $\overline{\cal C}$ of the two-fluid formalism [see Eqs.\ (\ref{js})] 
in units of $\mu^2/\lambda$ as a function of temperature for the same two couplings as in 
Fig.\ \ref{fignsnn}, $\lambda=0.005$ (solid lines) and $\lambda=0.05$ (dashed lines). Right panel: entrainment coefficient $\overline{\cal A}$ for low temperatures
and comparison with the analytical results from Eq.\ (\ref{ABCapprox}) for $\lambda=0.005$. As in Fig.\ \ref{fignsnn}, $\nabla\psi=m=0$.  }
\label{figABC}
\end{center}
\end{figure}

Next we compute the coefficients $\overline{\cal A}$, $\overline{\cal B}$, $\overline{\cal C}$ that relate the charge current $j^\mu$ and the entropy current $s^\mu$ with 
their corresponding momenta $\partial^\mu\psi$ and $\Theta^\mu$, see Eqs.\ (\ref{js}). We plot these coefficients in Fig.\ \ref{figABC}. Again, we have chosen the same 
two coupling strengths as in Fig.\ \ref{fignsnn}. We have normalized the coefficients not only by dividing by $\mu^2$ (such that they become dimensionless), 
but also by multiplying with a factor $\lambda$ such that the normalized $\overline{\cal B}$ is 1 at zero temperature for all couplings, which makes it easier to 
compare different couplings in a single plot. At zero temperature, $\overline{\cal A}=\overline{\cal C}=0$, which implies that there is no entropy current, 
$s^\mu=0$, as expected, and we have a single-fluid system. At finite temperature, both currents become nonzero and we have a two-fluid system.

The dependence on the coupling seems to be relatively weak for $\overline{\cal B}$, $\overline{\cal C}$, while 
the entrainment coefficient $\overline{\cal A}$ increases significantly with the coupling. We have checked that, for the case of the weaker coupling $\lambda=0.005$, $\overline{\cal A}$ behaves 
linearly in the temperature for all temperatures $T\gtrsim 0.5\,T_c$. For very low temperatures, we have \cite{2013PhRvD..87f5001A}
\bea \label{ABCapprox}
\overline{\cal A} &\simeq& \frac{4\pi^2T^3}{15\sqrt{3}\,\mu}-\frac{80\pi^4T^5}{63\sqrt{3}\,\mu^3}\, , \qquad 
\overline{\cal B} \simeq \frac{\mu^2}{\lambda}-\frac{4\pi^2T^4}{15\sqrt{3}\,\mu^2}+\frac{104\pi^4T^6}{63\sqrt{3}\,\mu^4} \, , \qquad 
\overline{\cal C} \simeq \frac{2\pi^2T^2}{15\sqrt{3}} +\frac{8\pi^4T^4}{63\sqrt{3}\,\mu^2} \, .
\eea
In the right panel of Fig.\ \ref{figABC} we compare the analytical low-temperature approximation for $\overline{\cal A}$ with the full result.
As for the superfluid and normal-fluid densities we see that we have to zoom in to very low temperatures compared to $T_c$ in order to find agreement between the 
approximation and the full result.

\subsection{Results III: sound modes}
\label{sec:results3}

\begin{figure}[t]
     \begin{center}
     \underline{$m=0$}\hspace{7cm} \underline{$m=0.6\,\mu$}

     \vspace{0.2cm}
        \subfigure{\includegraphics[height=0.29\textwidth]{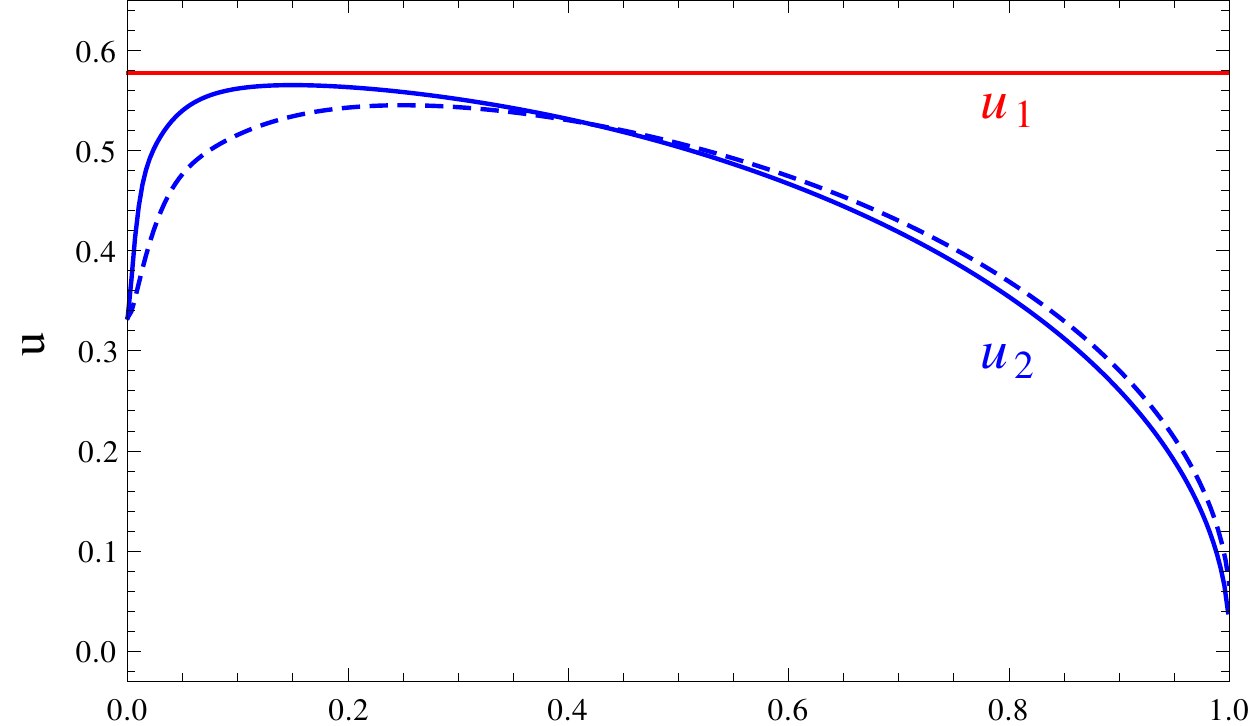}}\hspace{0.2cm}\subfigure{\includegraphics[height=0.29\textwidth]{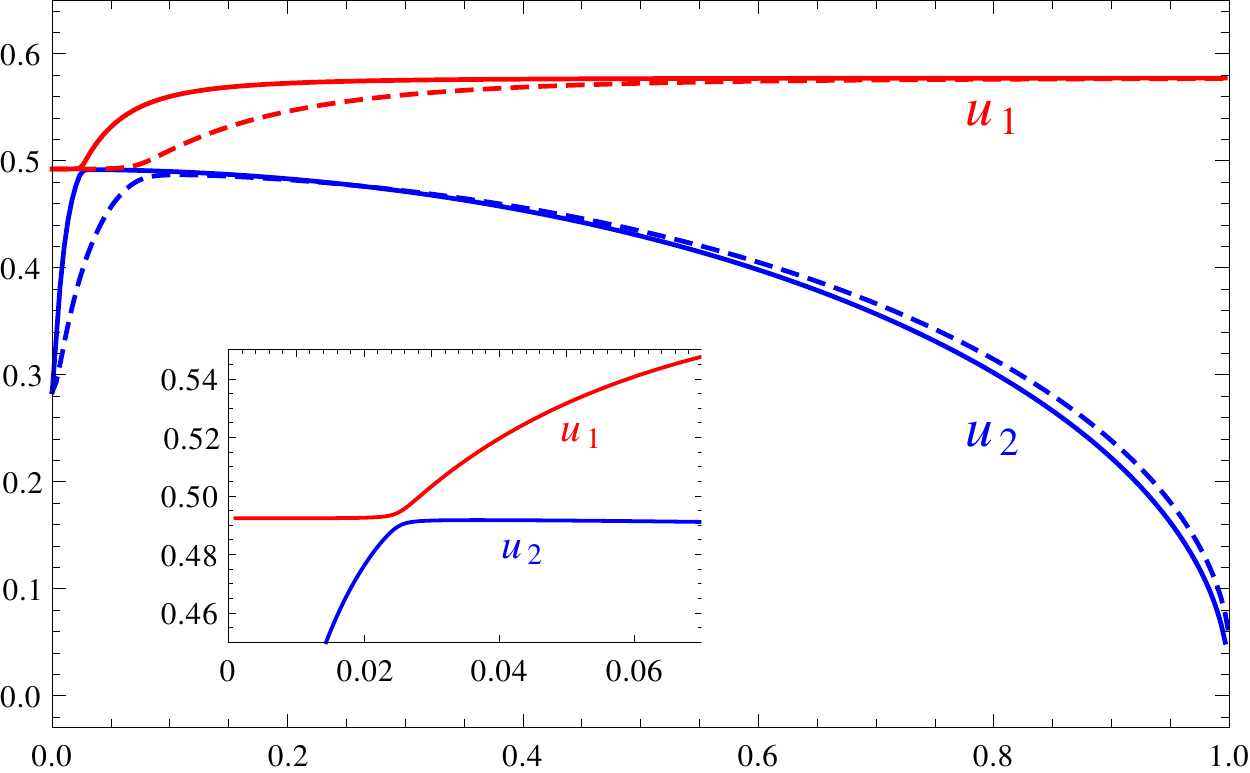}}
\\
        \subfigure{\includegraphics[height=0.332\textwidth]{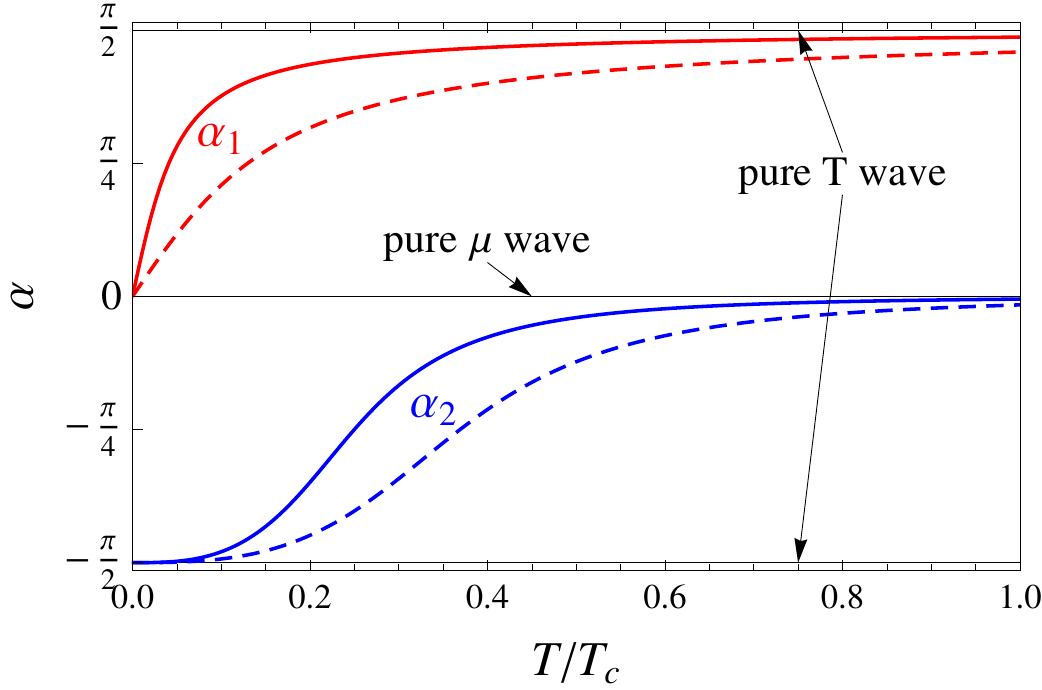}}\subfigure{\includegraphics[height=0.332\textwidth]{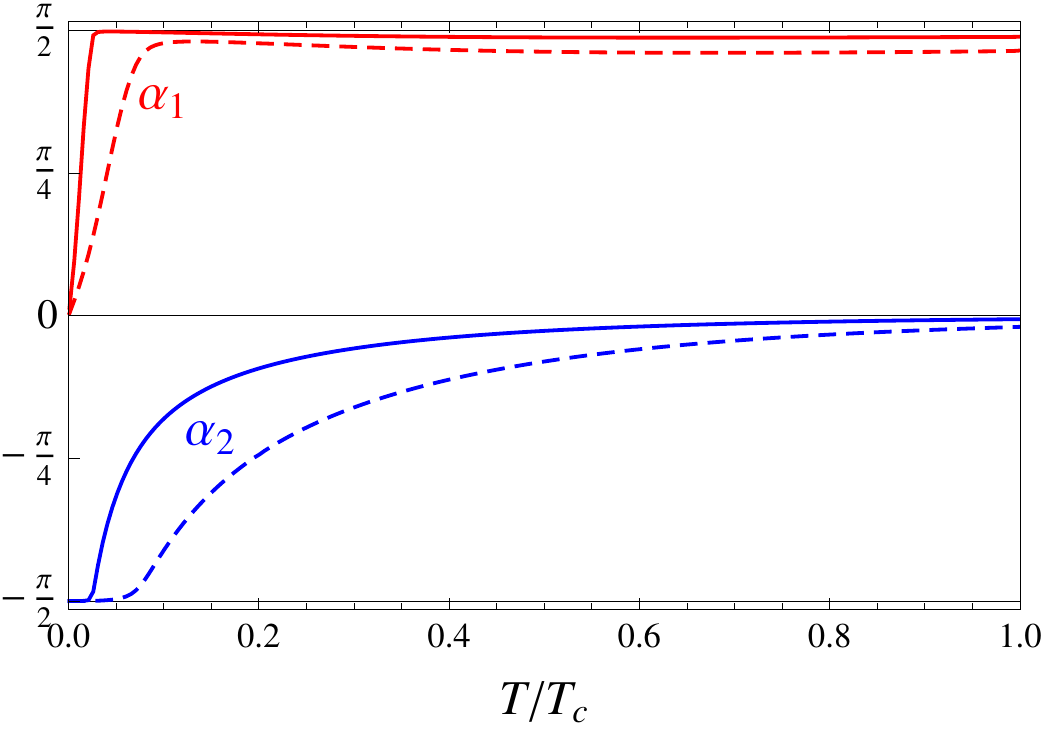}\hspace{0.07cm}}
    \end{center}
\caption{(Color online) Upper panels: speed of first and second sound in the absence of a superflow, $\nabla\psi=0$, 
 as a function of temperature for the ultra-relativistic limit $m=0$ (left panel) and (approaching) the non-relativistic limit $m=0.6\,\mu$ (right panel), 
as well as for two different coupling constants, $\lambda = 0.005$ (solid lines) and $\lambda = 0.05$ (dashed lines). The inset in the upper 
right panel magnifies the region of an avoided crossing between first and second sound for the lower coupling constant.  
Lower panels:
mixing angle $\alpha$ for the amplitudes in temperature and chemical potential [see Eq.\ (\ref{alpha})] associated to each sound wave, for the same values of 
$\lambda$ and $m$. 
Positive (negative) values of $\alpha$ correspond to in-phase (out-of-phase) oscillations, while $|\alpha|=\pi/2$ ($\alpha=0$) corresponds to a pure temperature
(chemical potential) wave.}
\label{figsound0}
\end{figure}

\begin{figure}[t] 
\begin{center}
\hbox{\includegraphics[width=0.5\textwidth]{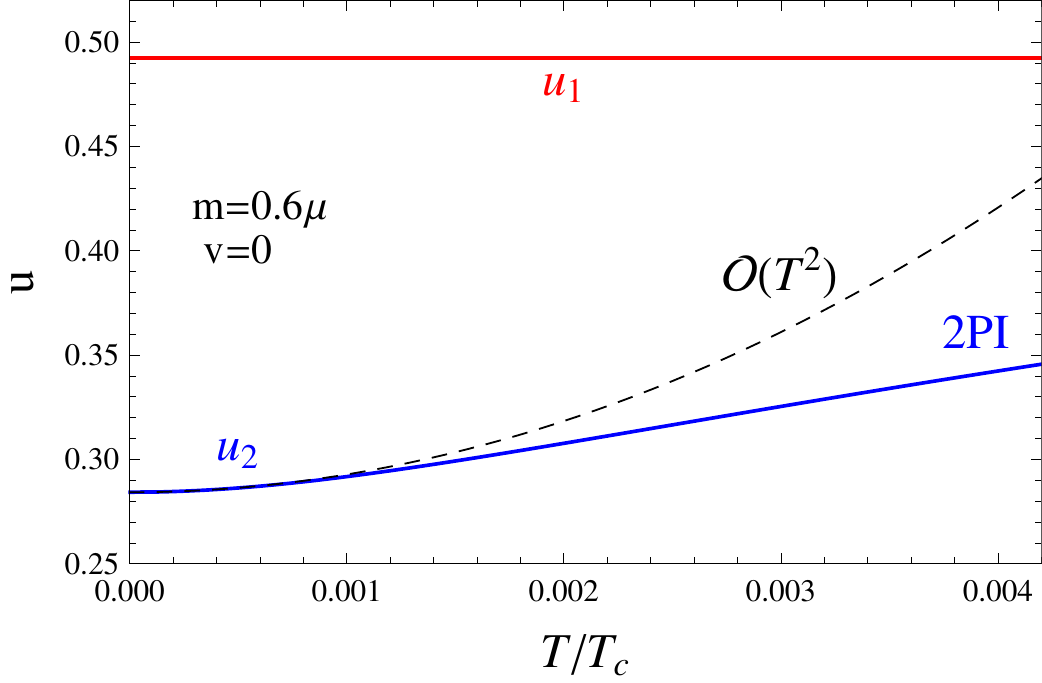}\includegraphics[width=0.5\textwidth]{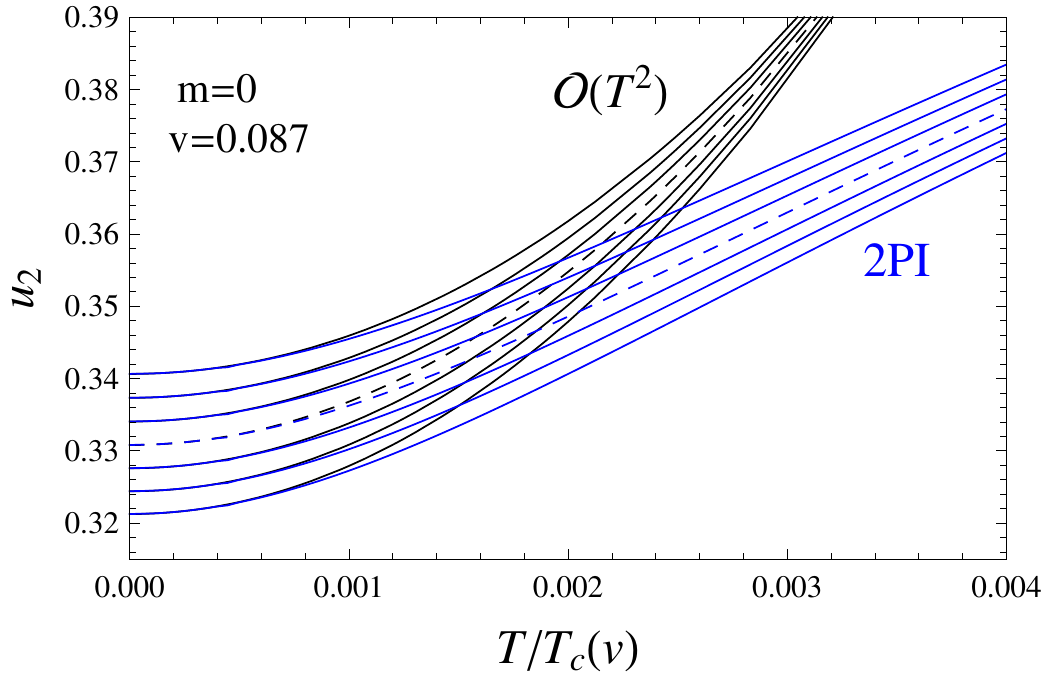}}
\caption{(Color online) Sound velocities for low temperatures and comparison with the analytical low-temperature approximations for vanishing
superflow and $m=0.6\mu$ (left panel) as well as for vanishing mass and $v=\frac{0.15}{\sqrt{3}}\simeq 0.087$ (right panel). 
In both panels, $\lambda=0.005$. The approximations
are given in Eqs.\ (85) of Ref.\ \cite{2013PhRvD..87f5001A} (for the right panel) and in Eqs.\ (\ref{utree}) of the present paper (for the left panel).
The various curves in the 
right panel correspond to different angles between the superflow and the sound wave, from parallel (uppermost curve) to anti-parallel (lowermost curve) with
the middle (dashed) line corresponding to the perpendicular case.}
\label{figsoundlowT}
\end{center}
\end{figure}

\begin{figure}[t]
     \begin{center}
\underline{$\lambda=0.05$}

     \underline{$m=0$}\hspace{7cm} \underline{$m=0.6\,\mu$}

     \vspace{0.2cm}
        \subfigure{\includegraphics[height=0.296\textwidth]{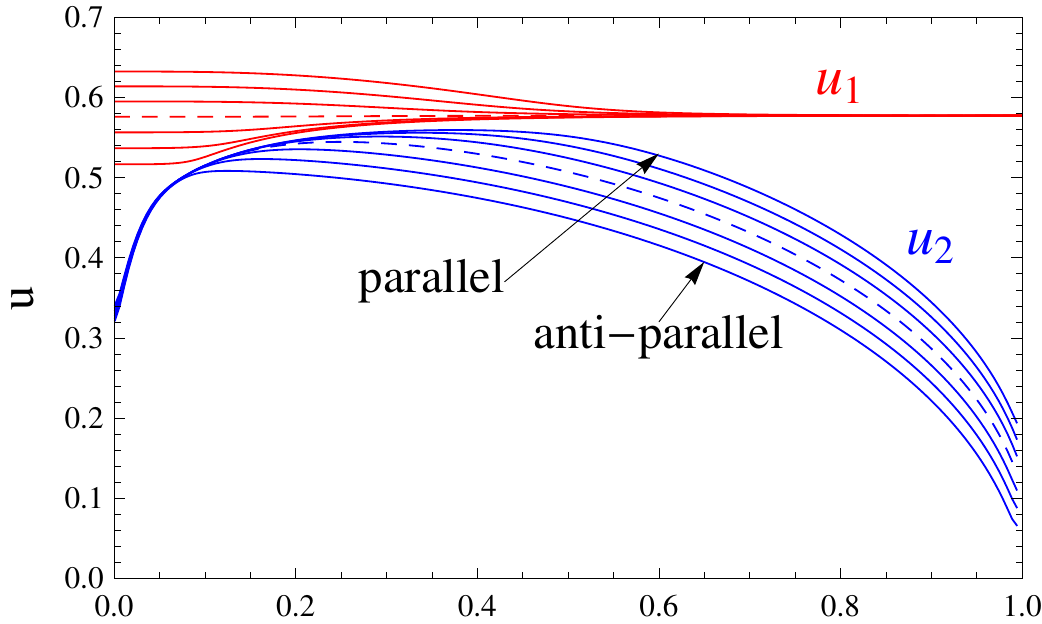}}\hspace{0.2cm}\subfigure{\includegraphics[height=0.296\textwidth]{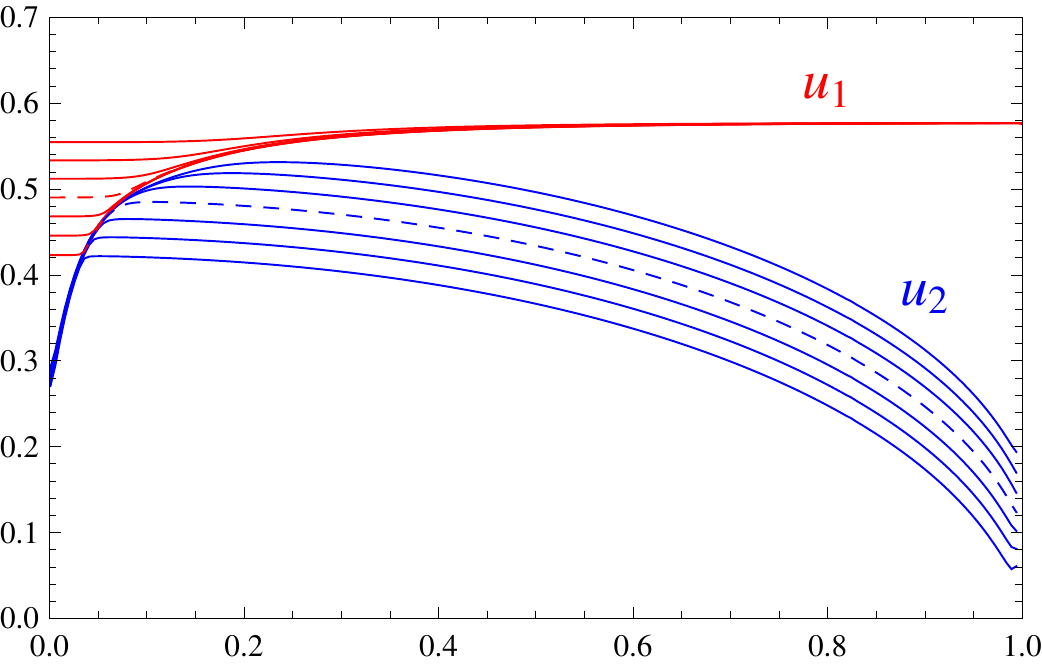}}
\\
        \subfigure{\includegraphics[height=0.332\textwidth]{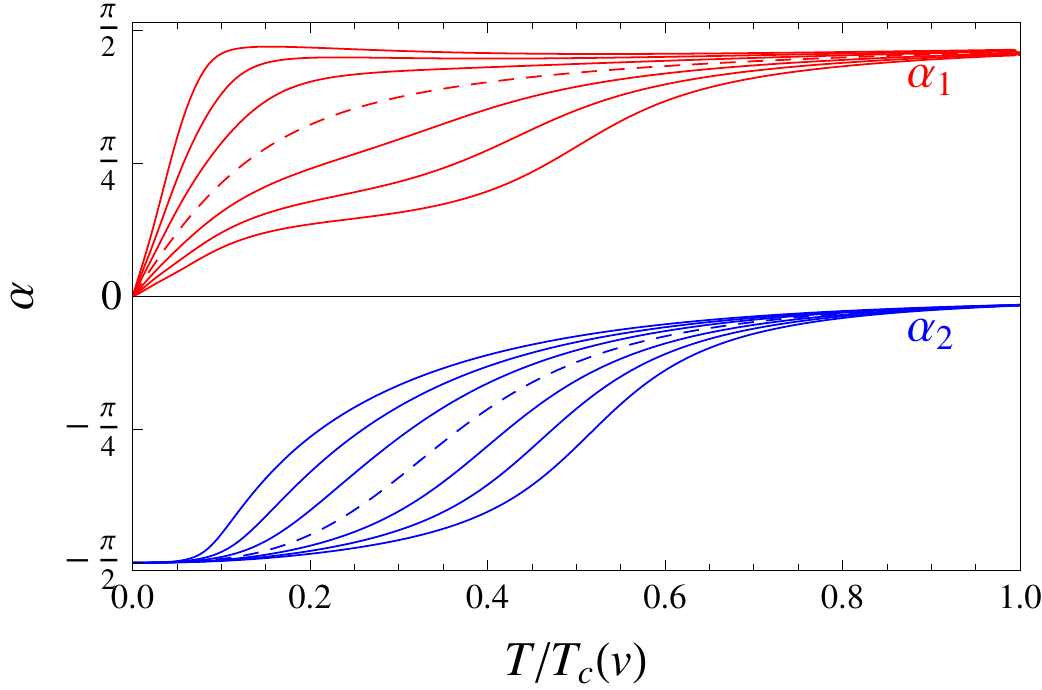}}\subfigure{\includegraphics[height=0.332\textwidth]{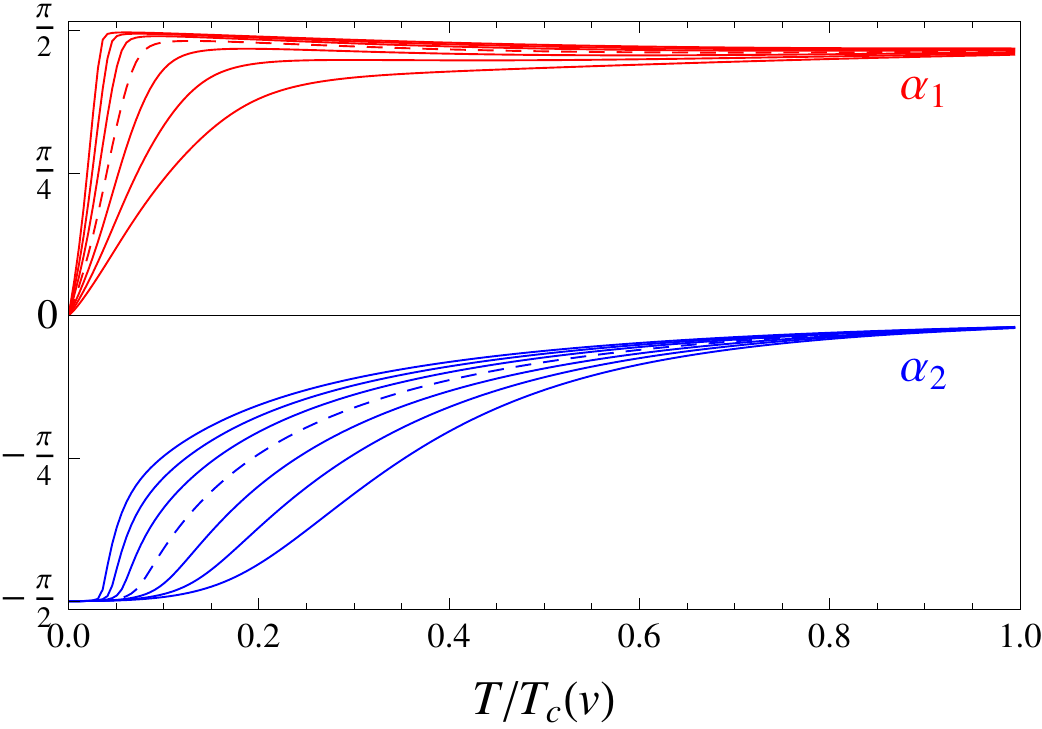}
\hspace{0.07cm}}
    \end{center}
\caption{(Color online) Same as Fig.\ \ref{figsound0}, but with a nonzero superfluid velocity, chosen to be $v=\frac{0.15}{\sqrt{3}}$, i.e., 15\% of the critical velocity at $m=T=0$.
Each plot shows the results for seven different angles between the propagation of the sound wave and the superflow, from parallel (uppermost curves) to anti-parallel
(lowermost curves) in equidistant steps of $\pi/6$ with the dashed lines corresponding to $\pi/2$. The coupling is chosen to be $\lambda=0.05$. The dashed 
lines, where the effect of the superflow is expected to be weakest, are comparable (however not exactly identical) to the dashed lines of Fig.\ \ref{figsound0}.}
\label{figsoundflow1}
\end{figure}

\begin{figure}[t]
     \begin{center}
\underline{$\lambda=0.005$}

     \underline{$m=0$}\hspace{7cm} \underline{$m=0.6\,\mu$}

     \vspace{0.2cm}
        \subfigure{\includegraphics[height=0.296\textwidth]{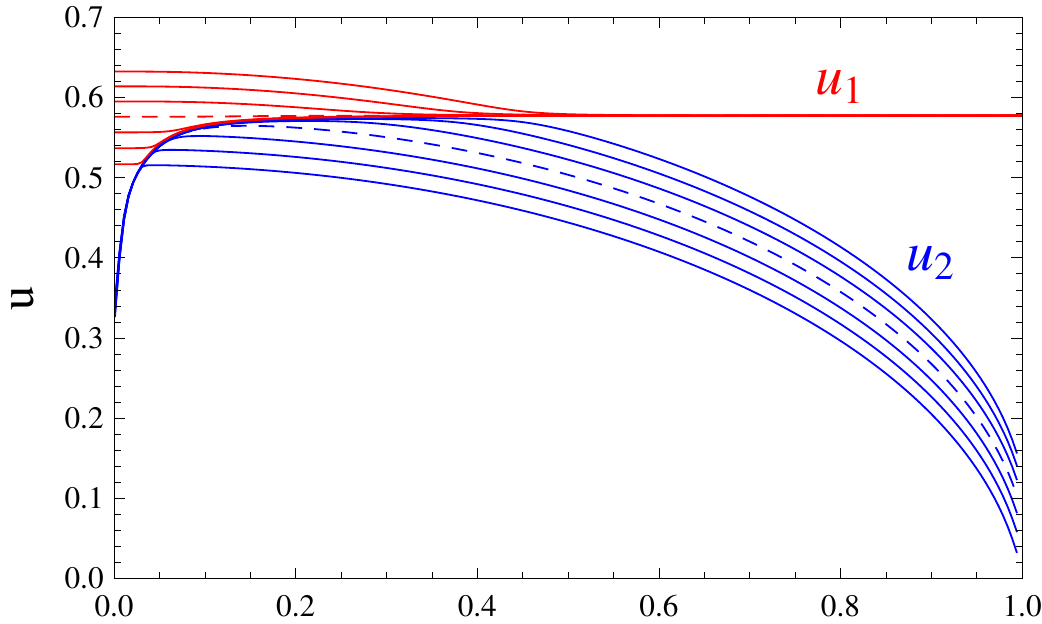}}\hspace{0.2cm}\subfigure{\includegraphics[height=0.296\textwidth]{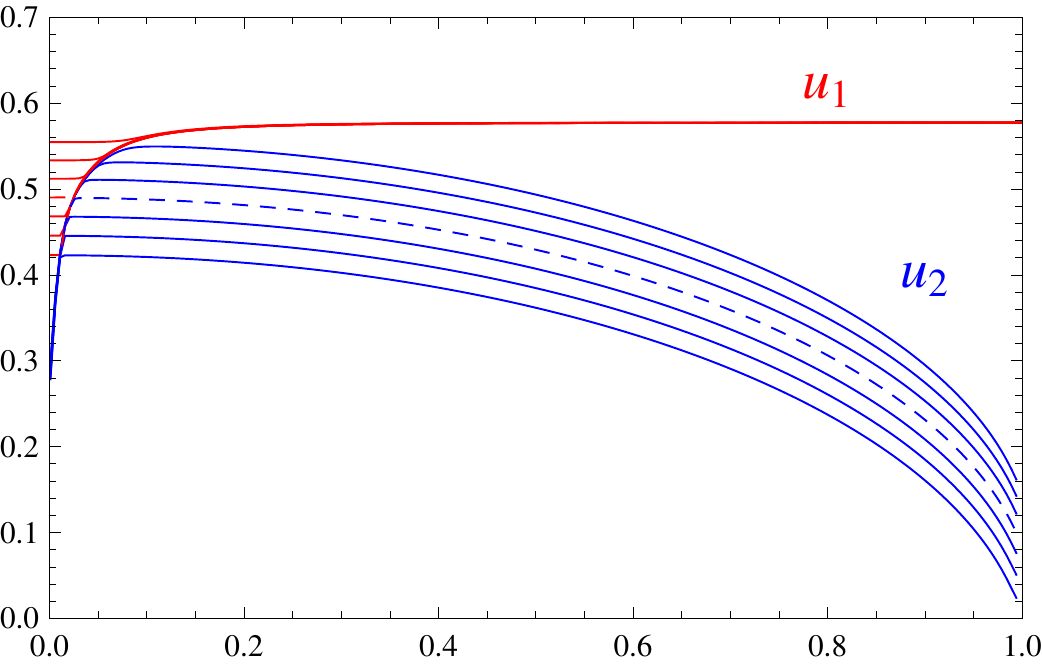}}
\\
        \subfigure{\includegraphics[height=0.332\textwidth]{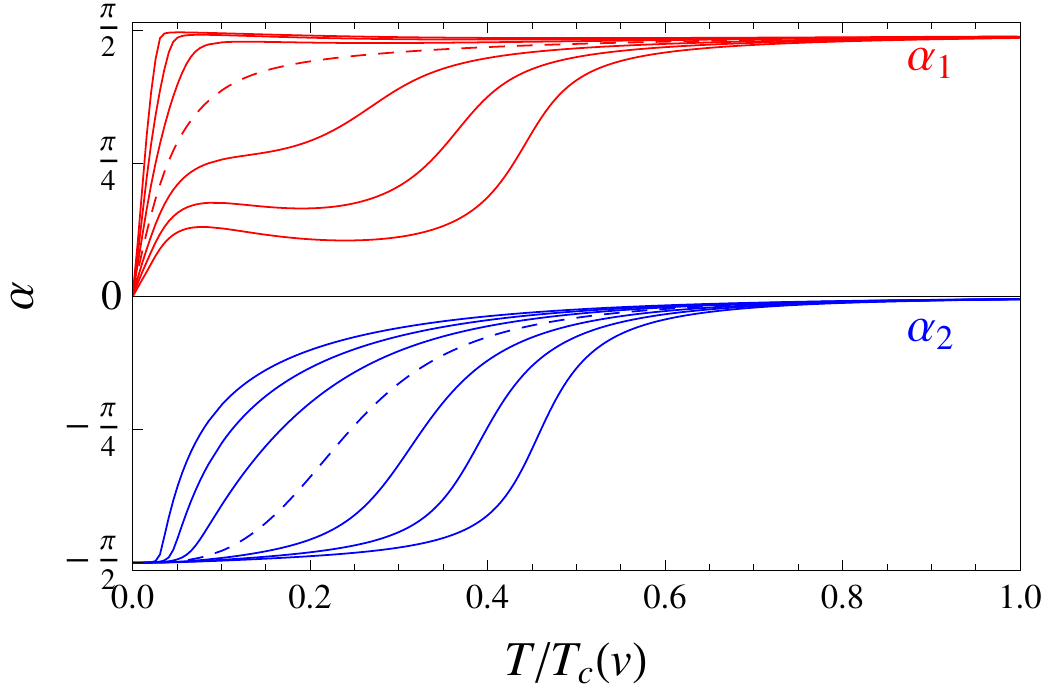}}\subfigure{\includegraphics[height=0.332\textwidth]{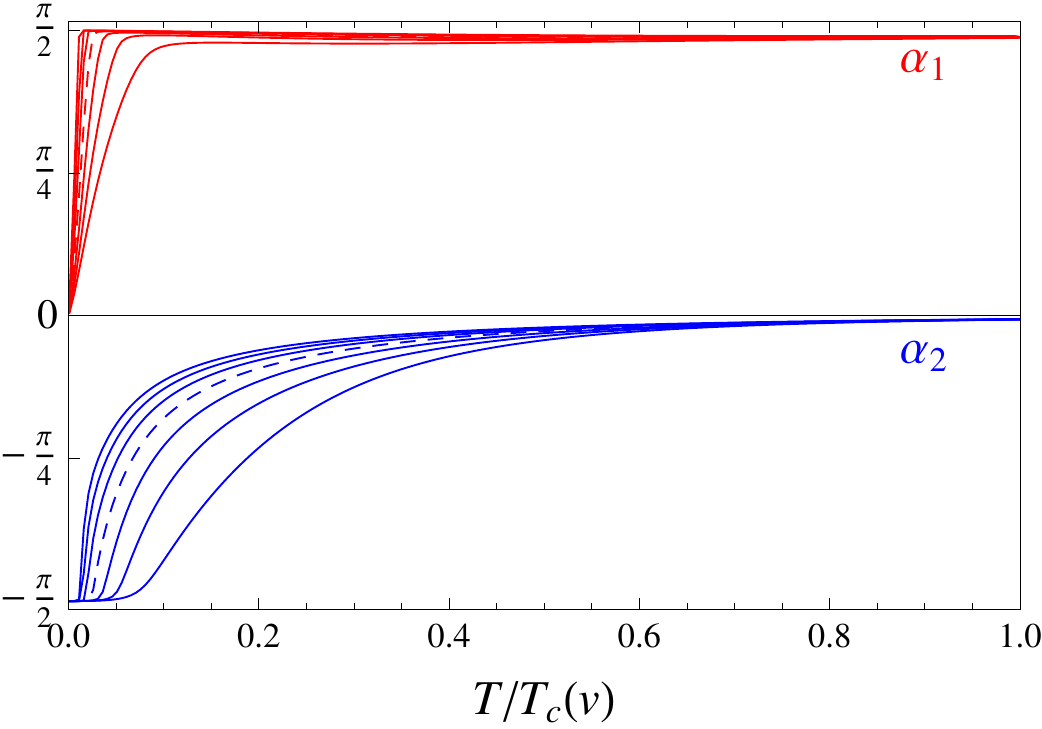}
\hspace{0.07cm}}
    \end{center}
\caption{(Color online) Same as Fig.\ \ref{figsoundflow1}, but at weaker coupling, $\lambda=0.005$.}
\label{figsoundflow2}
\end{figure}

Finally we compute the velocities of first and second sound $u_1$ and $u_2$, as laid out in Sec.\ \ref{sec:sound}. The results are shown in 
Fig.\ \ref{figsound0} (sound velocities and amplitudes for zero superflow), Fig.\ \ref{figsoundlowT} (sound velocities at very low temperatures and comparison 
with the analytical results) and Figs.\ \ref{figsoundflow1}, \ref{figsoundflow2} (sound velocities and amplitudes for nonzero superflow). We now discuss various 
aspects of the results separately.

\bigskip
{\it Speed of first sound and scale-invariant limit.} 
In the simplest case, with vanishing mass parameter and superflow, the speed of first sound is $u_1=\frac{1}{\sqrt{3}}$
for all temperatures. This is shown in the upper left panel of Fig.\ \ref{figsound0} and is in agreement with the analytical result (\ref{conformal}). 
For low temperatures, this sound speed is identical to the slope of the Goldstone dispersion. For higher temperatures, however, the slope deviates from 
the speed of first sound and approaches zero at the critical point, just like the speed of {\it second} sound. In other words, the Goldstone mode
is, in general, not a solution to the wave equations derived from the hydrodynamic conservation equations. Only in certain temperature limits do these waves 
coincide with the Goldstone mode. 

The upper right panel of Fig.\ \ref{figsound0}
shows that for a nonzero mass parameter $m$, the speed of first sound deviates from the scale-invariant value at low temperatures, but approaches this value for high temperatures 
$T\gg m$. Notice that we have chosen the same mass parameter in units of $\mu$ for both coupling strengths. 
As a consequence, the sound velocities for the two coupling strengths coincide at zero temperature, but the mass is different in units of $T_c$: 
for the smaller coupling (solid lines) we have $m\simeq 0.03\,T_c$, while for the larger coupling
(dashed lines) $m\simeq 0.1\,T_c$. This is the reason why $u_1$ appears to approach the scale-invariant value more slowly for the case of the larger coupling. 

\bigskip
{\it Speed of second sound.} In all cases we consider, the speed of second 
sound increases strongly at low temperatures. We can 
see this increase in the low-temperature approximation (\ref{utree}). 
As explained in Ref.\ \cite{2013PhRvD..87f5001A}, the positive $T^2$ contribution in this approximation 
originates from the $T^6$ term in the pressure which, in turn, originates from the $k^3$ contribution to the dispersion of the Goldstone mode. 
Even though Fig.\ \ref{figsoundlowT} shows that the analytic approximation is only valid for very low temperatures, we see that the strong increase continues 
beyond the validity of the analytical approximation 
(although it becomes less strong than the approximation suggests). One can see from Eq.\ (\ref{utree}) that the $T^2$ contribution 
does not, to leading order, depend on the coupling constant. Therefore, since smaller coupling strengths correspond to higher critical temperatures, the increase of $u_2$ can be made 
arbitrarily sharp (on the relative temperature scale $T/T_c$) by decreasing the coupling. This tendency is borne out in Fig.\ \ref{figsound0}.      

In the upper panels of Fig.\ \ref{figsound0} the 
velocity of second sound does not go to zero at the critical point. This 
is an artifact of our Hartree approximation: as we have discussed in Sec.\ \ref{sec:results1}, in our approach the phase transition is strictly speaking first order. 
Therefore, the condensate is not exactly zero at our critical point. The speed of second sound turns out to be sensitive to this effect, and therefore $u_2$
does not approach zero at $T_c$. The superfluid density appears to be less sensitive to this effect since it approaches zero to a very good accuracy, 
see Fig.\ \ref{fignsnn}.

\bigskip
{\it Role reversal of the sound modes.} To discuss the physical nature of the sound waves, we first 
notice that the speeds of sound show a feature that is reminiscent of an ``avoided level crossing'' in quantum mechanics. This
feature is most pronounced for small coupling and nonzero mass parameter $m$, see upper right panel of Fig.\ \ref{figsound0} and 
the zoomed inset in this panel. 
It suggests that there is a physical property that neither first nor second sound possesses for all temperatures, but that is 
rather ``handed over'' from first to second sound in the temperature region where the curves almost touch. We find this property by computing the amplitudes of 
the oscillations associated to the sound modes, as discussed in Sec.\ \ref{sec:sound}. 
In particular, we are interested in the mixing angle $\alpha$ defined in Eq.\ (\ref{alpha}) that indicates whether a given sound mode is 
predominantly an oscillation in chemical potential or in temperature or something in between. Our results show that $u_1$ always corresponds to $\alpha>0$ while $u_2$ 
always corresponds to $\alpha<0$. Therefore, the first sound is always an in-phase oscillation, while the 
second sound is always an out-of-phase oscillation. However, whether first or second sound is a density wave or an entropy wave is a temperature dependent statement,
as already discussed in the scale-invariant limit where there are simple expressions for the amplitudes, see Eq.\ (\ref{ampconf}). 
In all cases we consider, $u_1$ transforms from a pure density wave at $T=0$ to a pure entropy wave at $T=T_c$ and vice versa for $u_2$. This role reversal becomes 
sharper for larger $m$ and/or smaller $\lambda$, i.e., it is smoothest in the ultra-relativistic regime at strong coupling, see lower left panel of Fig.\ \ref{figsound0}.
(Remember that we compare two relatively weak coupling strengths, the ``strong coupling'' is $\lambda=0.05$.)
  
\bigskip
{\it Comparison to non-relativistic systems.}
We can view $m$ as a parameter with which we can go continuously from the ultra-relativistic limit $m=0$ to the non-relativistic limit of large $m$ (always keeping 
$m$ smaller than $\mu$ in order to allow for condensation). Therefore, the 
right panels of Fig.\ \ref{figsound0} are comparable to the results in non-relativistic calculations. Of course, $m=0.6\,\mu$, as chosen in the plots, 
is not actually a non-relativistic value; for instance, for this value of $m$ 
the speed of first sound at low temperatures is still about 50\% of the speed of light, while, for comparison, the speed of first
sound in superfluid helium is about $240\,{\rm m}/{\rm s}$, i.e., about $10^{-8}$ times the speed of light. Nevertheless, already for this moderate value of $m$
we find qualitative agreement with the non-relativistic results of Ref.\ \cite{2010NJPh...12d3040H}, see in particular Fig.\ 6 in this reference which also exhibits 
the avoided crossing and the sharp role reversal at a low temperature. As in this reference, we also find that a stronger
coupling smooths out both of these features. Our work generalizes the results of Ref.\ \cite{2010NJPh...12d3040H} 
to the relativistic regime and to the case of nonzero superflow, see Figs.\ \ref{figsoundflow1}, \ref{figsoundflow2}. (For a zero-temperature calculation of the sound 
velocities in the presence of a superflow in $^4$He see Ref.\ \cite{2009PhRvB..79j4508A}.)

Our results (and those of Ref.\ \cite{2010NJPh...12d3040H})
for the sound modes differ from the calculations and measurements for 
superfluid helium \cite{khala,pines,donnelly} and a (unitary) Fermi gas \cite{2009PhRvA..80e3601T,2010PhRvA..82f3619S,2013arXiv1302.2871S}. For instance, in neither of these experimentally 
accessible cases does the speed of second sound increase significantly at low temperatures. Another difference is that in superfluid $^4$He, second sound is 
predominantly a temperature wave 
for almost all temperatures, except for a regime close to the critical temperature. This shows that the behavior of the sound waves is very sensitive to the 
details of the underlying theory, i.e., the details of the interaction. We see from Eq.\ (\ref{conformal}) that even in the ultra-relativistic, scale-invariant 
limit the speed of second sound depends on thermodynamic functions that can be significantly different in different theories. 
Another feature of the second sound in $^4$He is a rapid decrease in a regime where rotons start to become important \cite{khala,donnelly}. Our model for 
a complex scalar field also gives rise to a massive mode whose mass is $\epsilon_{k=0}^-=\sqrt{6}\,\mu$ (the difference from rotons being that the minimum of the 
dispersion is at zero momentum). For instance for the case $m=v=0$ this means that the mass in units of the critical temperature is 
$\epsilon_{k=0}^-=0.1\,T_c$ (for the weaker coupling $\lambda=0.005$) and $\epsilon_{k=0}^-=0.3\,T_c$ (for the stronger coupling $\lambda=0.005$). Therefore, 
our sound velocities are dominated by the Goldstone mode only for temperatures $T \ll 0.1 \, T_c$ while for higher temperatures the massive mode plays an important 
role, even though there appears to be no characteristic drop in $u_2$ at the onset of that mode.

\bigskip
{\it Nonzero superflow.} In the presence of a nonzero, uniform superflow---a relative flow between superfluid and normal-fluid components, measured in the
normal-fluid rest frame---the sound velocities obviously become anisotropic. In Figs.\ \ref{figsoundflow1} and \ref{figsoundflow2} ($\lambda=0.05$ and $\lambda=0.005$, 
respectively) we plot the speeds of sound and the corresponding mixing angles for the amplitudes for seven different directions of the sound wave 
with respect to the superfluid velocity ${\bf v}$, from downstream propagation (uppermost curves in all panels) through perpendicular propagation 
(dashed curves in all panels) to upstream propagation (lowermost curves in all panels). We see that both sound speeds are faster in the 
forward direction, as was already observed in the low-temperature results of Ref.\ \cite{2013PhRvD..87f5001A}.  Since the superflow $\nabla\psi$,
like the mass parameter $m$,
introduces an additional energy scale, the speed of first sound $u_1$ 
deviates from the scale-invariant value, at least at low temperatures. 
In the ultra-relativistic limit, $u_1$ approaches the scale-invariant value at high temperatures 
from above (from below) for a downstream (upstream) sound wave. 
The value of the superflow used in the figures
corresponds to about 1\% of the 
critical temperature, $|\nabla\psi|\sim 0.01\,T_c(v)$ for the stronger coupling, Fig.\ \ref{figsoundflow1}, and to about about 0.4\% of the critical temperature for the
weaker coupling, Fig.\ \ref{figsoundflow2}. The critical temperatures are $T_c(v)\simeq 7.62\,\mu\, [6.04\,\mu]$ for $\lambda=0.05$ and $m=0\, [0.6\,\mu]$ and 
$T_c(v)\simeq 24.2\,\mu \,[19.23\,\mu]$ for $\lambda=0.005$ and $m=0 \,[0.6\,\mu]$.

The low-temperature behavior of the sound speeds can also be computed analytically for nonzero superflow. 
The expressions for the case where both $m$ and ${\bf v}$ are nonzero are
very complicated. But, for $m=0$ the dependence on the superfluid velocity ${\bf v}$ can be written in a relatively compact way, see Eqs.\ (85) of 
Ref.\ \cite{2013PhRvD..87f5001A}. We compare these analytical results with the full 2PI results in the right panel of Fig.\ \ref{figsoundlowT}. Even though
we only show the comparison for $u_2$, we have checked that the numerical results agree with the low-temperature approximation also for $u_1$. 

At the critical point, there is a sizable nonzero value of the speed of second sound for all 
angles. In contrast to the case without superflow, this is not only due to our use of the Hartree approximation. Remember from the 
discussion in Sec.\ \ref{sec:results1} that $T_c(v)$ is the point beyond which there is no stable uniform superfluid, see in particular the phase diagram in 
Fig.\ \ref{figorder}. At that critical point, the condensate is not zero (and is not expected to be zero in a more complete treatment), and therefore we do not 
expect $u_2$ to go to zero.

Comparing Figs.\ \ref{figsoundflow1} and \ref{figsoundflow2} we observe that a weaker coupling leads again to a more pronounced avoided crossing effect. 
This is particularly obvious from the upper right panel of Fig.\ \ref{figsoundflow2}, where we observe the avoided crossing effect now for each angle separately.
Like for vanishing superflow, a weaker coupling tends to shift the point of the role reversal to lower temperatures, even though this
statement is not completely general. Namely, in the ultra-relativistic limit we see that changing the coupling has a more complicated effect for the 
sound waves that propagate in the backward direction, see curves below the dashed one in the lower left panels of Figs.\ \ref{figsoundflow1} and \ref{figsoundflow2}. 
As a consequence, we find the following interesting phenomenon: depending on the external parameters, there can be a sizable temperature regime of intermediate 
temperatures where a second sound wave, sent out in the forward direction, is almost a pure chemical potential wave while sent out in the backward 
direction it is almost a pure temperature wave (and vice versa for the first sound). This effect is most pronounced for weak coupling and the ultra-relativistic limit. 
We have checked that it gets further enhanced by a larger value of the superflow. In other words, the role reversal in the sound modes
does not only occur by changing temperature (most pronounced in the {\it non-relativistic} case at weak coupling), but can also occur by changing the direction of
the sound wave (most pronounced in the {\it ultra-relativistic} case at weak coupling).

\section{Summary and outlook}
\label{sec:sum}

We have computed properties of a bosonic relativistic superfluid for all temperatures below the critical temperature within the 2PI formalism. 
As a microscopic starting point we have used a model Lagrangian for a complex scalar field with mass $m$ and a quartic interaction term with coupling 
constant $\lambda$. Our work builds on the connection between field theory and the two-fluid picture that was developed in our previous work 
\cite{2013PhRvD..87f5001A}. It addresses formal aspects of the 2PI approach such as renormalization and presents new physical results within that approach 
such as the velocities of first and second sound for all temperatures. 

\subsection{Formalism}

Even though the 2PI formalism is well suited to the treatment of systems with spontaneous symmetry breaking, in practice it has several difficulties, 
and we now describe how we have addressed them.

Firstly, the renormalization of the theory is nontrivial because there are ultraviolet divergences
in the action and stationarity equations which implicitly depend on the medium through the self-consistent masses. Presumably such unwanted dependences would 
be absent in a more complete treatment that takes into account the momentum dependence of the order parameter. We follow the approach adopted in
the existing literature, introducing counterterms on the level of the effective action to achieve renormalizability.
We have pointed out an additional ultraviolet divergence in this approach,
arising from nonzero superflow.

Secondly, the two-loop truncation of the 2PI effective action violates the Goldstone theorem by giving a small mass to 
the Goldstone mode. In the physics of a superfluid, however, the masslessness of the Goldstone mode is crucial since it determines the low-energy properties of the system. We have therefore built the Goldstone theorem into our calculation by hand, using a modification of the stationarity equations. 
This means that we do 
not work at the minimum of the potential, but at a point slightly away from that minimum. In particular, we have evaluated the effective action at that 
``Goldstone point''.

Thirdly, we have employed the Hartree approximation, meaning that we have
neglected the contribution to the effective action from the cubic interactions
that are induced by the condensate. This approximation is particularly simple
since the self-energy is then momentum-independent. The price one has to pay, however, is that the phase transition to the non-superfluid phase becomes
first order, while a complete treatment predicts a second order phase
transition. We control this problem by restricting our calculation
to weak coupling, in which case
the unphysical discontinuity of the order parameter at the critical point
is small, as is the sensitivity of our final results to the
arbitrary renormalization scale.

\subsection{Physical results}

One of our physical results is the critical velocity for superfluidity. The critical velocity manifests itself through the onset of an instability 
(negative energy) in the dispersion relation of the Goldstone mode. We have computed the critical velocity for all temperatures. At low temperatures, our critical 
velocity is in agreement with the original version of Landau's argument, which is based on a Lorentz (or Galilei) transformation of the dispersions 
at vanishing superflow. In general, however, the Goldstone dispersion at finite superflow is not just obtained by a Lorentz transformation. A superflow also 
affects the condensate which in turn influences the dispersion relation. This effect is taken into account in our self-consistent formalism and turns out to decrease the 
critical velocity sizably at intermediate temperatures. As a result of this calculation, we have presented a phase diagram in the plane of temperature and superfluid 
velocity. This phase diagram is incomplete in the sense that we have restricted ourselves to homogeneous phases. In particular, we have not constructed a superfluid phase
for velocities beyond the critical one. 

Next, we have computed the superfluid and normal-fluid densities. They are relevant if the charge current is decomposed into superfluid and normal parts, as 
in the original non-relativistic two-fluid approach. Alternatively, 
one can build the superfluid hydrodynamics on charge current and entropy current. We have also computed the relevant microscopic input for this approach. 
Most notably, this approach 
involves the so-called entrainment coefficient, which 
expresses the degree to which each current responds to the conjugate
momentum originally associated with the other current.
We have seen that the entrainment between the currents becomes larger with 
temperature and is also increased significantly by increasing the 
microscopic coupling $\lambda$. 

Finally, we have computed the velocities of first and second sound. We based this calculation on our previous 
work \cite{2013PhRvD..87f5001A}, which includes a nonzero superflow. (The calculation of the sound modes always requires at least an infinitesimal 
superflow; by nonzero superflow we mean larger than infinitesimal.) 
Within the 2PI formalism, the calculation requires us to compute first and second derivatives of the pressure with respect to the temperature, chemical potential, and 
superflow. To avoid numerical uncertainties we have computed these derivatives in a semi-analytical way. Our low-temperature results are in agreement 
with the analytical approximations of Ref.\ \cite{2013PhRvD..87f5001A}. We find that these approximations are valid only for a very small
temperature regime whose size depends on the value of the coupling. Even for the smallest coupling we have used, $\lambda=0.005$, the approximation deviates 
significantly from the full 2PI numerical result for all temperatures higher than about 0.1\% of the critical temperature. The main reason seems to be the 
temperature dependence of the condensate which was neglected in the approximation of Ref.\  \cite{2013PhRvD..87f5001A}. This approximation also neglects the 
massive mode that is present in our theory (and which is not unlike the roton in superfluid helium); the massive mode becomes important for temperatures higher than 
about 10\% of the critical temperature. 

We have investigated the dependence of the sound velocities on the coupling $\lambda$, the boson mass $m$, and the superflow $\nabla\psi$. 
For $m=\nabla\psi=0$ the speed of first sound assumes the universal 
value $\frac{1}{\sqrt{3}}$ for all temperatures, while an additional scale, provided by $m$ and/or $\nabla\psi$, leads to a deviation from this result. 
For temperatures higher than that scale but still lower than the critical temperature (if such a regime exists) the speed of first sound again approaches 
$\frac{1}{\sqrt{3}}$. The speed of second sound is more 
sensitive to details of the system and has a non-universal behavior even for  $m=\nabla\psi=0$. In our particular model
we found a strong increase for low temperatures before a decrease sets in, eventually leading to a vanishing speed of second sound at the critical 
point, if the superflow is zero. 

By computing the amplitudes in chemical potential and temperature of the sound waves for all temperatures, we have confirmed that first sound is always 
an in-phase oscillation of chemical potential and temperature (and thus also of density and entropy) and second sound is always an out-of-phase
oscillation, which can thus be viewed as their defining property. However, the degree to which a given sound wave is a density or entropy wave
depends on the temperature. We have shown that, with respect to this property, first and second sound typically reverse their roles as a function of temperature: the in-phase 
(out-of-phase) mode is a pure density (entropy) wave at low temperatures and becomes a pure entropy (density) wave at high temperatures. This observation is in 
agreement with non-relativistic studies \cite{2010NJPh...12d3040H}. While in the non-relativistic case this role reversal occurs 
rather abruptly in the very low temperature regime, it is more continuous in the ultra-relativistic case. We have also found that for certain parameters of 
the model and intermediate temperatures, 
there can be a role reversal at a fixed temperature: if there is a nonzero superflow, the first sound is (almost) a pure entropy wave parallel to the 
superflow and (almost) a pure density wave anti-parallel to the superflow, while the second sound behaves exactly opposite. This interesting effect is most pronounced 
in the ultra-relativistic limit at weak coupling.

\subsection{Outlook}

Our study leaves various open problems for the future. First of all, one might address the issues mentioned above regarding the 2PI formalism. For instance, one might
go beyond the Hartree approximation in order to correct the artifact of the first-order phase transition. This would also allow us to consider larger values of the 
coupling and see how for instance the sound velocities are affected. Of course, the necessary inclusion of the "sunset" diagram would render the 
calculation significantly more complicated due to the resulting momentum dependence of the self-energy. It would also be interesting to 
see how much our results depend on our specific 
strategy for fixing the violation of the Goldstone theorem. There are other suggestions in the literature which one 
could implement, see for instance Refs.\ \cite{Ivanov:2005yj,Pilaftsis:2013xna}. This would 
affect the low-temperature region because the violation of the 
Goldstone theorem appears to be sufficiently small to be negligible at high temperatures \cite{Andersen:2008tn}.

It would be very interesting to see whether our results for the sound modes are of direct relevance for the physics of compact stars. Sound modes 
in the inner crust of a star have recently been computed in a non-relativistic setup in Ref.\ \cite{Chamel:2012ix}. One can also build on our results for a 
hydrodynamic description of the CFL phase, possibly starting from a fermionic formalism and/or adding a second superfluid component 
for kaon condensation. Since the $U(1)$ that is spontaneously broken by kaon condensation is only an approximate symmetry, 
one first has to resolve some fundamental questions for such a ``broken superfluid'' \cite{future}. 

Besides the astrophysical applications, our work also raises some interesting questions regarding superfluids that are accessible in the laboratory. 
In view of recent measurements
of second sound in an ultra-cold fermionic gas \cite{2013arXiv1302.2871S}, it would be very interesting to see whether one may create a temporary superflow in 
these experiments and possibly observe the role reversal we have discussed as a function of the direction of the sound wave. Of course, as our results suggest, 
this effect tends to be weaker in the non-relativistic regime, and it might thus be difficult to see experimentally.

\begin{acknowledgments}
We are grateful to Amadeo Jim\'{e}nez-Alba and Karl Landsteiner for helpful comments and discussions. This work has been supported by the Austrian 
science foundation FWF under project no.~P23536-N16, and by U.S.~Department of Energy under contract
\#DE-FG02-05ER41375, 
and by the DoE Topical Collaboration 
``Neutrinos and Nucleosynthesis in Hot and Dense Matter'', 
contract \#DE-SC0004955.

\end{acknowledgments}

\appendix

\section{Renormalization}
\label{AppA}

In this appendix we discuss the renormalization of our 2PI approach.  The renormalization procedure is done on the level of the effective action. Its general 
form (\ref{Psifull}) can be written as
\be \label{Psicond}
\Psi = \frac{\rho^2}{2}(\mu^2-m^2) - \frac{\lambda}{4}\rho^4+ J +\frac{M^2-m^2-2\lambda\rho^2}{2}I^+ +\frac{\delta M^2-\lambda\rho^2}{2}I^- 
- \frac{\lambda}{2}(I^+)^2 - \frac{\lambda}{4}(I^-)^2\, , 
\ee
where we have abbreviated 
\be 
J \equiv  - \frac{1}{2}\frac{T}{V}\sum_K\Tr \ln \frac{S^{-1}}{T^2}  \, , \qquad I^\pm \equiv \frac{T}{V}\sum_K [S_{11}(K)\pm S_{22}(K)] \, .
\label{JI-def}
\ee 
For the $\Tr[S_0^{-1}S-1]$ term we have used the tree-level propagator (\ref{S0inv}) and the ansatz for the propagator (\ref{Sinv}), while for $V_2$ we have used 
the definition (\ref{V2}) and the fact that the propagator is antisymmetric, $S_{12}=-S_{21}$.

We shall add a counterterm $\delta\Psi$ to $\Psi$, such that the effective action becomes renormalized at the stationary point. It is instructive to start
with the non-superfluid case where there is no condensate, then discuss the case with condensate but without superflow, and then turn to the most
complicated case that includes condensate and superflow.

\subsection{Uncondensed phase with (spurious) background field $\nabla\psi$}
\label{sec:normal}

First we consider the high-temperature, non-superfluid, phase. We can formally include a background field $\nabla\psi$ also in this phase, although 
we shall see that the physics will turn out to be independent of $\nabla\psi$. If the condensate vanishes, there is no need to introduce two different 
self-consistent masses $M$ and $\delta M$, and the full propagator is given by 
\be
S^{-1}(K) = \left(\begin{array}{cc} -K^2+M^2-\sigma^2 & 2iK_\mu\partial^\mu\psi \\[2ex] -2iK_\mu\partial^\mu\psi & -K^2+M^2-\sigma^2\end{array}\right) \, .
\label{S-inv}
\ee   
From the poles of the propagator we obtain the dispersion relations  
\be \label{epskee}
\epsilon_{\bf k}^e = \sqrt{({\bf k}-e\nabla\psi)^2+M^2}-e\mu \, ,
\ee
where $\mu=\partial_0\psi$. These are simply the usual particle and anti-particle excitations, carrying one unit of positive and negative charge, respectively, 
but with the spatial momentum shifted by $\nabla\psi$, for particles and anti-particles in opposite directions.  

The effective action in the uncondensed phase is given by Eq.\ (\ref{Psicond}) with $\rho=0$. Moreover, since there is only one self-consistent mass $M$, we also have 
$\delta M=0$ and $I^-=0$, 
\be \label{psinormal}
\Psi = J +\frac{M^2-m^2}{2}I^+ - \frac{\lambda}{2}(I^+)^2 \, .
\ee
We now add counterterms to the effective action in order to cancel the infinities in $J$ and $I^+$, 
\bea \label{delPsi}
\delta \Psi &=& -\frac{\delta m^2}{2} I^+ - \frac{\delta\lambda}{2} (I^+)^2 \, . 
\eea
The recipe for finding these counterterms is very simple: we add counterterms $\delta m^2$ and $\delta\lambda$ to each mass squared and each coupling constant 
that appears in the action (\ref{psinormal}) (neither $J$ nor $I^+$ depend on $m$ or $\lambda$ explicitly). This will be a bit less straightforward in the 
condensed phase, where we shall need two different counterterms $\delta\lambda_1$ and $\delta\lambda_2$, see next subsection.
The mass counterterm $\delta m^2$ is of order $\lambda$, while $\delta \lambda$ is of order $\lambda^2$. 
The crucial point will be to show that all divergences can be cancelled with medium independent quantities $\delta m^2$ and $\delta\lambda$. Of course, the 
total counterterm $\delta\Psi$ does depend on the medium because $I^+$ and $J$ depend on $\mu$, $T$, and $\nabla\psi$. 
Let us first discuss the renormalized stationarity equation. In the uncondensed phase there is only one equation, for the self-consistent mass $M$,
\be \label{statnormal}
M^2=m^2+\delta m^2 + 2(\lambda+\delta \lambda) I^+   \, .   
\ee
In evaluating integrals like $I^+$ we will
use a notation where a subscripted argument
indicates subtraction of the function's value when that argument is zero,
\be
I_x(A) \equiv I(x,A) - I(0,A) \, .
\label{Isub-notation}
\ee
Using that notation, we split the integral $I^+$ into a zero 
temperature part that
depends on the cutoff $\Lambda$
and a part $I^+_T$ that depends on $T$ but goes to zero as $T\to 0$
and is cutoff-independent,
\be
I^+(T,{}\Lambda) = I^+(0,{}\Lambda) + I^+_T \, , 
\label{IsubT}
\ee
where the dependence on $\mu,M,\Del\psi$ is not explicitly shown.
Evaluating the Matsubara sum, we find
\be
I^+(0,{}\Lambda) = \frac{1}{2}\sum_{e=\pm}\int\frac{d^3{\bf k}}{(2\pi)^3}\frac{1}{\sqrt{({\bf k}-e\nabla\psi)^2+M^2}} \, ,\qquad 
I_T^+\equiv  
\sum_{e=\pm}\int\frac{d^3{\bf k}}{(2\pi)^3}\frac{f(\epsilon_{\bf k}^e)}{\sqrt{({\bf k}-e\nabla\psi)^2+M^2}} \, ,
\ee
where $f$ is the Bose distribution function. The
terms in the large-momentum expansion of the integrand
that lead to cutoff dependences are shown in Table \ref{table0}
(with  $\delta M=0$ for the uncondensed case).
\begin{table*}[t]
\begin{tabular}{|c|c|c|c|} 
\hline
\rule[-1.5ex]{0em}{6ex} 
 & $I^+$ & $I^-$ & $J$ \\[2ex]
\hline
\rule[-1.5ex]{0em}{6ex} 
$\;\;$
\parbox{6em}{UV behavior\\of integrand}$\;\;$ &$\;\;$ $\displaystyle{k-\frac{M^2}{2k} + \ldots }$$\;\;$ &$\;\;$ $\displaystyle{-\frac{\delta M^2}{2k}+ \ldots}$ $\;\;$ & 
$\;\;$$\displaystyle{-k^3-\left[M^2+\frac{2}{3}(\nabla\psi)^2\right]\frac{k}{2}+\frac{M^4+\delta M^4}{8k} + \ldots }$$\;\;$
\\[2ex] \hline

\end{tabular}
\caption{Ultraviolet divergent contributions to the various integrands of the three-momentum integrals, see definitions (\ref{JI-def}). 
The contributions are given for the most general case with condensation and superflow. The limit cases discussed in detail in this appendix are obtained by 
setting $\delta M=0$ (uncondensed case) and $\nabla\psi=0$ (condensed case without superflow). The divergent terms depend 
implicitly on temperature, chemical potential, and the superfluid velocity, the latter appearing even explicitly in the divergent terms of $J$.}
\label{table0}
\end{table*}
We evaluate the momentum integral $I^+(0,\Lambda)$ via proper time regularization \cite{Schwinger:1951nm}, using the general relation
\be
\frac{1}{x^a} = \frac{1}{\Gamma(a)}\int_0^\infty d\tau\,\tau^{a-1}e^{-\tau x}\,  ,
\ee
where, in this case, $x=({\bf k}-e\nabla\psi)^2+M^2$, and exchange the order of the ${\bf k}$ and $\tau$ integrals.
The ${\bf k}$ integral is now finite, so we can eliminate $\nabla\psi$
because it is simply a shift of the integration variable. The
ultraviolet cutoff $\Lambda$ is implemented by setting the lower limit
of the proper time integral to $1/\Lambda^2$.
This yields
\begin{subequations}
\bea\label{I0reg}
I^+(T,{}\Lambda) &=& 
\frac{\Lambda^2}{8\pi^2}-\frac{M^2}{8\pi^2}\ln\frac{\Lambda^2}{\ell^2}
+ I^+_{\rm finite}(T,\ell) \,  ,\\[2ex]
 I^+_{\rm finite}(T,\ell) &=&
\frac{M^2}{8\pi^2}\left(\gamma-1+\ln\frac{M^2}{\ell^2}\right) 
+ I^+_T \, ,
\eea
\end{subequations}
where we have introduced the renormalization scale $\ell$, and where $\gamma\simeq 0.5772$ is the Euler-Mascheroni constant. 
 
We can now insert the regularized integral into the stationarity equation (\ref{statnormal}), and separate it in to a cutoff-independent part
\be \label{statnormalR}
M^2 = m^2 +2\lambda I_{\rm finite}^+(T,\ell) \, , 
\ee
and a cutoff-dependent part
\be
0 = \delta m^2 + \frac{\lambda+\delta\lambda}{4\pi^2}\left(\Lambda^2-M^2\ln\frac{\Lambda^2}{\ell^2}\right) + 2\delta\lambda\, I^+_{\rm finite}(T,\ell)\, .
\ee
Note that the ambiguity in performing this separation corresponds to
choosing the renormalization scale $\ell$.
In order to determine $\delta m^2$ and $\delta\lambda$, we eliminate $I^+_{\rm finite}$ with the help of Eq.\ (\ref{statnormalR}). The resulting equation has two
contributions, one of which is medium independent and one of which is proportional to $M^2$. Both contributions have to vanish separately, and thus we obtain
two equations for $\delta m^2$ and $\delta\lambda$ whose solutions are
\be
\delta\lambda = \frac{\lambda^2}{4\pi^2}\ln\frac{\Lambda^2}{\ell^2}\left(1-\frac{\lambda}{4\pi^2}\ln\frac{\Lambda^2}{\ell^2}\right)^{-1} \, , \qquad
\delta m^2 = \delta\lambda\left(\frac{m^2}{\lambda}-\frac{\Lambda^2}{4\pi^2}\right)-\lambda\frac{\Lambda^2}{4\pi^2} \, .
\ee
If we introduce the bare mass $m_{\rm bare}^2 = m^2 + \delta m^2$ and the bare coupling, $\lambda_{\rm bare} = \lambda + \delta \lambda$, we can write
\be\label{lRmR}
\frac{1}{\lambda} = \frac{1}{\lambda_{\rm bare}}+\frac{1}{4\pi^2}\ln\frac{\Lambda^2}{\ell^2} \, , \qquad \frac{m^2}{\lambda} 
= \frac{m_{\rm bare}^2}{\lambda_{\rm bare}}+\frac{\Lambda^2}{4\pi^2} \, .
\ee
Next, we need to check whether the same counterterms cancel all divergences in the pressure $\Psi+\delta\Psi$.
Again, we write 
\be
J(T,{}\Lambda) = J(0,{}\Lambda) + J_T \, , 
\ee
where again we do not show the dependence on $\mu,M,\Del\psi$, and where, after performing the Matsubara sum and taking the thermodynamic limit,
we have
\be \label{J0JT}
J(0,{}\Lambda) = -\frac{1}{2}\sum_{e=\pm}\int\frac{d^3{\bf k}}{(2\pi)^3} \epsilon_{\bf k}^e \, ,  \qquad 
J_T = -T\sum_{e=\pm}\int\frac{d^3{\bf k}}{(2\pi)^3}\ln\left(1-e^{-\epsilon_{\bf k}^e/T}\right) \, . 
\ee 
Proper time regularization eliminates the cutoff-dependent term that
depended explicitly on the background field $\nabla\psi$,
and we find
\begin{subequations}
\bea \label{J01}
J(T,{}\Lambda) 
&=& \frac{\Lambda^4}{32\pi^2}-\frac{M^2\Lambda^2}{16\pi^2}+\frac{M^4}{32\pi^2}\ln\frac{\Lambda^2}{\ell^2} + J_{\rm finite}(T,\ell) \, , \\[2ex]
J_{\rm finite}(T,\ell) &=&\frac{M^4}{32\pi^2}\left(\frac{3}{2}-\gamma -\ln\frac{M^2}{\ell^2}\right) + J_T \, .
\eea
\end{subequations}
Inserting this into $\Psi+\delta\Psi$, using the stationarity equation (\ref{statnormal}) to eliminate $I^+$, and making use of the relations
(\ref{lRmR}), we find that indeed all medium dependent divergences in $\Psi+\delta\Psi$ are cancelled. We are left with 
\be \label{Psinormalren}
\Psi+\delta\Psi = \frac{\Lambda^4}{32\pi^2}-\frac{m^4}{8\lambda}+\frac{m_{\rm bare}^4}{8\lambda_{\rm bare}} +\frac{(M^2-m^2)^2}{8\lambda}+J_{\rm finite}(T,\ell) \, . 
\ee
The first three terms on the right-hand side are 
independent of the
thermodynamic parameters $\mu$, $T$, and $\nabla\psi$,  and hence have
no effect on the physics.

\subsection{Condensed phase without superflow}
\label{sec:cond-noflow}

As a next step, we consider the condensed phase, but first without supercurrent, $\nabla\psi=0$.
In this case, with $\mu=\partial_0\psi$, the inverse propagator is\footnote{Remember that $M^2\pm\delta M^2$ is just a notation for two different self-consistent masses,
as in Ref.\ \cite{Alford:2007qa}, i.e., $\delta M^2$ should not be confused with a counterterm.} 
\be 
S^{-1}(K) = \left(\begin{array}{cc} -K^2 + M^2+\delta M^2-\mu^2  & 2ik_0 \mu  \\[2ex]
-2ik_0\mu & -K^2 + M^2- \delta M^2-\mu^2 \end{array}\right) \, , 
\label{Sinv-noflow}
\ee
which leads to the dispersion relations 
\be \label{eps0}
\epsilon_k^e = \sqrt{E_k^2+\mu^2-e\sqrt{4\mu^2E_k^2+\delta M^4}} \, , 
\ee
where
\be
E_k \equiv \sqrt{k^2+M^2} \, .
\ee
Now, the counterterm (\ref{delPsi}) is generalized to  
\bea
\delta\Psi &=& -\frac{\delta m^2}{2}\rho^2 -\frac{2\delta\lambda_1+\delta\lambda_2}{4}\rho^4 -\frac{\delta m^2+2\delta\lambda_1\rho^2}{2} I^+
-\frac{\delta\lambda_2\rho^2}{2}I^- - \frac{\delta\lambda_1}{2} (I^+)^2 - \frac{\delta\lambda_2}{4} (I^-)^2 \, .
\eea
In the condensed phase it is necessary to introduce two different counterterms $\delta\lambda_1$ and $\delta\lambda_2$ for the two structures $I^+$ and $I^-$ 
\cite{Fejos:2007ec}\footnote{In the notation of Ref.\ \cite{Fejos:2007ec}, $\delta\lambda^A\equiv 2\delta\lambda_1-\delta\lambda_2$, 
$\delta\lambda^B\equiv \delta\lambda_2$.}. We could have put another different counterterm in front of the $\rho^4$ term, but we have already anticipated the
result that this counterterm is a particular linear combination of $\delta\lambda_1$ and $\delta\lambda_2$. 

The stationarity equations become, in agreement to Ref.\ \cite{Fejos:2007ec}, 
\begin{subequations} \label{123}
\bea
0&=& \mu^2-(m^2+\delta m^2)-(\lambda+2\delta\lambda_1+\delta\lambda_2)\rho^2-\left[2(\lambda+\delta\lambda_1)I^++(\lambda+\delta\lambda_2)I^-\right] \, , 
\label{one}\\[2ex]
M^2+\delta M^2&=& m^2+\delta m^2+(3\lambda+2\delta\lambda_1+\delta\lambda_2)\rho^2 + 2(\lambda+\delta\lambda_1)I^+ +(\lambda+\delta\lambda_2)I^- \, , \label{two}
\\[2ex]
M^2-\delta M^2&=& m^2+\delta m^2+(\lambda+2\delta\lambda_1-\delta\lambda_2)\rho^2 + 2(\lambda+\delta\lambda_1)I^+-(\lambda+\delta\lambda_2)I^- \, , \label{three}
\eea
\end{subequations}
where the first one is obtained from extremizing the action with respect to $\rho$ and the second and the third are the two nontrivial components of the Dyson-Schwinger 
equation.
Inserting Eq.\ (\ref{two}) into Eq.\ (\ref{one}) as well as adding and subtracting Eqs.\ (\ref{two}) and (\ref{three}) to/from each other yields the 
simpler system of equations
\begin{subequations} \label{simple}
\bea
M^2+\delta M^2 &=& \mu^2+2\lambda\rho^2\, ,  \label{simple1} \\[2ex]
M^2&=& m^2+\delta m^2+2(\lambda+\delta\lambda_1)(\rho^2 + I^+) \, , \label{simple2}\\[2ex]
\delta M^2&=& (\lambda+\delta\lambda_2)(\rho^2 + I^-) \, , \label{simple3} 
\eea
\end{subequations}
where the first equation already has its final, renormalized form.
Using the notation of Eq.~\eqn{Isub-notation}, we rewrite $I^\pm$ by
first separating off the $T$-dependent term, and then separating off the
$\mu$-dependence at $T=0$, leaving a $\mu=T=0$ vacuum term that
contains all the cutoff dependence,
\begin{subequations}\label{Ipm}
\bea
I^\pm(T,\mu,{}\Lambda) &=& I^\pm(0,\mu,{}\Lambda) + I^\pm_T(\mu{}) \, , 
\\[2ex]
I^\pm(0,\mu,{}\Lambda) &=& I^\pm(0,0,{}\Lambda) + I^\pm_\mu(0) \, ,
\label{Imu}
\eea
\end{subequations}
where each quantity has dependence on $(M,\delta M)$ which is not explicitly shown.
As in Ref.\ \cite{Fejos:2007ec}, when we set $\mu$ or $T$ to zero we keep unchanged the mass parameters
of the full propagator $M$ and $\delta M$, even though in reality they depend on $\mu$ and $T$. Evaluating \eqn{Ipm} with the help of \eqn{JI-def} and 
\eqn{Sinv-noflow}, the $T=0$ integrals are
\begin{subequations}
\bea 
I^+(0,\mu,{}\Lambda) &=&  \frac{1}{2}\sum_{e=\pm}\int\frac{d^3{\bf k}}{(2\pi)^3}\frac{1}{\epsilon_k^e}\left(1-\frac{2e\mu^2}{\sqrt{4\mu^2 E_k^2+\delta M^4}}\right) \, , \label{Ip-noflow}\\[2ex]
I^-(0,\mu,{}\Lambda) &=& -\frac{1}{2}\sum_{e=\pm}\int\frac{d^3{\bf k}}{(2\pi)^3}\frac{1}{\epsilon_k^e}\frac{e\delta M^2}{\sqrt{4\mu^2E_k^2+\delta M^4}} \, . \label{Im-noflow}
\eea
\end{subequations}
The thermal integrals $I^\pm_T(\mu)$ are simply given by $I^\pm(0,\mu,\Lambda)$ with an additional factor $2f(\epsilon_k^e)$ in the integrand,
which renders them cutoff-independent.
The vacuum contribution is
\be \label{I00Lam}
I^\pm(0,0,{}\Lambda) = \frac{1}{2}\int\frac{d^3{\bf k}}{(2\pi)^3} \left( \frac{1}{\sqrt{k^2+M^2+\delta M^2}}\pm \frac{1}{\sqrt{k^2+M^2-\delta M^2}}\right) \, ,
\ee
and its cutoff-dependence  arises from the terms given in Table \ref{table0} (after setting $\nabla\psi=0$).
This can be evaluated using proper-time regularization,
\begin{subequations}\label{I-vac-finite}
\bea
I^+(0,0,{}\Lambda) &=& \frac{\Lambda^2}{8\pi^2}- \frac{M^2}{8\pi^2}\ln\frac{\Lambda^2}{\ell^2} + I_{\rm vac,finite}^+(\ell{}) \, ,\qquad 
I^-(0,0,{}\Lambda) = - \frac{\delta M^2}{8\pi^2}\ln\frac{\Lambda^2}{\ell^2}+ I_{\rm vac,finite}^-(\ell{}) \, , \\[2ex]
I_{\rm vac,finite}^\pm(\ell{}) &\equiv& \frac{M^2}{8\pi^2}(\gamma-1) + \frac{M^2+\delta M^2}{16\pi^2}\ln\frac{M^2+\delta M^2}{\ell^2}
\pm \frac{M^2-\delta M^2}{16\pi^2}\ln\frac{M^2-\delta M^2}{\ell^2} \, .
\eea
\end{subequations}
The finite parts $I^\pm_{\rm finite}$ of $I^\pm$ are then given by
\be
I^\pm_{\rm finite}(T,\mu,\ell) = I_{\rm vac,finite}^\pm(\ell{}) 
+ I^\pm_\mu(0{}) + I^\pm_T(\mu{}) \, ,
\label{Ifinite}
\ee
where $I^\pm_\mu(0)$ is obtained via \eqn{Imu}, by numerically evaluating
$I^\pm(0,\mu,\Lambda)-I^\pm(0,0,\Lambda)$, combining them into one
cutoff-independent integral.
Now we can come back to the stationarity equations (\ref{simple}). The first of these equations does not contain any divergences anymore. With Eqs.\ (\ref{simple2}) 
and (\ref{simple3}) we proceed analogously as explained for the uncondensed phase: we insert Eqs.\ (\ref{I-vac-finite}) and separate finite and infinite contributions.
The finite contributions are the renormalized equations
\begin{subequations} \label{renorm12}
\bea
M^2&=& m^2+2\lambda[\rho^2 + I^+_{\rm finite}(T,\mu,\ell)] \, , \label{renorm121}\\[2ex]
\delta M^2&=& \lambda[\rho^2 + I^-_{\rm finite}(T,\mu,\ell)] \, . \label{renorm122} 
\eea
\end{subequations}
In the equations for the infinite contributions  we first 
eliminate $\rho$ and $I^\pm_{\rm finite}$ with the help of Eqs.\ (\ref{renorm12}) and then separate 
medium-independent terms from terms proportional to $M^2$ for Eq.\ (\ref{simple2}) and $\delta M^2$ for Eq.\ (\ref{simple3}). 
The requirement that all infinities cancel yields the conditions
\begin{subequations} \label{d1d2}
\bea
\delta\lambda_1 &=& \frac{\lambda^2}{4\pi^2}\ln\frac{\Lambda^2}{\ell^2}\left(1-\frac{\lambda}{4\pi^2}\ln\frac{\Lambda^2}{\ell^2}\right)^{-1} \, , \qquad
\delta m^2 = \delta\lambda_1\left(\frac{m^2}{\lambda}-\frac{\Lambda^2}{4\pi^2}\right)-\lambda\frac{\Lambda^2}{4\pi^2} \, , \label{dl1} \\[2ex]
\delta\lambda_2 &=& \frac{\lambda^2}{8\pi^2}\ln\frac{\Lambda^2}{\ell^2}\left(1-\frac{\lambda}{8\pi^2}\ln\frac{\Lambda^2}{\ell^2}\right)^{-1} \, , 
\eea
\end{subequations}
which confirms that $\delta\lambda_1$ and $\delta\lambda_2$ are indeed different. 
By introducing the two bare couplings $\lambda_{1/2,{\rm bare}} = \lambda+\delta\lambda_{1/2}$ and the bare mass $m_{\rm bare}^2=m^2+\delta m^2$ we 
can write this in a more compact way,
\be \label{renormml}
\frac{1}{\lambda} = \frac{1}{\lambda_{1,{\rm bare}}}+\frac{1}{4\pi^2}\ln\frac{\Lambda^2}{\ell^2} \, , \qquad \frac{m^2}{\lambda} = \frac{m_{\rm bare}^2}
{\lambda_{1,{\rm bare}}}+\frac{\Lambda^2}{4\pi^2}
\, , \qquad 
\frac{1}{\lambda} = \frac{1}{\lambda_{2,{\rm bare}}}+\frac{1}{8\pi^2}\ln\frac{\Lambda^2}{\ell^2} \, .
\ee
Finally, we need to check that all divergences in the pressure cancel. 
This requires evaluation of $J$ in Eq.\ (\ref{Psicond}). 
In analogy with our evaluation of $I^\pm$, we separate the $T$
and $\mu$ dependence from the vacuum term, writing
\begin{subequations}
\bea
J(T,\mu,{}\Lambda) &=& J(0,\mu,{}\Lambda) + J_T(\mu{}) \, , \\[2ex]
J(0,\mu,{}\Lambda) &=& J(0,0,{}\Lambda) + J_\mu(0{}) \, .
\eea
\end{subequations}
The $T=\mu=0$ ``vacuum'' integral is
\be
J(0,0,{}\Lambda) = -\frac{1}{2}\int\frac{d^3{\bf k}}{(2\pi)^3}\left(\sqrt{k^2+M^2+\delta M^2}+\sqrt{k^2+M^2-\delta M^2}\right) \, .
\ee
Evaluating this using a proper-time regulator we find
\begin{subequations}
\bea 
J(0,0,{}\Lambda) &=& 
\frac{\Lambda^4}{32\pi^2}-\frac{\Lambda^2 M^2}{16\pi^2}+\frac{M^4+\delta M^4}{32\pi^2}\ln\frac{\Lambda^2}{\ell^2} + J_{\rm vac,finite}(\ell{}) \, ,
\label{J0cond} \\[2ex]
J_{\rm vac,finite}(\ell{}) &\equiv&
\frac{M^4+\delta M^4}{64\pi^2}(3-2\gamma)-\frac{(M^2+\delta M^2)^2}{64\pi^2}
\ln\frac{M^2+\delta M^2}{\ell^2}-\frac{(M^2-\delta M^2)^2}{64\pi^2}\ln\frac{M^2-\delta M^2}{\ell^2} \, .
\eea
\end{subequations}
The finite part of $J$ is then the finite part of the vacuum contribution
plus the $\mu$ and $T$ dependence,
\be
J_{\rm finite}(T,\mu,\ell) = J_{\rm vac,finite}(\ell{}) 
+  J_\mu(0{}) + J_T(\mu{}) \, .
\label{J-finite}
\ee
By using Eqs.\ (\ref{simple2}) and (\ref{simple3}) to eliminate $I^+$ and $I^-$ we obtain
\be
\Psi + \delta \Psi = \frac{\rho^2}{2}(\mu^2-m^2) - \frac{\lambda}{4}\rho^4 + J -\frac{\delta m^2}{2}\rho^2 -\frac{2\delta\lambda_1+\delta\lambda_2}{4}\rho^4
+\frac{(M^2-m_{\rm bare}^2-2\lambda_{1,{\rm bare}}\rho^2)^2}{8\lambda_{1,{\rm bare}}} + \frac{(\delta M^2-\lambda_{2,{\rm bare}}\rho^2)^2}{4\lambda_{2,{\rm bare}}} \, .
\ee
With the help of Eqs.\ (\ref{renormml}) we rewrite the last two terms of this expression,
\begin{subequations}
\bea
\frac{(M^2-m_{\rm bare}^2-2\lambda_{1,{\rm bare}}\rho^2)^2}{8\lambda_{1,{\rm bare}}}
&=& \frac{(M^2-m^2-2\lambda\rho^2)^2}{8\lambda} -\frac{m^4}{8\lambda}+\frac{m^4}{8\lambda_1}
+\frac{M^2\Lambda^2}{16\pi^2}-\frac{M^4}{32\pi^2}\ln\frac{\Lambda^2}{\ell^2} +\frac{\delta m^2}{2}\rho^2 +\frac{\delta\lambda_1}{2}\rho^4 \, , \hspace{1cm}\\[2ex]
\frac{(\delta M^2-\lambda_{2,{\rm bare}}\rho^2)^2}{4\lambda_{2,{\rm bare}}}&=& 
\frac{(\delta M^2-\lambda\rho^2)^2}{4\lambda} -\frac{\delta M^4}{32\pi^2}\ln\frac{\Lambda^2}{\ell^2}
+\frac{\delta\lambda_2}{4}\rho^2 \, .
\eea
\end{subequations}
We see that the divergences appearing here cancel all divergences from $J$ in
Eq.\ (\ref{J0cond}) that depend on $M$ and $\delta M$, and we arrive at
the renormalized pressure
\be \label{psiren0}
\Psi + \delta\Psi = \frac{\Lambda^4}{32\pi^2}-\frac{m^4}{8\lambda}+\frac{m_{\rm bare}^4}{8\lambda_{1,{\rm bare}}} 
+ \frac{\rho^2}{2}(\mu^2-m^2) - \frac{\lambda}{4}\rho^4 
+ J_{\rm finite}(T,\mu,\ell) + \frac{(M^2-m^2-2\lambda\rho^2)^2}{8\lambda}+ \frac{(\delta M^2-\lambda\rho^2)^2}{4\lambda} \, .
\ee
The first three terms on the right-hand side are
independent of the thermodynamic parameters $\mu$ and $T$, and hence
have no effect on the physics;
the next two terms are the renormalized tree-level
potential; then, $J_{\rm finite}$ is the finite part of the $\Tr\ln S^{-1}$ term, while the last two terms are the renormalized version of the combined terms 
coming from $\Tr[S_0^{-1}S-1]$ and  $V_2$.

\subsection{Renormalization with Goldstone mode}
\label{AppGold}

As discussed in Sec.\ \ref{sec:gold}, the present formalism violates the Goldstone theorem, and since our discussion of the superfluid properties
requires an exact Goldstone mode we need to consider modified stationarity equations. We thus have to check how our modification affects the 
renormalization and what the renormalized pressure at the new ``Goldstone point'' is (which is slightly off the ``stationary point''). 
To this end, we emphasize that the renormalization procedure explained above is designed to work at the stationary point. In particular, 
Eq.\ (\ref{psiren0}) is the renormalized pressure at that point because we have used the stationarity equations (\ref{simple}) that include finite as well as 
infinite parts. It seems we would have to redo our whole analysis for the ``Goldstone point''. However, we may simply do the modification in the {\it finite} 
part of the stationarity equations, thus preserving all the results for the counterterms. This amounts to changing Eq.\ (\ref{renorm122}) to 
\be \label{modify}
\delta M^2= \lambda \rho^2  \, ,  
\ee
but keeping the two other renormalized equations (\ref{simple1}) and (\ref{renorm121}) as well as all infinite contributions in Eqs.\ (\ref{simple2}) and (\ref{simple3})
as they are. 
It is then obvious that the counterterms are still given by Eqs.\ (\ref{d1d2}). All we need to do is compute the finite part of the pressure;
by construction, all infinities in the pressure will still cancel. We can thus simply replace all integrals in Eq.\ (\ref{Psicond}) by their finite parts, 
and use Eqs.\ (\ref{renorm121}) and (\ref{modify}) to find 
\be 
\Psi+\delta\Psi = \frac{\Lambda^4}{32\pi^2}-\frac{m^4}{8\lambda}+\frac{m_{\rm bare}^4}{8\lambda_{1,{\rm bare}}} + 
\frac{\rho^2}{2}(\mu^2-m^2) - \frac{\lambda}{4}\rho^4
+ J_{\rm finite} +\frac{(M^2-m^2-2\lambda\rho^2)^2}{8\lambda}
- \frac{\lambda}{4}(I^-_{\rm finite})^2\, . 
\ee

\subsection{Condensed phase with superflow}
\label{sec:condsuper}

Following the procedure of Sec.~\ref{sec:cond-noflow}, we first separate the integrals $I^\pm$ and $J$
into their thermal parts $I^\pm_T(\mu)$ and $J_T(\mu)$ and the cutoff dependent 
integrals\footnote{For explicit numerical calculations, the identity 
\be
(\epsilon_{\bf k}^e+\epsilon_{-{\bf k}}^e)(\epsilon_{\bf k}^e+\epsilon_{-{\bf k}}^{-e})(\epsilon_{\bf k}^e-\epsilon_{\bf k}^{-e}) 
= 4\left\{\epsilon_{\bf k}^e\left[(\epsilon_{\bf k}^e)^2-k^2-M^2-(\partial_0\psi)^2-(\nabla\psi)^2\right]-2\partial_0\psi\,{\bf k}\cdot\nabla\psi\right\} \nonumber
\ee
can be useful, the right-hand side being simpler due to the fewer appearances of the complicated excitation energies.}
\begin{subequations}\label{Ipm-flow}
\bea
I^+(0,\mu,\Lambda) &=& 2\sum_{e=\pm}\int\frac{d^3{\bf k}}{(2\pi)^3}\frac{(\epsilon_{\bf k}^e)^2-k^2-M^2+\sigma^2}{(\epsilon_{\bf k}^e+\epsilon_{-{\bf k}}^e)
(\epsilon_{\bf k}^e+\epsilon_{-{\bf k}}^{-e})(\epsilon_{\bf k}^e-\epsilon_{\bf k}^{-e})} \, , \\[2ex]
I^-(0,\mu,\Lambda)  &=& 2\sum_{e=\pm}\int\frac{d^3{\bf k}}{(2\pi)^3}\frac{\delta M^2}{(\epsilon_{\bf k}^e+\epsilon_{-{\bf k}}^e)(\epsilon_{\bf k}^e
+\epsilon_{-{\bf k}}^{-e})(\epsilon_{\bf k}^e-\epsilon_{\bf k}^{-e})} \, ,
\eea
\end{subequations}
and
\be
J(0,\mu,\Lambda) = -\frac{1}{2}\sum_{e=\pm}\int\frac{d^3{\bf k}}{(2\pi)^3}\epsilon_{\bf k}^e \, , 
\ee
where $\epsilon_{\bf k}^e$ are the positive solutions to Eq.~(\ref{detS}), which depend on the angle between the 
momentum of the excitation and the superflow. Next, we need to regularize $I^\pm(0,\mu,\Lambda)$ and $J(0,\mu,\Lambda)$. 
The divergent contributions of these
integrals are shown in Table \ref{table0}. The integrals $I^\pm(0,\mu,\Lambda)$ show exactly the same divergences as for the case without superflow
discussed in Sec.\ \ref{sec:cond-noflow}. In $J(0,\mu,\Lambda)$, however, there is a divergent contribution that depends explicitly on $\nabla\psi$. 
This divergence is exactly the same as for the uncondensed case discussed in Sec.\ \ref{sec:normal}. In that case, the $\nabla\psi$ dependent divergence 
in the pressure was spurious because after regularization with the proper time method the integrals in pressure and self-energy did not depend on $\nabla\psi$
anymore. 

One might think that, in order to regularize the divergent integrals, we should subtract the same integrals at the point $\mu=T=\nabla\psi=0$.
However, this procedure would not take care of the $\nabla\psi$ dependent divergence. Thus we seem to be forced to subtract the integrals at the point
$\mu=T=0$ with $\nabla\psi$ kept fixed, i.e., $J(0,\mu,{}\Lambda)= J(0,0,{}\Lambda) + J_\mu(0{})$, which reads 
\be \label{Jalter}
J(0,\mu,\Lambda) = -\frac{1}{2}\sum_{e=\pm}\int\frac{d^3{\bf k}}{(2\pi)^3}\epsilon_{\bf k}^e(\mu=0)
-\frac{1}{2}\sum_{e=\pm}\int\frac{d^3{\bf k}}{(2\pi)^3}[\epsilon_{\bf k}^e
-\epsilon_{\bf k}^e(\mu=0)] \, , 
\ee
and analogously for $I^\pm(0,\mu,\Lambda)$. The $\mu=0$ dispersion turns out to be
\be \label{rootroot}
\epsilon_{\bf k}^e(\mu=0) = \sqrt{k^2+M^2+(\nabla\psi)^2\mp\sqrt{4(\nabla\psi\cdot{\bf k})^2+\delta M^4}} \, .
\ee
The presence of the two square roots in this expression renders a straightforward application of the proper time regularization very complicated
and one would have to proceed numerically.

We notice, however, that there is another way to treat the ultraviolet divergences, using the same proper time regularization. 
Since the structure of the divergences is a simple combination of the 
divergences of the cases discussed above, it is easy to ``guess'' a generalization of the subtraction terms to the present case,
\begin{subequations} \label{Ipm0muL}
\bea
I^+(0,\mu,\Lambda) &=& \frac{1}{2}\sum_{e=\pm}\int\frac{d^3{\bf k}}{(2\pi)^3}\frac{1}{\omega_{\bf k}^e} + \sum_{e=\pm}\int\frac{d^3{\bf k}}{(2\pi)^3}
\left[2\frac{(\epsilon_{\bf k}^e)^2-k^2-M^2+\sigma^2}{(\epsilon_{\bf k}^e+\epsilon_{-{\bf k}}^e)(\epsilon_{\bf k}^e+\epsilon_{-{\bf k}}^{-e})
(\epsilon_{\bf k}^e-\epsilon_{\bf k}^{-e})} - \frac{1}{2\omega_{\bf k}^e}\right] \, , \\[2ex]
I^-(0,\mu,\Lambda)  &=& \frac{1}{2}\sum_{e=\pm}\int\frac{d^3{\bf k}}{(2\pi)^3}\frac{e}{\omega_{\bf k}^e} + \sum_{e=\pm}\int\frac{d^3{\bf k}}{(2\pi)^3}
\left[2\frac{\delta M^2}{(\epsilon_{\bf k}^e+\epsilon_{-{\bf k}}^e)(\epsilon_{\bf k}^e+\epsilon_{-{\bf k}}^{-e})(\epsilon_{\bf k}^e-\epsilon_{\bf k}^{-e})}
- \frac{e}{2\omega_{\bf k}^e}\right] \, ,
\eea
\end{subequations}
and
\be \label{J0muL}
J(0,\mu,\Lambda) = -\frac{1}{2}\sum_{e=\pm}\int\frac{d^3{\bf k}}{(2\pi)^3}\omega_{\bf k}^e-\frac{1}{2}\sum_{e=\pm}\int\frac{d^3{\bf k}}{(2\pi)^3}(\epsilon_{\bf k}^e
-\omega_{\bf k}^e) \, . 
\ee
Here, 
\be \label{omegae}
\omega_{\bf k}^e \equiv \sqrt{({\bf k}+e\nabla\psi)^2+M^2+e\delta M^2}  
\ee
is simply the $\mu=0$ dispersion of the uncondensed phase in the presence of a $\nabla\psi$, see Eq.\ (\ref{epskee}), generalized to two different mass parameters 
$M^2+\delta M^2$ and $M^2-\delta M^2$. It is also the $\mu=0$ dispersion of the condensed phase without $\nabla\psi$, see Eq.\ (\ref{I00Lam}), with $\nabla\psi$ added 
as a simple shift of the three-momentum. According to the structure of the divergences, it is clear that the second integrals on the right-hand sides 
of Eqs.\ (\ref{Ipm0muL}) and (\ref{J0muL}) are finite. And, the first integrals can be regularized with the proper time method just as in the previous subsections: 
the $\nabla\psi$ dependence drops out since the proper time integrals ``ignore'' this dependence, and the resulting cutoff-dependent terms together with the finite 
parts $I^\pm(\ell)_{\rm vac,finite}$, $J(\ell)_{\rm vac,finite}$ are exactly the same as in Sec.\ \ref{sec:cond-noflow}. Therefore, the renormalization
works as above, with exactly the same medium independent counterterms as given in Eqs.\ (\ref{d1d2}).

The choice of the subtraction term corresponds to a renormalization condition, and usually this term is the vacuum contribution. The appearance of the 
superflow in the divergent contributions appears to make the choice ambiguous, and it is not a priori clear whether using (\ref{Jalter})-(\ref{rootroot}) 
or (\ref{Ipm0muL})-(\ref{omegae}) is the correct physical choice. We are rather led to the conclusion that the very existence of the 
$\nabla\psi$ dependent divergence is problematic, because we seem to have found two renormalization conditions that differ in their predictions of how physical 
observables depend on 
the superflow. Here we only point out this problem, and leave its
solution to further studies. It will not affect our physical results because we shall restrict ourselves to weak coupling strengths where these ambiguous terms are 
negligibly small,
see discussion in Sec.\ \ref{sec:results1}.

In the main part we summarize the results of the renormalization procedure using Eq.\ (\ref{Ipm0muL})-(\ref{omegae}), see Eq.\ (\ref{stat1ren}) for the 
stationarity equation and Eq.\ (\ref{Psistatren}) for the pressure.

\section{Sound velocities at tree-level for arbitrary $m$}
\label{AppB}

In this appendix we derive the result (\ref{utree}) for the sound velocities in the tree-level approximation in the limit of vanishing superflow.
At tree-level and zero temperature the condensate is  
\be \label{zerorho}
\rho^2 \simeq  \frac{\sigma^2-m^2}{\lambda} \, , 
\ee
and the pressure becomes 
\be
\Psi \simeq  \frac{(\sigma^2-m^2)^2}{4\lambda} -T\int\frac{d^3{\bf k}}{(2\pi)^3}\ln\left(1-e^{-\epsilon_{\bf k}/T}\right) \,, 
\ee
where $\epsilon_{\bf k}$ is the dispersion relation of the Goldstone mode, containing the superflow; 
the massive mode only becomes relevant at higher temperatures and can be neglected.  With the help of Eq.\ (\ref{ns1}) we write
\be \label{ff}
n_s = \mu\frac{\mu^2-m^2}{\lambda} -\mu \int\frac{d^3{\bf k}}{(2\pi)^3}\left\{\left(\frac{\partial\epsilon_{\bf k}}{\partial|\nabla\psi|}\right)^2
\frac{f(\epsilon_{\bf k})[1+f(\epsilon_{\bf k})]}{T}-\frac{\partial^2\epsilon_{\bf k}}{\partial|\nabla\psi|^2}f(\epsilon_{\bf k})\right\}_{|\nabla\psi|\to 0} \, .
\ee
Using that the dispersion $\epsilon_{\bf k}$ is given by the zeros of the 
determinant of the inverse tree-level propagator (\ref{S0inv}), 
\be
{\rm det}\,S_0^{-1} = -K^2[-K^2+2(\sigma^2-m^2)]-4(K_\mu\partial^\mu\psi)^2 \, ,
\ee
where Eq.\ (\ref{zerorho}) has been used, we find 
\begin{subequations}
\bea
\left.\frac{\partial\epsilon_{\bf k}}{\partial|\nabla\psi|}\right|_{\nabla\psi=0} &=& -\frac{2\mu k_\parallel}{\sqrt{4\mu^2k^2+(3\mu^2-m^2)^2}} \, , \\[2ex]
\left.\frac{\partial^2\epsilon_{\bf k}}{\partial|\nabla\psi|^2}\right|_{\nabla\psi=0} &=& 
\frac{\epsilon_k^2-2k_\parallel^2-k^2}{\epsilon_k\sqrt{4\mu^2k^2+(3\mu^2-m^2)^2}}
+ \frac{8\mu^2k_\parallel^2}{\epsilon_k[4\mu^2k^2+(3\mu^2-m^2)^2]} + \frac{4\mu^2k_\parallel^2(3\epsilon_k^2-k^2-3\mu^2+m^2)}
{\epsilon_k[4\mu^2k^2+(3\mu^2-m^2)^2]^{3/2}} \, .
\eea
\end{subequations}
Here, $k_\parallel$ is the longitudinal component of the momentum with respect to the superflow, and
\be
\epsilon_k = \sqrt{k^2+3\mu^2-m^2-\sqrt{4\mu^2k^2+(3\mu^2-m^2)^2}} 
\ee
is the dispersion of the Goldstone mode at vanishing superflow. Now the only nontrivial angular integration is the one over $k_\parallel^2$,
\be
\int\frac{d\Omega}{4\pi} k_\parallel^2 = \frac{k^2}{3} \, .
\ee
For low temperatures, we can expand the integrand in Eq.\ (\ref{ff}) for small $k$. We find 
\be
\mu^2\int\frac{d\Omega}{4\pi}\left(\frac{\partial\epsilon_{\bf k}}{\partial|\nabla\psi|}\right)^2_{\nabla\psi=0} \simeq q_1 k^2 + \frac{q_2}{\mu^2}k^4 \, , \qquad 
\mu^2\int\frac{d\Omega}{4\pi}\left(\frac{\partial^2\epsilon_{\bf k}}{\partial|\nabla\psi|^2}\right)_{\nabla\psi=0} \simeq p_1 k + \frac{p_2}{\mu^2}k^3 \, , \qquad 
\epsilon_k \simeq c_1 k + \frac{c_2^2}{\mu^2}k^3 \, , 
\ee
with the dimensionless coefficients 
\begin{subequations}\label{abc}
\bea
q_1&=& \frac{4\mu^4}{3(3\mu^2-m^2)^2} \, , \qquad q_2 = -\frac{16\mu^8}{3(3\mu^2-m^2)^4} \, , \\[2ex]
p_1&=& -\frac{2\mu^2(4\mu^2-m^2)}{3(\mu^2-m^2)^{1/2}(3\mu^2-m^2)^{3/2}} \, ,\qquad p_2 = \frac{2\mu^6(5\mu^4-6\mu^2m^2+2m^4)}{(\mu^2-m^2)^{3/2}(3\mu^2-m^2)^{7/2}}
\, , \\[2ex]
c_1&=& \frac{(\mu^2-m^2)^{1/2}}{(3\mu^2-m^2)^{1/2}} \, , \qquad c_2 = \frac{\mu^6}{(\mu^2-m^2)^{1/2}(3\mu^2-m^2)^{5/2}} \, . 
\eea
\end{subequations}
This yields 
\bea \label{ns2}
n_s \simeq \mu\frac{\mu^2-m^2}{\lambda} -\frac{\pi^2T^4}{6\mu c_1^4}\left[\frac{1}{5}\left(\frac{4q_1}{c_1}-p_1\right)+\frac{8\pi^2T^2}{\mu^2c_1^2}\left(\frac{2}{7}
\frac{q_2+p_1c_2}{c_1}-\frac{2q_1c_2}{c_1^2}-\frac{p_2}{21}\right)\right]\, ,
\eea
where the integrals 
\bea
&&\int_0^\infty dy\,\frac{y^3}{e^y-1}=\frac{\pi^4}{15} \, , \qquad  \int_0^\infty dy\,\frac{y^4e^y}{(e^y-1)^2}=\frac{4\pi^4}{15}\, , \\[2ex]  
&&\int_0^\infty dy\,\frac{y^5}{e^y-1}=\frac{8\pi^6}{63} \, , \qquad \int_0^\infty dy\,\frac{y^6e^y}{(e^y-1)^2}=\frac{16\pi^6}{21}
 \, , \qquad \int_0^\infty dy\,\frac{y^7e^y(e^y+1)}{(e^y-1)^3}=\frac{16\pi^6}{3}
\eea
for the dimensionless variable $y=c_1k/T$ have been used. Inserting the coefficients from Eqs.\ (\ref{abc}) into Eq.\ (\ref{ns2}) 
yields the result for the superfluid density 
\bea
n_s \simeq \mu\frac{\mu^2-m^2}{\lambda} - \frac{\pi^2T^4}{9\mu}\left[\frac{\mu^2(12\mu^2-m^2)(3\mu^2-m^2)^{1/2}}
{5(\mu^2-m^2)^{5/2}}-\frac{8}{7}\left(\frac{\pi T}{\mu}\right)^2\frac{\mu^6(57\mu^4-24\mu^2m^2+2m^4)}{(\mu^2-m^2)^{9/2}(3\mu^2-m^2)^{1/2}}\right] \, .
\eea
The $m=0$ limit of this result is in agreement with Eq.\ (79a) of Ref.\ \cite{2013PhRvD..87f5001A}. 
The pressure, evaluated at $|\nabla\psi|=0$ becomes (see appendix C of  
Ref.\ \cite{2013PhRvD..87f5001A})
\bea 
\Psi \simeq \frac{(\mu^2-m^2)^2}{4\lambda} + \frac{\pi^2 T^4}{90}\left[\frac{(3\mu^2-m^2)^{3/2}}{(\mu^2-m^2)^{3/2}}-
\frac{40}{7}\left(\frac{\pi T}{\mu}\right)^2\frac{\mu^6(3\mu^2-m^2)^{1/2}}{(\mu^2-m^2)^{7/2}}\right] \, .
\eea
This is all we need to compute the sound velocities: we can now straightforwardly take all relevant derivatives of the pressure, compute the normal fluid 
density via $n_n=n-n_s$, and insert the results into the equation for the sound velocities (\ref{au}). The result for $u_1$ and $u_2$ is given in Eqs.\ (\ref{utree})
in the main text.

\bibliography{refs}

\end{document}